\documentclass[12pt]{article}
\usepackage{graphicx, amssymb, amsthm, amsbsy, amsmath}
\usepackage{xcolor}
\usepackage{url}
\usepackage[colorlinks=true,allcolors=blue]{hyperref}%
\usepackage{algorithm}
\usepackage{multirow}
\usepackage{algorithmic}
\usepackage{natbib}
\usepackage{epsfig}
\usepackage{subfigure}

\usepackage[margin=1in]{geometry}

\DeclareGraphicsExtensions{.pdf,.eps.gz,.eps,.epsi.gz,.epsi,.ps,.ps.gz}

\newcommand{\mE}{\mathcal{E}}
\newcommand{\mX}{\mathcal{X}}
\newcommand{\dd}{\mathrm{d}}
\newcommand{\bm}{\boldsymbol}
\newcommand{\bbeta}{{\bm\beta}}
\newcommand{\bOmega}{{\bm\Omega}}
\newcommand{\bomega}{{\bm\omega}}
\newcommand{\bgamma}{{\bm\gamma}}
\newcommand{\bLambda}{{\bm\Lambda}}

\newcommand{\blambda}{{\bm\lambda}}

\newcommand{\balpha}{{\bm\alpha}}
\newcommand{\bphi}{{\bm\phi}}
\newcommand{\bSigma}{{\bm\Sigma}}

\newcommand{\bnu}{{\bm\nu}}

\newcommand{\bPsi}{{\bm\Psi}}

\def\A{{\bf A}}
\def\N{{\bf N}}

\def\B{{\bf B}}
\def\C{{\bf C}}

\def\G{{\bf G}}
\def\V{{\bf V}}
\def\g{{\bm g}}

\def\f{{\bf f}}

\def\M{{\bf M}}
\def\b{{\bf b}}
\def\I{{\bf I}}

\def\mS{{\mathcal S}}
\def\u{{\bf u}}

\def\V{{\bf V}}

\def\X{{\bf X}}
\def\x{{\bf x}}

\def\1{{\bf 1}}
\def\0{{\bf 0}}
\def\R{{\bf R}}
\def\eop{\hfill $\Box$}

\newtheorem{thm}{Theorem}%
\newtheorem{lemma}{Lemma}%
\newtheorem{cor}{Corollary}%
\newtheorem{assumption}{{\bf Assumption}}
\newtheorem{prop}{Proposition}

\begin{document}
\title{Latent Network Structure Learning from High Dimensional Multivariate Point Processes
\vspace{0.25in}}
\date{}
\author{%
{Biao Cai, Jingfei Zhang and Yongtao Guan}%
\vspace{1.6mm}\\
\fontsize{11}{10}\selectfont\itshape
\,Department of Management Science, Miami Herbert Business School, \\ 
\fontsize{11}{10}\selectfont\itshape
University of Miami, Coral Gables, FL, 33146. \\
}
\maketitle

\begin{abstract}
Learning the latent network structure from large scale multivariate point process data is an important task in a wide range of scientific and business applications. For instance, we might wish to estimate the neuronal functional connectivity network based on spiking times recorded from a collection of neurons. To characterize the complex processes underlying the observed data, we propose a new and flexible class of nonstationary Hawkes processes that allow both excitatory and inhibitory effects. We estimate the latent network structure using an efficient sparse least squares estimation approach.
Using a thinning representation, we establish concentration inequalities for the first and second order statistics of the proposed Hawkes process. 
Such theoretical results enable us to establish the non-asymptotic error bound and the selection consistency of the estimated parameters. 
Furthermore, we describe a least squares loss based statistic for testing if the background intensity is constant in time. We demonstrate the efficacy of our proposed method through simulation studies and an application to a neuron spike train data set.

\end{abstract}

\newpage
\section{Introduction}
Large-scale multivariate point process data are fast emerging in a wide range of scientific and business applications.
Learning the latent network structure from such data has become an increasingly important task. For instance, one may wish to estimate the neuronal functional connectivity network based on spiking times (i.e., times when a neuron fires) recorded from a collection of neurons \citep{farajtabar2015coevolve}, or to estimate the financial network based on trading times recorded for a collection of stocks \citep{linderman2014discovering}. Both the neuron spiking times and the trading times can be viewed as realizations from multivariate point processes. To characterize the latent interactions between the different point processes, a useful class of models is the multivariate Hawkes process \citep{hawkes1971spectra}. The multivariate Hawkes process is a mutually-exciting point process, in which the arrival of one event in one point process may trigger those of future events across the different processes. Because of its flexibility and interpretability, the multivariate Hawkes process has been widely used in many applications, such as social studies \citep{zhou2013learning}, criminology \citep{linderman2014discovering}, finance \citep{bacry2013modelling} and neuroscience \citep{okatan2005analyzing}. In the network setting, each component point process of the multivariate Hawkes process is viewed as a node. A directed edge connecting two nodes indicates an event in the source point process increases the probability of occurrence of future events in the target point process. Recently, work such as \cite{xu2016learning} connected such networks to the notion of Granger causality.

Despite the popularity of the multivariate Hawkes process, there is a need of new statistical theory and methodology for its broader applications. \textit{First}, most existing theoretical results for the Hawkes process are derived using a cluster process representation of the process. This cluster process representation by its definition depends on the mutually excitation assumption, that is, the arrival of one event increases the probability of occurrence of future events \citep{hawkes1974cluster,hansen2015lasso}. However, such an assumption may not be valid in certain applications.  For example, it is well known that the firing activity of one neuron can inhibit the activities of other neurons \citep{amari1977dynamics}. A more flexible model should allow both excitatory and inhibitory effects, which renders the cluster process representation infeasible. \textit{Second}, most existing models assume that the background intensities, i.e., the baseline arrival rates of events from the different component processes, are constant in time. Under this assumption, the multivariate Hawkes process satisfies a stationary condition \citep[e.g.,][]{bremaud1996stability}. However, assuming constant background intensities may also be too restrictive in practice. For example, stock trading activities tend to be much higher during market opens and closes \citep{engle1998autoregressive}, and the associated background intensities are therefore not constant in time. A multivariate Hawkes process with constant background intensities may not fit such data well \citep{chen2013inference}. 
A more flexible approach instead should allow the background intensities to be time-varying. For such nonstationary models, new development on both theory and methodology is needed, as most existing results are established assuming the underlying process to be stationary.

Some existing work have considered broadening the class of Hawkes process models.
Specifically, \cite{bremaud1996stability,chen2017multivariate,costa2018renewal} considered a class of nonlinear Hawkes processes that allows both excitatory and inhibitory effects. A thinning process representation was used to investigate the properties of the proposed process. However, these work focused on processes with constant background intensities and the thinning representation technique depended critically on the stationarity condition. They derived the concentration inequality for the second order statistics of the process, and subsequently considered the latent network structure estimation. Similar to the above contributions, the proposed method and derived results also required the stationarity condition.
Some recent work also considered nonstationary Hawkes processes. \cite{lewis2011nonparametric, chen2013inference,roueff2016locally} considered Hawkes processes with time varying background intensities. However, they only considered univariate processes, and only with excitatory effects. 
\cite{lemonnier2014nonparametric} considered a multivariate Hawkes process with time varying background intensities. However, they focused on an approximate optimization algorithm for model estimation, and did not provide any theoretical results.

In this article, we propose a flexible class of multivariate Hawkes process that admits time-varying background intensities and allows both excitatory and inhibitory effects. {We show the existence of a thinning process representation of this nonstationary process. Such a result has not yet been established in the literature, and it enables our subsequent theoretical analysis.} 
To estimate the network structure, we consider a computationally efficient penalized least squares estimation, in which both the background intensities and the transfer functions are approximated using basis functions. 
We establish theoretical properties of the penalized least squares estimator in the high-dimensional regime, where the dimension of the multivariate process $p$ can grown much faster than the length of the observation window $T$. 
Specifically, we investigate the following properties in our analysis:
\begin{description}
\item1. (Concentration inequalities.) We establish concentration inequalities for the first and second order statistics of the proposed Hawkes process. {Such inequalities are established using a new thinning process representation result for nonlinear and nonstationary Hawkes processes.} 
\item2. (Non-asymptotic error bound.) Under certain regularity conditions, we establish, in the high-dimensional regime, the non-asymptotic error bound of the intensity functions estimated using basis approximations. {Specifically, we verify that the design matrix satisfy a restricted eigenvalue condition and a bounded eigenvalue condition for the diagonal blocks; these bounds on eigenvalues depend on the number of basis functions.}
\item3. (Network recovery.) We show that, under certain regularity conditions, our proposed estimation method can consistently identify the true edges in the network with probability tending to one. {Moreover, we propose a consistent generalized information criterion (GIC) for regularizing parameter selection.}
\item4. (Test for background intensity.) We propose a least squares based statistic for testing if the background intensity is constant in time. {Specifically, we show that the null distribution of the test statistic is asymptotically $\chi^2$ and the test is powerful against alternatives.}
\end{description} 

It is worth mentioning that there is another class of approaches that estimate the latent network structure from high dimensional multivariate point process data \citep{zhang2016statistical, vinci2016separating, vinci2018adjusted}. These methods divide the observation window into a number of bins, and model the number of events in each bin. The network structure is estimated using methods such as correlation of event counts \citep{vinci2016separating}, regularized generalized linear models \citep{zhang2016statistical}, or Gaussian graphical models \citep{vinci2018adjusted}. Such a type of approaches do not involve estimating the multivariate intensity function underlying the observed point patterns, and are potentially more computationally efficient. However, the heuristic binning procedure may lose important information. For example, short-term excitatory effects may be overlooked if the bins are chosen to be too wide.
Choosing an appropriate binning procedure and understanding its effect on the subsequent model estimation and analysis remain a challenging task.

The rest of the article is organized as follows. 
Section~\ref{sec::model} introduces the proposed model, and Section~\ref{sec::est} describes the model estimation and selection. The aforementioned theoretical results are detailed in Sections~\ref{sec::theory}. Section~\ref{sec::sim} includes simulation studies. The detailed analysis of a neuron spike train dataset is presented in Section~\ref{sec::spike}. A short discussion section concludes the article.


\section{Model}
\label{sec::model} 
\subsection{Notation}
We start with some notation. Given a function $f$ on $\mX\in\mathbb{R}$, let $\Vert f\Vert_{\infty,\mX}=\sup_{t\in\mX}|f(t)|$ and $\Vert f\Vert_{2,\mX}=\{\int_{t\in\mX}f(t)^2\dd t\}^{1/2}$(or, respectively, $\Vert f\Vert_{\infty}$ and $\Vert f\Vert_{2}$, when there is no ambiguity).
Let $f^{(k)}$ denote the $k$th derivative of a function $f$ when such a derivative exists. 
For a matrix $\A\in\mathbb{R}^{m\times n}$, we use $\Vert\A\Vert_2$, $\Vert\A\Vert_{\max}$ and $\Vert\A\Vert_{\infty}$ to denote its spectral norm, maximum entry-wise $\ell_1$ norm and maximum row-wise $\ell_1$ norm, respectively.
We write $[n]=\{1,2,\ldots,n\}$ and let $\lfloor x \rfloor$ denote the largest integer less than $x$.
For a set $\mS$, we use $\vert\mS\vert$ to denote the cardinality of $\mS$.
We write $\mathbf{1}_n$ to denote a length-$n$ vector of 1, $\I_{n\times n}$ to denote a $n\times n$ identity matrix, $\text{diag}\{d_1,\ldots,d_n\}$ to denote a $n\times n$ diagonal matrix with diagonal elements $d_1,\ldots,d_n$, and use $\sigma_{\min}(\cdot)$ and $\sigma_{\max}(\cdot)$ to denote the smallest and largest eigenvalues of a matrix, respectively. For two positive sequences $a_n$ and $b_n$, write $a_n=\mathcal{O}(b_n)$ if there exist $c>0$ and $N>0$ such that $a_n<cb_n$ for all $n>N$, write $a_n\asymp b_n$ if $a_n=\mathcal{O}(b_n)$ and $b_n=\mathcal{O}(a_n)$, and $a_n=o(b_n)$ if $a_n/b_n\rightarrow 0$ as $n\rightarrow\infty$. 
For a sequence of random variables $Y_n$ and a positive sequence $a_n$, we write $Y_n=\mathcal{O}_p(a_n)$ if for any $\epsilon>0$, there exist $M>0$ and $N>0$ such that $\mathbb{P}(\vert Y_n/a_n\vert> M)<\epsilon$ for any $n>N$; we write $Y_n=o_p(a_n)$ if $\lim_{n\rightarrow\infty}\mathbb{P}(\vert Y_n{/a_n}\vert\ge\epsilon)=0$ for any $\epsilon>0$.

\subsection{The multivariate Hawkes process model}\label{subsec::model} 
Consider a directed network with $p$ nodes. For each node $j\in[p]$, we observe its event locations $\{t_{j,1}, t_{j,2},\ldots\}$ in the time interval $[0,T]$ such that $0<t_{j,1}<t_{j,2}<\cdots\le T$.
For node $j$, let the associated counting process be {$N_j(t)=\vert\{i: t_{j,i}\le t\}\vert$, $t\in[0,T]$.} 
Write $\N=(N_j)_{j\in[p]}$ as the $p$-variate counting process. 
Let $\mathcal{H}_t$ denote the entire history of $\N$ up to time $t$, and write $N_j\left([t,t+\dd t)\right)$ as $\dd N_j(t)$.
The $p$-variate intensity function $\blambda (t)=(\lambda_{1}(t),\ldots,\lambda_{p}(t))^\top$ of $\N$ is defined as
$$
\lambda_{j}(t)\dd t=\mathbb{P}(\dd N_j(t)=1|\mathcal{H}_t),\quad j\in[p]. 
$$
We propose a flexible class of Hawkes processes with intensity functions defined as
\begin{equation}\label{intensity}
\lambda_{j}(t)=h\left\{\nu_{j}(t)+\sum_{k=1}^{p}\int_{0}^{t}\omega_{j,k}(t-u)\dd N_{k}(u)\right\}, \quad j\in[p],
\end{equation}
where $h(\cdot):\mathbb{R}\rightarrow \mathbb{R}^+$ is a link function and it is assumed to be $\theta$-Lipschitz (see Assumption \ref{ass1}), $\nu_{j}(\cdot):\mathbb{R}^+\rightarrow \mathbb{R}^+$ is the time-varying background (or baseline) intensity function of the $j$th process, and $\omega_{j,k}(\cdot):\mathbb{R}^{+}\rightarrow \mathbb{R}$ is the transfer function that characterizes the effect of the $k$th process on the $j$th process. 
Specifically,  
\begin{description}
\item[i.] $\omega_{j,k}(s)>0$ corresponds to \textit{excitatory} effect, that is, an event in process $k$ increases the probability of event occurrence in process $j$ at a time distance of $s$.
\item[ii.] $\omega_{j,k}(s)<0$ corresponds to \textit{inhibitory} effect, that is, an event in process $k$ decreases the probability of event occurrence in process $j$ at a time distance of $s$.
\item[iii.] $\omega_{j,k}(s)=0$ corresponds to no effect, that is, an event in process $k$ has no effect on the event occurrence in process $j$ at a time distance of $s$.
\end{description}
The proposed model in \eqref{intensity} considers a time dependent background intensity function instead of the constant background intensity considered in existing multivariate Hawkes process models \citep{hansen2015lasso,chen2013inference,bacry2020sparse,wang2020statistical}. Consequently, the proposed Hawkes process model is nonstationary, and its analysis requires new theoretical tools, which will be introduced in Section \ref{sec::theory}.

Let the directed network $\mathcal{G}(\mathcal{V},\mathcal{E})$ summarize the relationships between the $p$ component processes. 
Specifically, let $\mathcal{V}=\{1,2,\ldots,p\}$ be the set of $p$ nodes and $\mathcal{E}$ be the set of edges such that 
$$
\mathcal{E}=\{(j,k):\omega_{j,k}\neq 0,\; j,k\in[p]\},
$$
where $\omega_{j,k}$'s are the transfer functions in \eqref{intensity}. 
Therefore, $(j,k)\in \mathcal{E}$ if and only if the $k$th process has an excitatory or inhibitory effect on the $j$th process.

Next, we introduce a set of regularity conditions on the background intensities and transfer functions in \eqref{intensity}. 
\begin{assumption}\label{ass1}
Let $\bOmega$ be a $p\times p$ matrix with $\Omega_{jk}=\int_{0}^{\infty}|\omega_{j,k}(t)|\dd t$, $j,k\in[p]$ and assume that $\sigma_{max}(\bOmega^\top\bOmega)\leq\sigma_{\bOmega}<1$. Moreover, {assume that $h(\cdot)$ is a $\theta$-Lipschitz link function with $\theta\le 1$}, and the background intensity functions are bounded, i.e., $0<\nu_j(t)\le\nu$, $j\in[p]$, for some positive constant $\nu$. 
\end{assumption}
\noindent 
{Assuming the Lipschitz constant $\theta$ to satisfy $\theta\le 1$ is not restrictive. For example, if $h(\cdot)$ is $K_0$-Lipschitz for some $K_0>1$, we can reparameterize \eqref{intensity} by setting $\tilde h(x)=h(x/K_0)$, $\tilde \nu_j(t)=K_0\nu_j(t)$ and $\tilde\omega_{j,k}(t)=K_0\omega_{j,k}(t)$. In this reparameterized model, $\tilde h(\cdot)$ is 1-Lipschitz. The Lipschitz condition on the link function was also considered in \cite{massoulie1998stability} and \cite{chen2017multivariate}.}
Assumption \ref{ass1} implies that $h\{\nu_j(t)\}$ is bounded as Lipschitz functions are bounded on bounded supports.

Define the mean intensity of \eqref{intensity} as $\bar\lambda_j(t)=\mathbb{E}\{\dd N_j(t)\}/\dd t$. Under Assumption \ref{ass1}, the mean intensity $\bar\lambda_j(t)$ is upper bounded. 
{
This can be shown in three steps. 
First, define a $p$-dimensional Hawkes process $\N^*=(N^*_j)_{j\in[p]}$ with intensity function
\begin{equation}\label{ibound}
\lambda^*_{j}(t)=\nu^*+\sum_{k=1}^{p}\int_{0}^{t}|\omega_{j,k}(t-u)|\dd N^*_{k}(u), \quad j\in[p],
\end{equation}
where $\nu^*$ is a positive constant such that $\nu^*\ge h\{\nu_j(t)\}$ for any $j$ and $t$, and $\omega_{j,k}$'s are as defined in \eqref{intensity}. 
By \citet{bremaud1996stability}, the point process defined in \eqref{ibound} satisfies a stationary condition under Assumption \ref{ass1}.
Next, write the mean intensity of \eqref{ibound} as $\bLambda^*=(\Lambda^*_1, \ldots, \Lambda^*_p)^\top$, where $\Lambda^*_j=\mathbb{E}\{\dd N^*_j(t)\}/\dd t$. Correspondingly, we have
\begin{equation}\label{mean}
\bLambda^*=\bnu^*+\left\{\int_{0}^{\infty}|\bomega(t)|\dd t\right\}\bLambda^*,
\end{equation}
where $\bnu^*=(\nu^*,\ldots,\nu^*)^\top\in\mathbb{R}^p$ and $\bomega(t)\in\mathbb{R}^{p\times p}$, with $\{\bomega(t)\}_{jk}=\omega_{j,k}(t)$. 
The mean intensity $\bLambda^*$ in \eqref{mean} can be rewritten as $\bLambda^*=\sum_{k=0}^\infty\bOmega^k\bnu^*$, which is upper bounded given $\sigma_{\bOmega}<1$ in Assumption \ref{ass1}. 
Lastly, it can be shown that the mean intensity of \eqref{intensity}, i.e., $\bar\lambda_j(t)$, is upper bounded by $\Lambda^*_j$ (see Lemma \ref{lemma:dom} and its proof in the Supplementary Materials). Consequently, $\bar\lambda_j(t)$ is also upper bounded under Assumption \ref{ass1}.
}


\section{Estimation}
\label{sec::est} 
From the observed event locations in $[0,T]$, our objective is to estimate the intensity functions $\lambda_j(t)$, $j\in[p]$. 
Furthermore, by identifying the non-zero transfer functions $\omega_{j,k}$'s in the estimated intensity functions, we can {estimate the structure of the} directed network $\mathcal{G}(\mathcal{V},\mathcal{E})$. 
In this section, we consider the link function to be $h(x)=\max(0,x)$ in \eqref{intensity}. It is seen that this link function is a 1-Lipschitz function.
To ease notation, we write
\begin{equation}\label{estimate}
\psi_{j}(t)=\nu_{j}(t)+\sum_{k=1}^{p}\int_{0}^{t}\omega_{j,k}(t-u)\dd N_{k}(u),
\end{equation}
and $\lambda_{j}(t)=\max\{0,\psi_{j}(t)\}$, $j\in[p]$.

To estimate the intensity functions, one may consider a likelihood function based approach \citep{ogata1981lewis,chen2013inference,zhou2013learning}. 
However, the surface of the negative loglikelihood function of the Hawkes processes can be complex, especially when $p$ is large, and highly nonconvex \citep{wang2016isotonic}. Moreover, minimizing the negative loglikelihood function requires a very involved and computationally intensive iterative procedure \citep{veen2008estimation}. To overcome the challenge of nonconvexity and improve the estimation efficiency, we consider a least squares loss based estimation approach. That is, we consider the following loss function 
\begin{equation}\label{sloss}
\frac{1}{T}\sum_{j=1}^p\int_{0}^T \{\psi_{j}^2(t)\dd t-2\psi_{j}(t)\dd N_j(t)\}.
\end{equation}
The least squares loss comes from the empirical risk minimization principle \citep{geer2000empirical} and has been fairly commonly considered in estimating point process models \citep{bacry2020sparse,hansen2015lasso,chen2017multivariate}.
We later show that \eqref{sloss} can be separated into $p$ convex objective functions that can be estimated individually, which significantly reduces the computation cost.

{We consider a nonparametric estimation of $\nu_{j}(t)$ and $\omega_{j,k}(t)$, $j,k\in[p]$, using B-spline approximations. 
Given $[a_1,a_2]\subset\mathbb{R}$ and a set of $K$ knots $a_1=\zeta_0<\zeta_1<\cdots<\zeta_{K+1}=a_2$ such that $\max_{1\le l\le K+1}|\zeta_k-\zeta_{k-1}|=\mathcal{O}(K^{-1})$, let $\mS_{K,l}$ be the space of polynomial splines of degree $l\ge 1$ consisting of functions satisfying: (i) restricting to each interval $[\zeta_i,\zeta_{i+1}]$, $i\in[K]$, the function is a polynomial of degree $l-1$; (ii) for $l\ge 2$ and $0\le l'\le l-2$, the function is $l'$ times continuously differentiable \citep{stone1985additive}. Such a space $\mS_{K,l}$ is of dimension $m=K+l$ \citep{schumaker2007spline} and as such, let $\{\phi_1(t),\ldots,\phi_m(t)\}$ be the normalized B-spline basis of $\mS_{K,l}$. When $l=1$, the basis is a set of $K+1$ step functions with jumps at knots \citep{stone1985additive}.}
In our procedure, we approximate the background intensity $\nu_{j}(t)$ with an $m_0$-dimensional normalized B-spline basis $\bphi_0(t)=(\phi_{0,1}(t),\ldots,\phi_{0,m_0}(t))^\top$, such that $\nu_{j}(t)=\bbeta_{j,0}\bphi_{0}(t)+r_{j,0}(t)$, where $\bbeta_{j,0}\in\mathbb{R}^{m_0}$ and $r_{j,0}(\cdot)$ denotes the approximation residual. Furthermore, we approximate the transfer functions $\omega_{j,k}(t)$ with an $m_1$-dimensional normalized B-spline basis $\bphi_1(t)=(\phi_{1,1}(t),\ldots,\phi_{1,m_1}(t))^T$, such that $\omega_{j,k}(t)=\bbeta_{j,k}\bphi_1(t)+r_{j,k}(t)$, where $\bbeta_{j,k}\in\mathbb{R}^{m_1}$ and $r_{j,k}(\cdot)$ denotes the approximation residual. 
The dimensions and degrees of the bases $\bphi_0(t)$ and $\bphi_1(t)$ are allowed to be different for more flexibility in characterizing the background intensities and transfer functions. For example, one may use cubic B-splines to approximate the background intensities and step functions to approximate the transfer functions \citep{hansen2015lasso}. 
The choices for the number and locations of knots are discussed in Section \ref{tuning}.

Write $\bbeta_j=(\bbeta_{j,0},\bbeta_{j,1},\ldots,\bbeta_{j,p})^\top$. 
We define $\balpha_j=(\balpha^{(j,0)},\balpha^{(j,1)},\ldots,\balpha^{(j,p)})^\top$ such that $\balpha^{(j,0)}\in\mathbb{R}^{m_0}$ with 
$$
\alpha^{(j,0)}_{l}=\frac{1}{T}\int_{0}^T \phi_{0,l}(t)\dd N_j(t), \quad l\in[m_0],
$$
and $\balpha^{(j,k)}\in\mathbb{R}^{m_1}$, $k\in[p]$ with
$$
\alpha^{(j,k)}_{l}=\frac{1}{T}\int_{0}^T \int_{0}^{t}\phi_{1,l}(t-u)\dd N_{k}(u)\dd N_j(t),\quad l\in[m_1].
$$
Moreover, we define $\G\in\mathbb{R}^{(m_{0}+pm_1)\times(m_{0}+pm_1)}$ such that 
\begin{equation}
\label{Gmatrix}
\G= \begin{pmatrix}
  \G^{(0,0)} & \G^{(0,1)} & \ldots & \G^{(0,p)} \\
  \G^{(1,0)} & \G^{(1,1)} & \ldots & \G^{(1,p)} \\
  \vdots  & \vdots  & \ddots & \vdots  \\
  \G^{(p,0)} & \G^{(p,1)}& \ldots & \G^{(p,p)}
 \end{pmatrix},
\end{equation}
where the component $\G^{(k_1,k_2)}$ is defined as
$$
\G^{(k_1,k_2)} =
\begin{cases}
  \frac{1}{T}\int_{0}^T \bphi_{0}(t)\bphi^\top_{0}(t)\dd t,   &\text{if } k_1=0, k_2=0, \\
  \frac{1}{T}\int_{0}^T \bphi_{0}(t)\left\{\int_0^t\bphi^\top_{1}(t-u)\dd N_{k_2}(u)\right\}\dd t, &\text{if } k_1=0, k_2\neq 0,  \\
  \frac{1}{T} \int_{0}^T \left\{\int_0^t\bphi_{1}(t-u)\dd N_{k_1}(u)\right\}\bphi^\top_{0}(t)\dd t, &\text{if } k_1\neq 0, k_2=0,  \\
  \frac{1}{T}\int_{0}^T \left\{\int_0^t\bphi_{1}(t-u)\dd N_{k_1}(u)\right\}\left\{\int_0^t\bphi^\top_{1}(t-u)\dd N_{k_2}(u)\right\}\dd t, &\text{if } k_1\neq 0, k_2\neq 0.  \\
\end{cases}
$$
With the above expressions for $\bbeta_j$, $\balpha_j$ and $\G$, we define 
\begin{equation}\label{loss}
\ell_j(\bbeta_j)\overset{\Delta}{=}-2\bbeta_j^\top\balpha_j+\bbeta_j^\top\G\bbeta_j.
\end{equation} 
Some straightforward algebra shows that the loss function in \eqref{sloss} can be written as $\sum_{j=1}^p\ell_j(\bbeta_j)$. 
We note that both $\balpha_j$ and $\G$ are calculated based on the observed event locations and the pre-specified basis functions, i.e., $\bphi_0(t)$ and $\bphi_1(t)$. Therefore, to estimate the background intensities and transfer functions, we can directly optimize \eqref{loss} with respect to $\bbeta_j$.
Since the loss function $\sum_{j=1}^p\ell_j(\bbeta_j)$ can be decomposed into $p$ separate convex loss functions, i.e., $\ell_1(\bbeta_1), \ldots,\ell_p(\bbeta_p)$, we can optimize each loss function separately.

Define $\mathcal{E}_j=\{k:\omega_{j,k}\neq 0, k\in[p]\}$. To estimate $\mathcal{E}_j$, we consider $\mathcal{\hat E}_j=\{k:\hat\omega_{j,k}\neq 0, k\in[p]\}$, where $\hat\omega_{j,k}$'s are the estimated transfer functions.
Estimating $\omega_{j,k}$'s from \eqref{loss} will result in a densely connected network, as $\hat\omega_{j,k}$ may not be exactly zero, even when process $k$ has no effect on process $j$.
Note that if $\omega_{j,k}=0$, then all coefficients associated with $\omega_{j,k}$ are zero (i.e., $\bbeta_{j,k}=0$). 
Thus, to encourage sparsity in the estimated network, we impose a standardized group lasso penalty on $\bbeta_j$, in which the coefficients in $\bbeta_{j,k}$ are grouped together, $k\in[p]$. 
Specifically, we consider the following optimization problem
\begin{equation}\label{obj}
\min_{\bbeta_j\in\mathbb{R}^{m_0+pm_1}}-2\bbeta_{j}^\top\balpha_{j}+\bbeta_{j}^\top \G\bbeta_{j}+\eta_{j}\sum_{k=1}^{p}\left(\bbeta_{j,k}^\top\G^{(k,k)}\bbeta_{j,k}\right)^{1/2}.
\end{equation} 
The penalty term $\sum_{k=1}^{p}\left({\bbeta_{j,k}}^\top\G^{(k,k)}\bbeta_{j,k}\right)^{1/2}$ is an extension of the standardized group lasso penalty \citep{simon2013sparse}.  
This optimization problem in \eqref{obj} is convex and can be efficiently solved using a block coordinate descent algorithm \citep{simon2013sparse}. 
The terms $\balpha_1,\ldots,\balpha_p$, and $\G$ can be computed using standard numerical integration methods and such calculations can be carried out before implementing the block coordinate descent algorithm.

\subsection{Tuning parameter selection}\label{tuning}
{Our proposed estimation procedure involves a number of tuning parameters, including the numbers of B-splines (i.e., $m_0$ and $m_1$) for approximating the background and transfer functions, respectively, knots locations for the B-splines, and tuning parameter $\eta_j$'s in the penalized least squares estimation in \eqref{obj}. Cross-validation procedures may not be appropriate for tuning parameter selections under our setting as the proposed process is nonstationary due to the time-varying background intensity in \eqref{intensity}. As such, the data cannot be divided into training and validation sets in a straightforward manner.

Given $m_0$ and $m_1$, we let the knots be evenly distributed \citep{ravikumar2009sparse,huang2010variable}. 
For $m_0$ and $m_1$, theoretical conditions in Theorem \ref{thm3} can guide their empirical choices. In Section \ref{sec:m0m1}, we describe a heuristic procedure for selecting $m_0$ and $m_1$; a similar procedure was considered in \citet{kozbur2020inference}. In Section \ref{sec:varym}, we show that this heuristic procedure achieves good performance; additionally, we demonstrate that the estimation accuracy is not overly sensitive to the choices of $m_0$ and $m_1$. Once $\bphi_0(t)$ and $\bphi_1(t)$ are determined, we then move to select $\eta_j$'s.

The tuning parameter $\eta_j$ in \eqref{obj} controls the sparsity of $\bbeta_j$, which in turn controls the sparsity of the estimated network. 
To select $\eta_j$, we propose a generalized information criterion (GIC) defined as
\begin{equation}\label{bic}
\text{GIC}(\eta_j)=\ell_j(\hat\bbeta_j)\cdot\kappa_j+(\alpha_T/T)\cdot |\hat{\mE}_j|
\end{equation}
where $\ell_j(\cdot)$ is as defined in \eqref{loss}, $\kappa_j=T/N_j\{(0,T]\}$ is a scaling parameter, $\hat\bbeta_j$ is estimated from \eqref{obj} with $\eta_j$, and $\alpha_T>0$ is a parameter that scales with $T$ and $p$. 
As $\ell_j(\hat\bbeta_j)$ is the squares loss and not the loglikelihood function, the GIC is not directly comparable to the likelihood based selection criteria such as the BIC or extended BIC \citep{schwarz1978estimating,chen2008extended}. In Theorem \ref{thm4.2}, we show that the proposed GIC is consistent given appropriate choices of $\alpha_T$, such as $\mathcal{O}((\log p)^2\log T)$. In Section \ref{sec::sim}, we evaluate the efficacy of the proposed GIC and show it achieves satisfactory performance. }


\section{Theoretical Properties}
\label{sec::theory}
In this section, we first show the existence of a thinning process representation of the proposed nonlinear and nonstationary Hawkes process. 
We then establish concentration inequalities for the first and second order statistics of the proposed point process. 
These results are useful in the subsequent analysis of the estimated intensity functions. 
The concentration inequalities are also of independent interest, as they provide important theoretical tools for analyzing statistical methods (e.g., regression analysis) that are applied to the proposed processes.
Next, we establish the non-asymptotic error bound of the intensity functions estimated using the proposed method and show that our method can consistently identify the true edges in the network. 
Lastly, we propose a test statistic for testing if the background intensities are constant in time. We derive its asymptotic null distribution and show the test is powerful against alternatives. All proofs are collected in the Supplementary Materials.

\subsection{Concentration inequalities}
\label{sec:con}
Studying the theoretical properties of the proposed class of Hawkes processes is challenging as most existing techniques are not applicable.
For example, many existing theoretical analyses rely on the cluster process representation of the Hawkes process \citep{hawkes1974cluster,bacry2020sparse,hansen2015lasso}, which assumes the transfer functions to be nonnegative.
In this case, the Hawkes process can be viewed as a sum of independent processes (or clusters), and theoretical properties of the Hawkes process, such as the concentration inequality of the first and second order statistics, can be investigated by studying the properties of the independent processes.
When negative transfer functions are permitted, however, the cluster process representation is no longer applicable. 
To overcome this challenge, \cite{bremaud1996stability} employed a thinning process representation of the Hawkes process, which did not require the transfer functions to be nonnegative. With this representation and a coupling result from \cite{dedecker2004coupling}, \cite{chen2017multivariate} bounded the temporal dependence of the Hawkes process and obtained a concentration inequality for the second order statistics. 
Both \cite{bremaud1996stability} and \cite{chen2017multivariate} require the stationarity condition of the process. 
Hence, the results and techniques in \cite{bremaud1996stability} and \cite{chen2017multivariate} are not directly applicable to our setting.  
In what follows, we first show the existence of a thinning process representation of the proposed nonlinear and nonstationary Hawkes process in \eqref{intensity}.

Let $\overline\N=(\overline{N}_{j})_{j\in[p]}$ be a $p$-variate homogeneous Poisson process on $\mathbb{R}^2$ with intensity 1.
Let $\lambda_{j}^{(0)}(t)=0$, $j\in[p]$, and $N_{j}^{(0)}=\varnothing$. 
For $n\ge1$, construct recursively $\blambda^{(n)}(t)=(\lambda^{(n)}_1(t),\ldots,\lambda^{(n)}_p(t))^\top$ and $\N^{(n)}=(N^{(n)}_{j})_{j\in[p]}$ as follows:
\begin{equation}
\label{thin}
\begin{split}
&\lambda_{j}^{(n+1)}(t)=h\left\{\nu_{j}(t)+\sum_{k=1}^{p}\int_{0}^{t}\omega_{j,k}(t-u)\dd N_{k}^{(n)}(u)\right\},\\
&\dd N_{j}^{(n+1)}(t)=\overline{N}_{j}\left(\left[0,\lambda_{j}^{(n+1)}(t)\right]\times\dd t\right),\quad j\in[p],
\end{split}
\end{equation}
where $h(\cdot)$, $\nu_j$ and $w_{j,k}$ are as defined in \eqref{intensity}, and $\overline{N}_{j}([0,\lambda_{j}^{(n+1)}(t)]\times\dd t)$ denotes the number of points for $\overline{N}_{j}$ in the area $[0,\lambda_{j}^{(n+1)}(t)]\times[t,t+\dd t]$.
It follows from Lemma \ref{thinexp} that $\lambda_{j}^{(n)}(t)$ is the intensity function of the point process $N^{(n)}(t)$.
Next, we show that the sequence $\{\N^{(n)}\}_{n=1}^{\infty}$ in \eqref{thin} converges in distribution to the Hawkes process $\N$ with intensity function \eqref{intensity}.

\begin{thm}
\label{thm1}
Let $\blambda(t)$ be as defined in \eqref{intensity} satisfying Assumption \ref{ass1}.
Let $\{\blambda^{(n)}(t)\}_{n=1}^{\infty}$ and $\{\N^{(n)}\}_{n=1}^{\infty}$ be sequences as defined in \eqref{thin}. 
Then, it holds that
\begin{description}
\item{(i)} $\blambda^{(n)}(t)$ converges to $\blambda(t)$ almost surely for any $t$,
\item{(ii)} $\left\{\N^{(n)}\right\}_{n=1}^{\infty}$ converges in distribution to $\N$ with intensity \eqref{intensity}.
\end{description}
\end{thm}
\noindent
Theorem~\ref{thm1} shows the existence of a thinning process representation of the proposed nonstationary Hawkes Process. This new result is critical in our subsequent theoretical analysis.
{Comparing with the thinning construction in \citet{chen2017multivariate}, at $n=1$, \citet{chen2017multivariate} starts the sequence with a stationary counting process $\dd N_{j}^{(1)}(t)=\overline{N}_{j}\left(\left[0,\mu_j\right]\times\dd t\right)$, $j\in[p]$, where $\mu_j$ is the constant background intensity, and \eqref{thin} starts from a nonstationary counting process $\dd N_{j}^{(1)}(t)=\overline{N}_{j}\left(\left[0,h\{\nu_{j}(t)\}\right]\times\dd t\right)$, $j\in[p]$. Formulas in the ensuing iterative construction remain similar for both representations, with the only difference being the background intensity in \cite{chen2017multivariate} is constant but is time-varying in \eqref{thin}.
Convergence of the iteratively constructed sequence in \cite{chen2017multivariate} is established in \cite{massoulie1998stability}, which ensures the validity of the thinning process representation. However, the result in \cite{massoulie1998stability} is established under a stationarity condition, which does not hold for our proposed nonstationary processes. Theorem \ref{thm1} thus provides a theoretical guarantee, analogous to \cite{massoulie1998stability}, for nonstationary multivariate Hawkes processes.}

Combining the result  in Theorem \ref{thm1} and the coupling technique in \cite{chen2017multivariate}, we are able to establish concentration inequalities for the first and second order statistics of the proposed Hawkes process. To that end, we introduce the following conditions.

\begin{assumption}\label{ass2}
Assume that there exists $\lambda_{\max}>0$ such that $\lambda_{j}(t)\leq\lambda_{\max}$ for any $t$ and $j$ and $\omega_{j,k}$, $j,k\in[p]$ are bounded functions with a bounded support $[0,b]$ for some $b>0$.
\end{assumption}
\noindent This condition first assumes that the intensities are bounded above by a constant. 
One example of such processes is when the link function $h(\cdot)$ is upper bounded by a positive constant.
Assumption \ref{ass2} also assumes that the transfer functions $\omega_{j,k}$'s have a bounded support. 
The bounded support assumption has been fairly commonly considered in the analysis of multivariate Hawkes process \citep{hansen2015lasso,costa2018renewal}.

\begin{assumption}\label{ass3}
There exists $\rho_{\bOmega}\in(0,1)$ such that $\sum_{k=1}^{p}\Omega_{j,k}\leq\rho_{\bOmega}$, $j\in[p]$.
\end{assumption}
\noindent This assumption requires that $\bOmega$ has bounded column sums, which prevents the intensity function from concentrating on any single process.

Recall that $\mathcal{H}_{t}$ denotes the history of $\N$ up to time $t$.
For $\mathcal{H}_{t}$-predictable functions $f_1(\cdot)$ and $f_2(\cdot)$, define
$$
y_{k}=\frac{1}{T}\int_{0}^T f_1(t)\dd N_{k}(t),
$$
$$
y_{j,k}=\frac{1}{T}\int_{0}^T \int_{0}^T f_2(t-t')\dd N_{k}(t')\dd N_{j}(t).
$$

\begin{thm}
\label{thm2}
Consider a Hawkes process on $[0,T]$ with intensity as defined in \eqref{intensity} satisfying Assumptions \ref{ass1}-\ref{ass3}.
Let $f_1(t)$ be a bounded function and $f_2(t)$ be a bounded function on a bounded support.
Then, for $k\in[p]$, it holds that
\begin{equation}\label{firstorder}
\mathbb{P}(|y_{k}-\mathbb{E}y_{k}|\geq c_{1} T^{-3/5})\leq c_{2}T\exp(-c_3T^{1/5}),
\end{equation}
where $c_{1}$, $c_{2}$ and $c_3$ are positive constants.
For any $j,k\in[p]$, it holds that
\begin{equation}\label{secondorder}
\mathbb{P}(|y_{j,k}-\mathbb{E}y_{j,k}|\geq c_{1}'T^{-2/5})\leq c_{2}' T \exp(-c_{3}'T^{1/5}),
\end{equation}
where $c_{1}'$, $c_{2}'$ and $c_{3}'$ are positive constants.
\end{thm}
\noindent
The proof of Theorem~\ref{thm2} is provided in the Supplementary Materials. This result is used frequently in deriving the non-asymptotic error bound of the estimated intensity functions and establishing edge selection consistency. In the proof, we first define a coupling process of $\N$ using results from Theorem~\ref{thm1}. This coupling process is used to bound the temporal dependence of $\N$. Finally, a Bernstein type inequality for weakly dependent sequences \citep{merlevede2011bernstein} is used to obtain the desired results.

Quantities such as $y_k$ and $y_{j,k}$ appear in many statistical problems such as regression analysis \citep{massart2000some} and clustering analysis \citep{chen2017multivariate}. Therefore, establishing concentration inequalities for $y_k$ and $y_{j,k}$ are important in the theoretical analysis of such methods. In the next corollary, we give an example of the application of Theorem~\ref{thm2}.

\begin{cor}\label{cor1}
Consider a Hawkes process on $[0,T]$ with intensity as defined in \eqref{intensity} satisfying Assumptions \ref{ass1}-\ref{ass3}.
Considering the matrix $\G$ defined in \eqref{Gmatrix}, we have
\begin{equation*}
\mathbb{P}\left[\bigcap_{i\neq j}\left\{\left|\G_{ij}-\mathbb{E}(\G_{ij})\right|\le c_{4}T^{-2/5}\right\}\right]\ge 1-c_{5}(p+1)^{2}T \exp(-c_{6}T^{1/5}),
\end{equation*}
where $c_{4}$, $c_{5}$ and $c_{6}$ are positive constants.
\end{cor}
\noindent
The result in Corollary~\ref{cor1} is a direct consequence of Theorem~\ref{thm2}, once we show that the entries in $\G$ are first and second order statistics of the proposed Hawkes process.

\subsection{Non-asymptotic error bound}
\label{sec:bound}
In this section, we derive the non-asymptotic error bound of the estimated intensity function in the diverging $p$ regime. 
To simplify notation, we define $\bPsi(t)=(\bPsi^\top_0(t),\bPsi^\top_1(t),\ldots,\bPsi^\top_p(t))^\top$, where $\bPsi_{0}(t)=\bphi_{0}(t)$ and $\bPsi_k(t)=\int_{0}^{t}\bphi_1(t-u)\dd N_{k}(u)$, $k\in[p]$. Correspondingly, it holds that $\G=\frac{1}{T}\int_0^T\bPsi(t)\bPsi^\top(t)\dd t$ and $\G^{(k,k)}=\frac{1}{T}\int_0^T\bPsi_k(t)\bPsi^\top_k(t)\dd t$.
Let $s=\max_j|\mE_j|$, where $\mE_j=\{k:  \omega_{j,k}\neq 0, k\in[p]\}$. 

{Recall the first order mean intensity function $\bar{\lambda}_{k}(u)$ is defined as $\bar{\lambda}_k(u)=\mathbb{E}(\dd N_k(u))/\dd u$, $k\in[p]$. 
For $k_1\neq k_2\in[p]$ and $k_1=k_2\in[p], \,u_1\neq u_2$, define the second order mean intensity function $\bar{\lambda}^{(2)}_{k_1,k_2}(u_1,u_2)$ as
\begin{equation}\label{eqn:2nd}
\bar{\lambda}^{(2)}_{k_1,k_2}(u_1,u_2)=\mathbb{E}\{\dd N_{k_1}(u_1)\dd N_{k_2}(u_2)\}/(\dd u_1\dd u_2).
\end{equation}
Denote the $p\times p$ covariance function as $\C^0(u_1,u_2)$, such that, for $k_1\neq k_2\in[p]$ and $k_1=k_2\in[p], \,u_1\neq u_2$, the $(k_1,k_2)$th entry is defined as
\begin{equation}\label{eqn:c}
C_{k_1,k_2}^0(u_1,u_2)=\bar{\lambda}^{(2)}_{k_1,k_2}(u_1,u_2)-\bar{\lambda}_{k_1}(u_1)\bar{\lambda}_{k_2}(u_2).
\end{equation}
When $k_1=k_2$ and $u_1=u_2$, it holds that $\mathbb{E}\left\{\dd N_{k}(u)\dd N_{k}(u)\right\}=\mathbb{E}\left\{\dd N_{k}(u)\right\}$ \citep{hawkes1971spectra}. Thus, the complete covariance matrix can be written as
$$
\C(u_1,u_2)=\delta(u_1-u_2)\bar{\bm{\Lambda}}(u_1)+\C^0(u_1,u_2),
$$
where $\delta(\cdot)$ is the Dirac function, $\bar{\bm{\Lambda}}(u_1)=\text{diag}\left\{\bar{\lambda}_1(u_1),\ldots,\bar{\lambda}_p(u_1)\right\}$ and $C_{k,k}^0(u_1,u_2)$ is continuous at $u_1=u_2$, $k\in[p]$ \citep{hawkes1971spectra}.}

{\begin{assumption}\label{ass4}
Assume that there exist constants $\Lambda_{\min}, \Lambda_{\max}>0$ such that, $\bar{\lambda}_{k}(t)\geq \Lambda_{\min}$ and $\bar{\lambda}^{(2)}_{k_1,k_2}(u_1,u_2)\le\Lambda_{\max}$. Additionally, assume that $\C^0(u_1,u_2)$ is non-negative definite, i.e., $\int\int \f(u_1)^\top\C^0(u_1,u_2)\f(u_2)\dd u_1\dd u_2\geq 0$ for any square-integrable functions $\f=(f_1,\ldots,f_p)$.
\end{assumption}
\noindent 
This condition assumes that the first and second order mean intensities are bounded. 
The non-negative definite assumption of $\C^0(u_1,u_2)$ holds true for many commonly used univariate point process models \citep{guan2013var}. In the stationary multivariate Hawkes process case, \cite{bacry2016second} showed that $\C^0(u_1,u_2)$ is directly related to the solution to an integral equation involving the transfer functions; the integral equation can be numerically solved and an estimate of $\C^0(u_1,u_2)$ can therefore be obtained. In our nonstationary multivariate Hawkes process setup, $\C^0(u_1,u_2)$ may instead be estimated through parametric bootstrap. Validity of the non-negative definite assumption of $\C^0(u_1,u_2)$ can be subsequently assessed using an estimated $\C^0(u_1,u_2)$.}

\begin{assumption}\label{ass6}
Assume that there exist $\widetilde\bbeta_j=(\widetilde\bbeta_{j,0},\widetilde\bbeta_{j,1},\ldots,\widetilde\bbeta_{j,p})^\top\in\mathbb{R}^{m_0+pm_1}$, $j\in[p]$, and a smoothness parameter $d\ge2$ such that, for some positive constants $C_1$, $C'_2$ and $C'_3$,
{
\begin{equation}\label{berror}
\frac{1}{T}\int_{0}^T \left\{\bPsi^\top(t)\widetilde\bbeta_{j}-\lambda_{j}(t)\right\}^{2}\dd t\leq C_1(s+1)^2m_1^{-2d},
\end{equation}
with probability at least $1-C'_2pT\exp(-C'_3T^{1/5})$}, where $m_1/m_0=\mathcal{O}(1)$ and $\widetilde{\bbeta}_{j,k}=\0$ for $k\notin\mE_j$.
\end{assumption}

\noindent 
This condition assumes that the true intensity function can be well approximated by the basis functions, in that residuals from the truncated basis approximation decreases at a polynomial rate of the number of basis functions. 
{The $d\ge2$ is a smoothness parameter for the background intensities $\nu_j(t)$'s and transfer functions $\omega_{j,k}(t)$'s. 
While this parameter may differ between $\nu_j(t)$'s and $\omega_{j,k}(t)$'s, it is assumed to be the same to simplify notations in our analysis.
Condition \eqref{berror} can be verified when, for example, $h(x)=x$ and the approximation errors satisfy $\frac{1}{T}\|\b_{j,0}\bphi_0(t)-\nu_j(t)\|^2_{2,[0,T]}=\mathcal{O}(m_0^{-2d})$ and $\|\b_{j,k}\bphi_1(t)-\omega_{j,k}(t)\|^2_{2,[0,b]}=\mathcal{O}(m_1^{-2d})$ for some $\b_{j,0}\in\mathbb{R}^{m_0}$ and $\b_{j,k}\in\mathbb{R}^{m_1}$, $j\in[p]$, where $b$ is as defined in Assumption \ref{ass2}; see a detailed proof of this statement in Section \ref{proof:ass6}. 
Such approximation errors hold for B-spline basis \citep{stone1985additive} or trigonometric basis \citep{tsybakov2008introduction} when the target functions belong to certain function classes. For example, when $\omega_{j,k}$ is $d$-smooth \citep{chen2007large}, i.e., $|\omega_{j,k}^{(l)}(t)-\omega_{j,k}^{(l)}(s)|\le c|t-s|^{d-l}$, where $l=\lfloor d\rfloor$ and $c$ is some positive constant, there exists $\b_{j,k}\in\mathbb{R}^{m_1}$ for normalized B-spline basis $\bphi_1(t)$ of dimension $m_1$ such that $\|\b_{j,k}\bphi_1(t)-\omega_{j,k}(t)\|^2_{2,[0,b]}=\mathcal{O}(m_1^{-2d})$ \citep{stone1985additive}. We refer to \citet{chen2007large,tsybakov2008introduction} for thorough reviews of basis approximations and truncation errors.}

Next we establish the non-asymptotic error bound of the estimated intensity functions. 
\begin{thm}\label{thm3}
Consider a Hawkes process on $[0,T]$ with intensity as defined in \eqref{intensity} satisfying Assumptions \ref{ass1}-\ref{ass6}.
For $j\in[p]$, let $\hat{\lambda}_{j}(t)=\bPsi^\top(t)\hat\bbeta_{j}$, where $\hat\bbeta_j$ is estimated from \eqref{obj}.
Given $\eta_{j}=(C_2\log p/T)^{1/2}$, $s=o(T^{2/5})$, $\log p=\mathcal{O}(T^{1/5})$ and {$sm_1=\mathcal{O}(T^{4/5})$}, we have, for $j\in[p]$,
\begin{equation}\label{bound}
\frac{1}{T}\int_{0}^T \left\{\hat{\lambda}_{j}(t)-\lambda_{j}(t)\right\}^{2}\dd t\leq 32\left\{{C_1(s+1)^{2}m_1^{-2d}}+9s\lambda_{\max}\frac{\log p}{T}\right\},
\end{equation}
holds with probability at least $1-C_{3}p^{-2}-C_{4}p^{2}T\exp(-C_{5}T^{1/5})$, where $C_2$, $C_3$, $C_4$, and $C_5$ are positive constants, and $C_1$ is as defined in \eqref{berror}.
\end{thm}
\noindent
Theorem~\ref{thm3} shows the error bound of the intensity functions estimated from minimizing the penalized least squares loss in \eqref{obj}.
The error bound on the right hand side of \eqref{bound} consists of two terms. 
{The first term comes from the B-spline basis approximation error (i.e., bias from approximating the nonparametric background and transfer functions using basis functions) and the second term comes from the statistical error (i.e., stochastic error in estimating the intensity functions). 
It is seen that when $sT/\log p=o(m_1^{2d})$, the bias term $C_1(s+1)^{2}m_1^{-2d}$ would become negligible when compared to the statistical error term.
When, for example, $s=\mathcal{O}(1)$, $d=2$ and $m_1\asymp T^{1/5}$, the error bound in \eqref{bound} reduces to $\mathcal{O}(T^{-4/5}+\log p/T)$, which is comparable with the estimation error in sparse additive regressions \citep{raskutti2012minimax}.}

{Two key ingredients in the proof of Theorem \ref{thm3} are establishing an upper and lower bounded eigenvalue condition for $\G^{(k,k)}$ (see Lemma \ref{lemma:block}) employed in the standardized group lasso penalty in \eqref{obj} and a restricted eigenvalue condition for $\G$ under the group lasso setting (see Lemma \ref{lemma:re}). Establishing these two conditions under the proposed nonstationary process is nontrivial; it requires a delicate analysis that combines properties of the basis functions and concentration inequalities of first and second order statistics of the proposed process. Combining these two ingredients and a martingale central limit theorem for counting processes \citep{van1995exponential}, we are able to derive the result in Theorem \ref{thm3}.
We note that if the basis approximation error condition in Assumption \ref{ass6} is not satisfied, we may replace the first term in the error bound \eqref{bound}, i.e., $C_1(s+1)^{2}m_1^{-2d}$ with 
$$
R_{m_0,m_1}=\min_{\widetilde\bbeta_{j}\in\mathbb{R}^{m_0+pm_1}}\frac{1}{T}\int_{0}^T \left\{\bPsi^\top(t)\widetilde\bbeta_{j}-\lambda_{j}(t)\right\}^{2}\dd t
$$ and Theorem \ref{thm3} holds with an error bound of $32(R_{m_0,m_1}+9s\lambda_{\max}\log p/T)$.}

\subsection{Network structure recovery}
\label{sec:select}
In this section, we establish the network structure recovery consistency. Specifically, we show that our proposed method can consistently identify the true edges in the network with probability tending to one. To this end, we introduce two assumptions.
\begin{assumption}\label{irr}
For all $j\in[p]$, we assume that
\begin{equation*}
\max_{k\notin \mathcal{E}_{j}}\left\|\left\{\mathbb{E}\int_{0}^T \bPsi_{k}(t)\bPsi_{\bar{\mathcal{E}}_{j}}^\top (t)\dd t\right\}\left\{\mathbb{E}\int_{0}^T \bPsi_{\bar{\mathcal{E}}_{j}}(t)\bPsi_{\bar{\mathcal{E}}_{j}}^\top (t)\dd t\right\}^{-1} \right\|_{2}\leq\frac{\gamma_{\min}}{6\sqrt{s}\gamma_{\max}}, 
\end{equation*}
where $\bar\mE_{j}=\mE_{j}\cup\{0\}$, $\bPsi_{\bar{\mathcal{E}}_{j}}(t)\in \mathbb{R}^{m_0+m_1\cdot |\mathcal{E}_{j}|}$ is the concatenation of vectors $\{\bPsi_{k}(t):k\in\bar{\mathcal{E}}_{j} \}$, and $\gamma_{min}$ and $\gamma_{max}$ are constants as defined in Lemma \ref{lemma:block}.
\end{assumption}

\noindent 
{This is the irrepresentable condition \citep{Zhao2006irr} under our setting and it is a condition on covariances between the component processes. A similar condition has also been employed in \cite{ravikumar2009sparse}. 
Considering the $j$th component process, this condition stipulates that the $\bPsi_{k}(t)$ of the irrelevant component processes (i.e., $k\notin \mE_{j}$) has small covariances with $\bPsi_k(t)$ of the relevant processes (i.e., $k\in\bar\mE_j$) in the sense that $\|\{\mathbb{E}\int_{0}^T \bPsi_{k}(t)\bPsi_{\bar{\mathcal{E}}_{j}}^\top (t)\dd t\}\{\mathbb{E}\int_{0}^T \bPsi_{\bar{\mathcal{E}}_{j}}(t)\bPsi_{\bar{\mathcal{E}}_{j}}^\top (t)\dd t\}^{-1} \|_{2}$ is small for $k\notin \mE_{j}$. Assumption \ref{irr} implies
$$
\max_{k\notin \mathcal{E}_{j}}\left\|\left\{\mathbb{E}\int_{0}^T \bPsi_{k}(t)\bPsi_{\bar{\mathcal{E}}_{j}}^\top (t)\dd t\right\}\left\{\mathbb{E}\int_{0}^T \bPsi_{\bar{\mathcal{E}}_{j}}(t)\bPsi_{\bar{\mathcal{E}}_{j}}^\top (t)\dd t\right\}^{-1} \right\|_{\infty}\leq\frac{\sqrt{m_1}\gamma_{\min}}{6\gamma_{\max}},
$$
as $\frac{1}{\sqrt{n}}\|A\|_{\infty}\le\|A\|_2\le \sqrt{m}\|A\|_{\infty}$ for an $m\times n$ matrix $A$.
This condition can be further relaxed if the adaptive lasso \citep{zou2006adaptive,huang2010variable} penalty term is considered and we plan to investigate this extension in our future work.}
The next condition is a minimal signal condition. 

\begin{assumption}
\label{betamin}
There exists a constant $\beta_{min}>0$ such that $\|\tilde{\bbeta}_{j,k}\|_{2}\geq \beta_{\min}$ for $k\in\mE_j$, where $\tilde{\bbeta}_j$ is as defined in Assumption \ref{ass6}.
\end{assumption}
\noindent
Note that this condition is not placed on $\tilde\bbeta_{j,0}$ since $\bbeta_{j,0}$ is not included in the penalty term.
With these two new assumptions and Theorem~\ref{thm3}, we can now state the edge selection consistency property of our estimator.

\begin{thm}\label{thm4}
Consider a Hawkes process on $[0,T]$ with intensity as defined in \eqref{intensity} satisfying Assumptions \ref{ass1}-\ref{betamin}. 
Assume that $\eta_{j}=(C_2\log p/T)^{1/2}$, {$s^2T/\log p=\mathcal{O}(m_1^{2d})$}, $s=\mathcal{O}(T^{1/5})$, $\log p=\mathcal{O}(T^{1/5})$, {$s^2m_1=o(T^{4/5})$}. It holds that, for $j\in[p]$,
$$
\hat{\mathcal{E}}_{j}=\mE_j,
$$ 
with probability at least $1-2C_3p^{-2}-3C_4p^2T\exp(-C_5T^{1/5})$, where $C_2$, $C_{3}$, $C_{4}$ and $C_{5}$ are the same constants as in Theorem~\ref{thm3}.
\end{thm}
\noindent 
This result establishes selection consistency. 
The condition $s^2T/\log p=\mathcal{O}(m_1^{2d})$ is needed in selection to ensure the bias term $\mathcal{O}(s^2m_1^{-2d})$ does not dominate the group-wise estimation error $\mathcal{O}(\log p/T)$. 
The conditions in Theorem \ref{thm4} are satisfied when, for example, $s=\mathcal{O}(1)$, $d=2$, $\log p\asymp T^{1/5}$ and $m_1\asymp T^{1/5}$.
We note that selection consistency was also studied in \cite{chen2017multivariate}, under a stationary Hawkes process setting. In comparison, our result is established under the more flexible nonstationary setting. Moreover, our result significantly relaxes a restrictive condition in \cite{chen2017multivariate}. Specifically, Assumption 7 (second equation) in \cite{chen2017multivariate}, after some simplification, would require $T$ to be upper bounded. This result on the selection consistency has an important implication in practice, as it ensures that our method can correctly identify the true edges in the latent network. 

{
Next, we investigate the selection consistency of the proposed GIC in \eqref{bic}.
We use $\hat\mE_j^{\eta_j}$ to denote the estimated $\mE_j$ with tuning parameter $\eta_j$.
Let $\eta_{\max}$ and $\eta_{\min}$ be, respectively, the upper and lower limits of the tuning parameter $\eta_j$, where $\eta_{\max}$ can be easily chosen such that $\hat{\mE}_j^{\eta_{\max}}$ is empty and $\eta_{\min}$ can be chosen such that $\hat{\mE}_j^{\eta_{\min}}$ is sparse, and the corresponding model size $s_0=|\hat{\mE}_j^{\eta_{\min}}|$ satisfies conditions in Theorem \ref{thm4.2}.
We partition the interval $[\eta_{\min}, \eta_{\max}]$ into two subsets
\begin{equation*}
\begin{aligned}
&\Gamma_{-}=\left\{\eta_j\in[\eta_{\min},\eta_{\max}]:\hat{\mE}^{\eta_j}_j\not\supset \mE_j\right\},	\\
&\Gamma_{+}=\left\{\eta_j\in[\eta_{\min},\eta_{\max}]:\hat{\mE}^{\eta_j}_j\supset \mE_j\,\,\text{and}\,\,\hat{\mE}^{\eta_j}_j\neq \mE_j\right\},
\end{aligned}	
\end{equation*}
corresponding to $\eta_j$'s that result in under-fitted and over-fitted models, respectively. The next result states that the proposed GIC is consistent in model selection. 

\begin{thm}\label{thm4.2}
Consider a Hawkes process on $[0,T]$ with intensity as defined in \eqref{intensity} satisfying Assumptions \ref{ass1}-\ref{ass6},\ref{betamin} and that $s^2T/\log p=\mathcal{O}(m_1^{2d})$, $s=\mathcal{O}(T^{1/6})$, $\log p=\mathcal{O}(T^{1/6})$ and $m_1=\mathcal{O}(T^{1/3})$. 
Assume that there exists $\eta_j^\ast\in[\eta_{\min}, \eta_{\max}]$ such that $\hat{\mE}^{\eta^*_j}_{j}=\mE_j$.
Consider the GIC function defined in \eqref{bic}.
When $s_0=|\hat{\mE}_j^{\eta_{\min}}|=o(T^{1/2})$, $s=o(s_0)$, $sm_1\alpha_T/T=o(1)$ and $s\log\,p/\alpha_T=o(1)$, it holds that
\begin{equation*}
\mathbb{P}\left(\inf_{\eta_j\in\Gamma_{-}\cup\Gamma_{+}}\emph{GIC}(\eta_j)-\emph{GIC}(\eta_j^\ast)>0\right)\rightarrow 1.
\end{equation*}
\end{thm}
\noindent 
The assumption that there exists $\eta_j^\ast\in[\eta_{\min}, \eta_{\max}]$ such that $\hat{\mE}^{\eta^*_j}_{j}=\mE_j$ is satisfied, for example, by the result in Theorem 4, which would additionally require Assumption \ref{irr}. This is true by noting $|\hat{\mE}_j^{\eta_{\max}}|=0$, $|\hat{\mE}_j^{\eta_{\min}}|=s_0$, $s=o(s_0)$ and the size of the selected model decreases as $\eta_j$ increases \citep{zhang2010regularization}.
The main challenge in establishing Theorem \ref{thm4.2} is the large number of candidate models in the over-fitted case, which increases combinatorially fast with $p$. To overcome this challenge, we introduce a proxy criterion on a support of size $s_0<p$ \citep{zhang2010regularization}; see proof details in Section \ref{sec:gic}. 
The two conditions on $\alpha_T$ specify a range that ensures consistency. Specifically, $s\log\,p/\alpha_T=o(1)$ suggests that $\alpha_T$ should diverge adequately fast such that the true model is not dominated by over-fitted models. On the other hand, $sm_1\alpha_T/T=o(1)$ restricts the rate of divergence for $\alpha_T$ based on the size of the true model $s$ and observation window length $T$. 
The first condition on $\alpha_T$ can be written as $\alpha_T=o(T/(sm_1))$. 
If we take, for example, $\alpha_T=\mathcal{O}((\log p)^2\log T)$, both conditions on $\alpha_T$ are met. While other choices of $\alpha_T$ can also satisfy both conditions, we have chosen a uniform choice $\alpha_T=\mathcal{O}((\log p)^2\log T)$ in our empirical investigations. Moreover, we take $s_0=\mathcal{O}(\sqrt{T}/\log\log T)$ and choose $\eta_{\max}$ such that that $|\hat{\mE}_j^{\eta_{\max}}|=0$ and $\eta_{\min}$ such that $|\hat{\mE}_j^{\eta_{\min}}|=\mathcal{O}(\sqrt{T}/\log\log T)$.}

\subsection{Test of background intensity}\label{sec:test}
In this section, we consider the problem of testing if the background intensities of the proposed Hawkes process are constant in time. If the background intensities $\nu_j(t)$'s in \eqref{intensity} are constant, under Assumption 1, a stationary process whose intensity follows \eqref{intensity} exists \citep{bremaud1996stability}. 
In this case, our proposed model reduces to that in \cite{chen2017multivariate}. However, if the background intensities are not constant, then the corresponding multivariate Hawkes process is nonstationary.

In our model, the background intensity $\nu_j(t)$ for the $j$th process is represented as $\bphi_0(t)\bbeta_{j,0}$. 
Without loss of generality, let the first term in the basis $\bphi_0(t)$, i.e., $\phi_{01}(t)$, be the constant term. 
Testing if $\nu_j(t)$ is constant in time can then be formulated as testing the following hypotheses:
\begin{equation}\label{test}
\text{H}_{0}: \A\bbeta_{j,0}=\0 \qquad vs. \qquad \text{H}_{1}: \A\bbeta_{j,0}\neq \0
\end{equation} 
where $\A = 
\begin{bmatrix}
  0 & \0_{m_0-1} \\
  \0^\top_{m_0-1} & \I_{(m_0-1)\times (m_0-1)} \\
\end{bmatrix}\in\mathbb{R}^{m_0\times m_0}$.
The test in \eqref{test} can detect any fixed departure in $\nu_j(t)$ from a constant provided that $m_0$ is sufficiently large \citep{fan2001generalized}.
Recall that $\hat\bbeta_j$ is obtained from 
$$
\hat\bbeta_j=\arg\min_{\bbeta_j\in\mathbb{R}^{m_0+m_1p}}\left\{\ell_j(\bbeta_j)+\eta_{j}\sum_{k=1}^{p}\left(\bbeta_{j,k}^\top\G^{(k,k)}\bbeta_{j,k}\right)^{1/2}\right\},
$$
where $\ell_j(\bbeta_j)$ is defined as in \eqref{loss}. Note that the coefficients $\bbeta_{j,0}$ does not appear in the penalty term, that is, the coefficients from the background intensity are not penalized. 

{Next, letting $\text{supp}_1(\bbeta_j)=\{k\in [p]:\bbeta_{j,k}\neq \0\}$, and define the refitted estimator $\hat\bbeta_j^1$ and the restricted estimator $\hat\bbeta^{H_0}_j$ under $H_0$ as 
$$
\hat\bbeta^{1}_j=\arg\min_{\substack{\b_j\in\mathbb{R}^{m_0+m_1p} \\ \text{supp}_1(\b_j)=\text{supp}_1(\hat\bbeta_j)}}\ell_j(\b_j),\quad\quad \hat\bbeta^{H_0}_j=\arg\min_{\substack{\b_j\in\mathbb{R}^{m_0+m_1p}:\A\b_{j,0}=\0 \\ \text{supp}_1(\b_j)=\text{supp}_1(\hat\bbeta_j)}}\ell_j(\b_j).
$$
Finally, we define the test statistic as
$$
S_j=T\left\{\ell_j(\hat\bbeta^{H_0}_j)-\ell_j(\hat\bbeta^1_j)\right\},
$$
where $\ell_j(\bbeta_j)$ is defined as in \eqref{loss}.}
The following theorem states the asymptotic null distribution of the test statistic. 
\begin{thm}\label{thm5}
Assume that all conditions in Theorem \ref{thm4} are satisfied. {Additionally, assume that $s\log p=\mathcal{O}(T^{1/5})$ and $s^3T=o(m_1^{2d+1})$. 
Under $H_0$, we have that
\begin{equation*}
S_{j}/\bar\lambda_j\stackrel{\mathcal{D}}{\rightarrow}\chi_{m_0-1}^2,
\end{equation*}
where $\bar\lambda_j=\mathbb{E}\{\dd N_j(t)\}/\dd t$ is a constant under $H_0$.}
\end{thm}
\noindent 
This result suggests that a Wilks type of result \citep{fan2001generalized} holds for our test; that is, the asymptotic null distribution is independent of the nuisance parameters (i.e., $\{\bbeta_{j,k}\}_{k\in[p]}$) and is $\chi^2$-distributed, after rescaled by a constant $\bar\lambda_j$. {As $m_0$ increases, we may alternatively write the limiting distribution as $(S_{j}/\bar\lambda_j-m_0+1)/(2m_0-2)^{1/2}\rightarrow^d\mathcal{N}(0,1)$. As such, the standard normal limiting distribution is not dependent of $m_0$.
In Theorem \ref{thm5}, the condition $s^3T=o(m_1^{2d+1})$ is needed such that the bias term from approximating the nonparametric background and transfer functions using basis functions is asymptotically negligible relative to variance of the test statistic. Such a condition is usually referred to as under-smoothing \citep{chen2007large} and is common in nonparametric regression testing problems. 
For instance, if $m_1\asymp T^{1/5}$ is used in the network structure estimation (see discussion of Theorem \ref{ass4}), noting $s^2T/\log p=\mathcal{O}(m_1^{2d})$ as assumed in Theorem \ref{ass4}, the under-smoothing condition $s^3T=o(m_1^{2d+1})$ will be satisfied if we multiply $m_1$ by, for example, a factor of $T^{1/20}$.}
 
 {In practice, given the $m_0$ and $m_1$ used in network structure estimation (see Section \ref{sec:m0m1}), we multiple both of them by an under-smoothing factor (e.g., $T^{1/20}$), which results in the under-smoothed $m_0$ and $m_1$ used in the testing procedure.
The tuning parameter $\eta_j$ used to calculate $\bbeta_j$ is selected following the proposed GIC in \eqref{bic}, where selection consistency of the GIC function still holds under the conditions in Theorem \ref{thm5}.}
Based on the result in Theorem~\ref{thm5}, we would reject the null $H_{0}: \A\bbeta_{j,0}=\0$ if $S_j/\bar\lambda_j\ge z_{1-\alpha}$, where $z_{\alpha}$ is the $\alpha$th quantile of $\chi_{m_0-1}^2$.
The constant $\bar\lambda_j$ needs to be estimated and we estimate it with 
$$
\bar\lambda_j=\frac{1}{T}\int_{0}^{T}\hat\lambda_j(t)\dd t.
$$

{Next, we discuss the asymptotic power of our proposed test against alternatives. The following theorem provides a lower bound on the growth rate of the test statistic under alternatives.
\begin{prop}\label{pop1}
Assume that all conditions in Theorem \ref{thm4} are satisfied. For any alternative $H_1$ such that $\|\A\tilde\bbeta_{j,0}\|_2/(s^2m_1\log p/T)^{1/2}\rightarrow\infty$, we have 
\begin{equation*}
\mathbb{P}(S_j>M_1s\log p)\rightarrow 1,
\end{equation*}
for some constant $M_1>0$.
\end{prop}
\noindent
This result shows that the growth rate of $S_j$ under the alternative is at least $s\log p$, while $p$ diverges.  
The asymptotic null distribution in Theorem \ref{thm5} and the growth rate under the alternative together suggest that the null and the alternative hypotheses are well separated, and our proposed test is asymptotically powerful against alternatives. Moreover, the test is locally powerful, i.e., $\|\A\tilde\bbeta_{j,0}\|_2$ is allowed to tend to 0 as long as it decreases no faster than the rate of $(s^2m_1\log p/T)^{1/2}$.}
In Section \ref{sec::sim}, we carry out simulation studies to evaluate the size and power of the proposed test.

\section{Simulation Studies}
\label{sec::sim}
In this section, we carry out simulation studies to investigate the finite sample performance of our proposed method, and to compare with existing solutions. 
We consider three simulation settings.
In Simulation 1, we simulate data from the proposed Hawkes process and investigate the estimation accuracy of our proposed method; in Simulation 2, we simulate data from the proposed Hawkes process and investigate the network edge selection accuracy; in Simulation 3, we evaluate the size and power of our proposed test of hypothesis.
We refer to our proposed method for nonstationary Hawkes processes as \texttt{NStaHawkes}.
In Simulations 1 and 2, we compare our method with \cite{chen2017multivariate}, referred to as \texttt{StaHawkes}, which was proposed for stationary Hawkes processes. 
We also compare with a binning based approach in \cite{zhang2016statistical} on selection accuracy. 
Specifically, \cite{zhang2016statistical} considers a binning approach that divides the observation window into several bins and models the number of events in each bin; the network structure is estimated using a regularized generalized linear model framework; we refer to this method as \texttt{BinGLM}.

In all simulations, we use the criterion in \eqref{bic} with $\alpha_T=(\log p)^2\log T/2$ to select the tuning parameter for \texttt{NStaHawkes}. 
The tuning parameters in \texttt{StaHawkes} and \texttt{BinGLM} are selected using the BIC-type functions recommended, respectively, in \cite{chen2017multivariate} and \cite{zhang2016statistical}.

\subsubsection*{Simulation 1}
\begin{figure}[t!]
\centering
\includegraphics[scale=0.3]{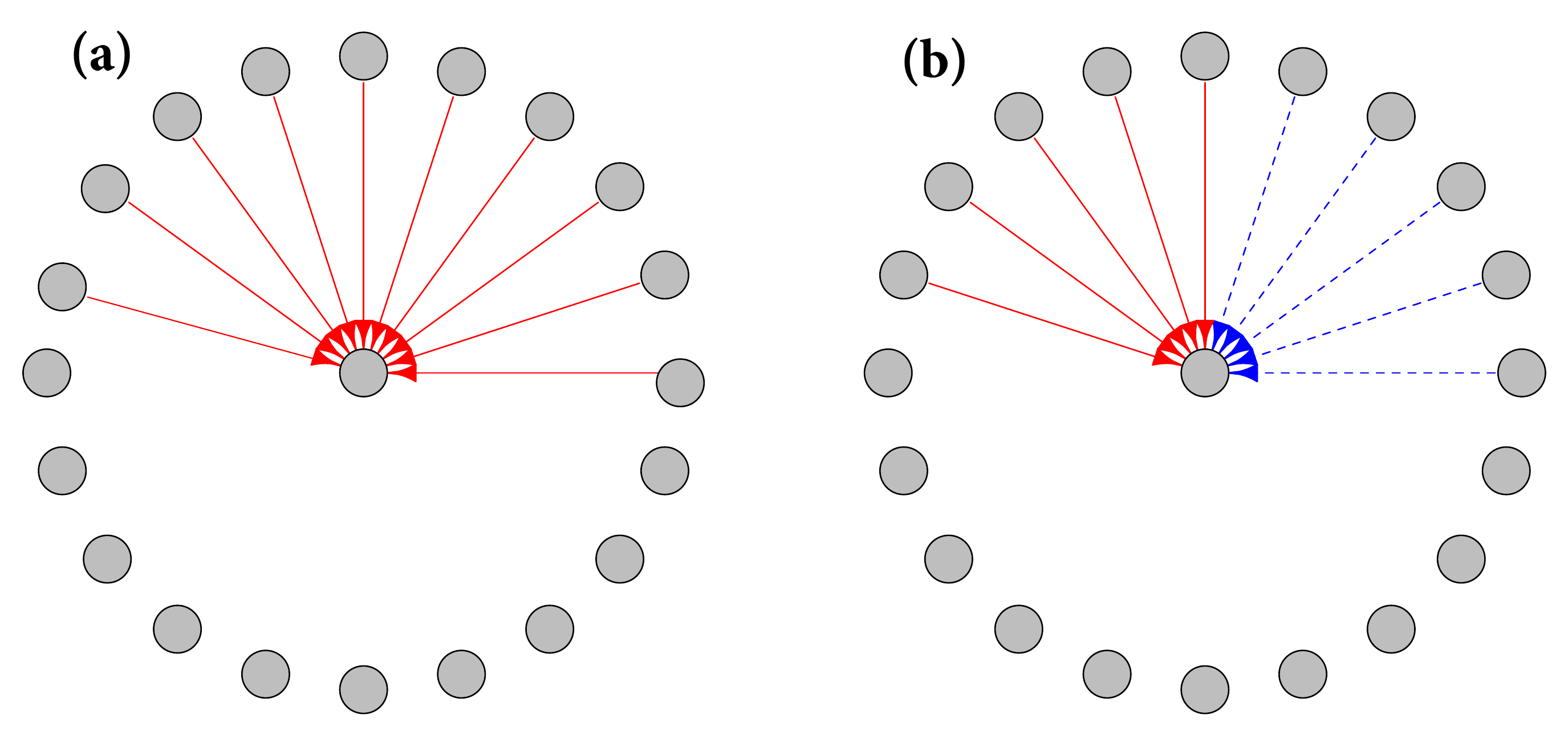}
\caption{Directed network structures in Simulation 1. The network in (a) is considered in Setting 1.1; the network in (b) is considered in Setting 1.2. Red (solid) edges represent excitatory effects and blue (dashed) edges represent inhibitory effects.}
\label{sim1}
\end{figure}

In this simulation, we consider two different network settings. 
The first setting considers the network in Figure~\ref{sim1}(a), where all transfer functions are positive, corresponding to excitatory effects.
The second setting considers the network in Figure~\ref{sim1}(b), with both positive and negative transfer functions, corresponding to excitatory and inhibitory effects, respectively. Let the network edge set be $\mathcal{E}=\{(k,1), k=2,\ldots,11\}$.
The background intensity functions and transfer functions for each setting are as follows
\begin{description}
\item \textbf{Setting 1.1:}
\begin{eqnarray*}
&&\nu_1(t)=60+50\times\sin(2\pi t/T), \quad \nu_{j} (t)=\alpha_{j}+\alpha_{j}\times\sin(2\pi t/T),\quad j=2,\ldots,21,\\
&&\omega_{1,k}=20000(x+0.001)\exp(1-500x),\quad k=2,\ldots,11,
\end{eqnarray*}
\item \textbf{Setting 1.2:}
\begin{eqnarray*}
&&\nu_1(t)=60+50\times\sin(2\pi t/T), \quad \nu_{j} (t)=\alpha_{j}+\alpha_{j}\times\sin(2\pi t/T),\quad j=2,\ldots,21,\\
&&\omega_{1,k}=20000(x+0.001)\exp(1-500x),\quad k=2,\ldots,6,\\
&&\omega_{1,k}=-15000(x+0.001)\exp(1-500x),\quad k=7,\ldots,11,
\end{eqnarray*}
\end{description}
where $\alpha_j$ is generated from $N(30,5^2)$. We let the supports of all transfer functions be $[0,0.01]$, and simulate events in $[0,T]$ with the intensity function \eqref{intensity} under Settings 1.1-1.2. 
{To estimate the background intensities and transfer functions, we use cubic B-splines with equally spaced knots.
To select the numbers of B-splines $m_0$ and $m_1$, we first perform selection using the proposed procedure in Section \ref{sec:m0m1} over 20 data replications.
The numbers of B-splines are then fixed at the respective averages of the 20 selected values for $m_0$ and $m_1$. 
It is worth noting that the estimation and selection accuracy are not overly sensitive to the number of B-splines used in the estimation (see additional results in Section \ref{sec:varym}).} We have also considered larger ranges for the transfer functions and the results are very similar. We thus focus on the current setting when reporting our simulation results. To evaluate the estimation accuracy, we report the mean squared errors. 
\begin{table}[!t]
	\setlength{\tabcolsep}{3pt}
	\centering
{	\begin{tabular}{c|c|rcccc}\hline
		\multicolumn{6}{c}{\textbf{Setting 1}}&\\\hline
		$T$ & Method       & MSE$(\nu)$\quad\quad & MSE$(\omega)$ & FNR & FPR & F$_1$ score \\ \hline
		\multirow{2}{*}{10} & \texttt{NStaHawkes} & 7.921 (0.071)  & 0.332 (0.003) & 0.008 (0.003) & 0.011 (0.001) & 0.816 (0.007)       \\
		&  \texttt{StaHawkes}  & 25.576 (0.073) & 1.363 (0.026) & 0.000 (0.000) & 0.305 (0.008) & 0.139 (0.003)  \\ 
		&  \texttt{BinGLM} & -  &  - & 0.003 (0.002) & 0.058 (0.001) & 0.443 (0.002) \\ \hline
		\multirow{2}{*}{20} & \texttt{NStaHawkes}  &  7.389 (0.038)  &  0.279 (0.003)  & 0.002 (0.001) & 0.005 (0.000) & 0.908 (0.006)     \\
		&  \texttt{StaHawkes}   &  24.574 (0.034) & 0.899 (0.010)   & 0.000 (0.000) & 0.321 (0.005)  & 0.129 (0.002) \\ 
		&  \texttt{BinGLM} & -  &  - & 0.000 (0.000) & 0.058 (0.001)  & 0.447 (0.002)    \\ \hline
		\multicolumn{6}{c}{\textbf{Setting 2}}&\\\hline
		$T$ & Method         & MSE$(\nu)$\quad\quad & MSE$(\omega)$ & FNR & FPR & F$_1$ score \\ \hline
		\multirow{2}{*}{10} & \texttt{NStaHawkes}  & 6.111 (0.038) & 0.279 (0.003) & 0.113 (0.012) & 0.005 (0.000) & 0.835 (0.007)     \\
		&   \texttt{StaHawkes} & 25.549 (0.083) & 1.297 (0.028) & 0.042 (0.012) & 0.305 (0.009) & 0.135 (0.004)     \\ 
		&   \texttt{BinGLM} & -  &  -  & 0.342 (0.019) & 0.041 (0.001)  & 0.376 (0.007)    \\ \hline
		\multirow{2}{*}{20} & \texttt{NStaHawkes} & 5.545 (0.026) & 0.252 (0.002)  & 0.094 (0.011) & 0.001 (0.000) & 0.924 (0.006)     \\
		&  \texttt{StaHawkes} & 24.648 (0.047) & 0.938 (0.014) & 0.414 (0.010) & 0.316 (0.007) & 0.080 (0.002) \\ 
		&  \texttt{BinGLM} & -  &  - & 0.359 (0.018) & 0.094 (0.004) & 0.236 (0.007)  \\ \hline    
	\end{tabular}}
	\caption{Comparison of the three methods with varying observation window length $T$ in Simulation 1. \texttt{NStaHawkes} refers to the proposed method, \texttt{StaHawkes} refers to \cite{chen2017multivariate} and \texttt{BinGLM} refers to \cite{zhang2016statistical}. Standard errors are shown in parentheses.}
	\label{tab1}
\end{table}
For the background intensity, it is calculated as $\text{MSE}(\nu)=\frac{1}{p}\sum_{j\in[p]}\text{MSE}(\nu_j)$, where 
$$
\text{MSE}(\nu_j)=\left\{\frac{1}{T}\int_0^T (\hat\nu_j(t)-\nu_j(t))^2\dd t\right\}^{1/2}
$$ 
and $\hat\nu_j(t)$ is the estimate of $\nu_j(t)$.
For the transfer functions, it is calculated as $\text{MSE}(\omega)=\frac{1}{p}\sum_{j\in[p]}\text{MSE}(\omega_{j,\cdot})$,
where 
$$
\text{MSE}(\omega_{j,\cdot})=\left\{\sum_{k=1}^p\int_0^b (\hat\omega_{j,k}(t)-\omega_{j,k}(t))^2\dd t\right\}^{1/2}
$$ 
and $\hat\omega_{j,k}(t)$ is the estimate of $\omega_{j,k}(t)$. To evaluate the selection accuracy, we report the the false positive rate (FNR), the false positive rate (FPR) and the F$_1$ score, calculated as 2TP$/$(2TP+FP+FN), where TP is the true positive count, FP is the false positive count, and FN is the false negative count. 
{The highest F$_1$ score is 1 indicating perfect selection. This measure is commonly used in machine learning to measure selection and classification accuracy; see for example, \citet{forman2003extensive,ho2012multiscale}.}
For both \texttt{NStaHawkes} and \texttt{StaHawkes} estimators, we report the estimation and selection accuracy. 
For the \texttt{BinGLM} estimator, we only report the selection accuracy, as this method cannot be used to estimate the intensity functions. 
Table~\ref{tab1} reports the average criteria from the three methods, with standard errors in the parentheses, over 100 data replications. 
The proposed method \texttt{NStaHawkes} is seen to achieve the best performance, both in terms of the estimation accuracy and edge selection accuracy, and this holds true for different observation window length $T$. Moreover, it is seen that the estimation error of \texttt{NStaHawkes} decreases as $T$ increases. Such an observation agrees with our theoretical result in Theorem~\ref{thm3}.

{
\subsubsection*{Simulation 2}
\begin{figure}[t!]
\centering
\includegraphics[scale=0.5]{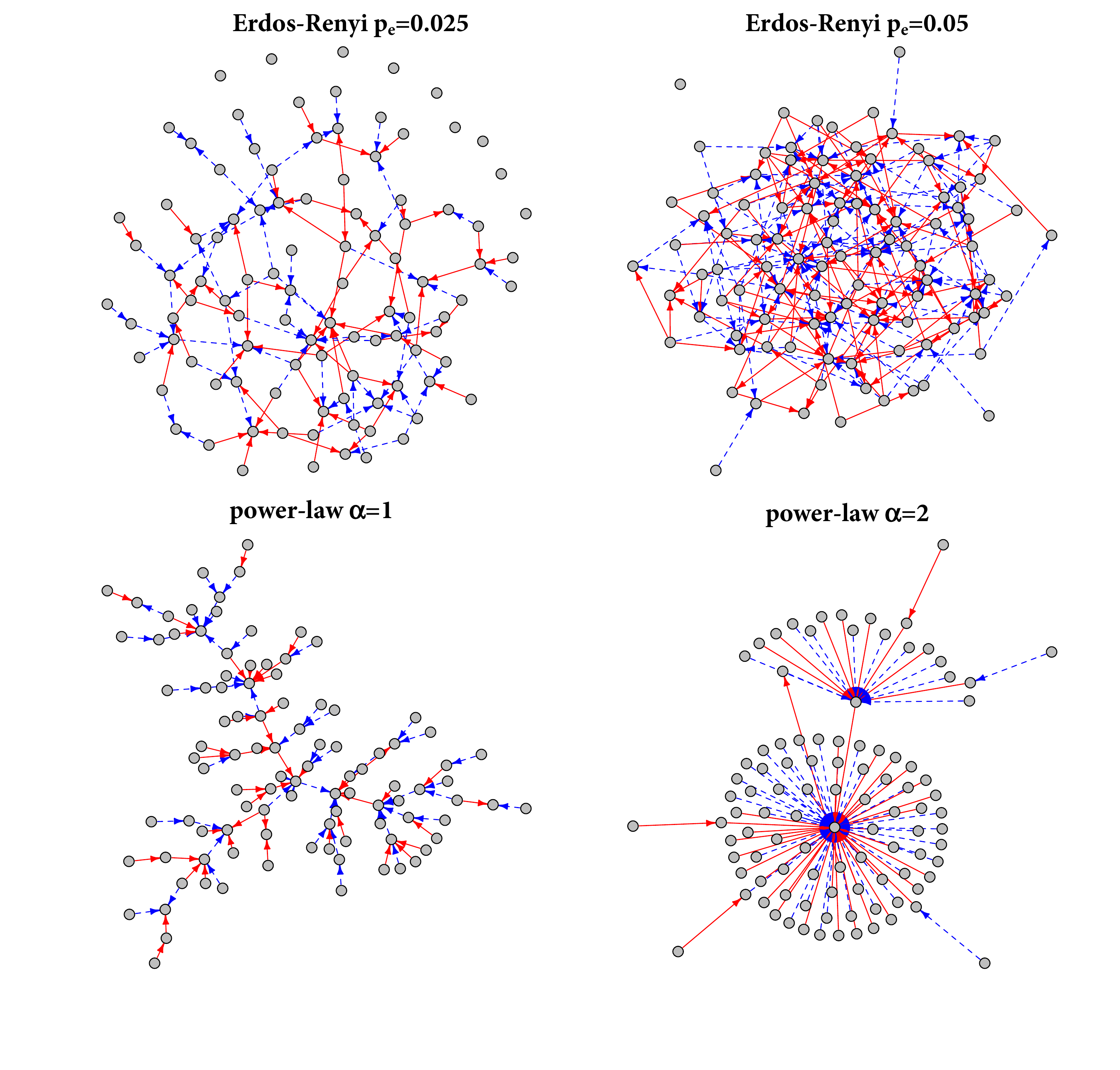}
\caption{Directed network structures in Simulation 2. Red (solid) edges represent excitatory effects and blue (dashed) edges represent inhibitory effects.}
\label{sim2}
\end{figure}

\begin{table}[!t]
	\centering
	{\renewcommand{\arraystretch}{0.85}
{\begin{tabular}{c|ccc|ccc} \hline
                 \multicolumn{7}{c}{\textbf{Erdos-Renyi network}}\\\hline
	         & \multicolumn{3}{c}{$p_e=0.025$} \vline & \multicolumn{3}{c}{$p_e=0.05$}\\ \cline{2-7}
		&FNR &FPR &F$_1$ score  &FNR &FPR &F$_1$ score\\ \hline
		\texttt{NStaHawkes} & 0.003 & 0.002 & 0.935  & 0.005 & 0.005 & 0.909\\ 
		& (0.001) & (0.000) & (0.002) & (0.000) & (0.000) & (0.001) \\ \hline 
		\texttt{StaHawkes}    & 0.000 & 0.859 & 0.027 & 0.000 & 0.882 & 0.053 \\ 
		 & (0.000) & (0.001)  & (0.000)  & (0.000) & (0.001)  & (0.000) \\ \hline
		\texttt{BinGLM}    & 0.394 & 0.133 & 0.097    & 0.369 & 0.171 & 0.148  \\ 
		  & (0.003) & (0.001)  & (0.001)  & (0.002) & (0.001)  & (0.000) \\ \hline 
		 \multicolumn{7}{c}{\textbf{power-law network}}\\\hline
		& \multicolumn{3}{c}{$\alpha=1$} \vline & \multicolumn{3}{c}{$\alpha=2$}\\ \cline{2-7}
		&FNR &FPR &F$_1$ score  &FNR &FPR &F$_1$ score\\ \hline
		\texttt{NStaHawkes}  & 0.005 & 0.002 & 0.928  & 0.003 & 0.003 & 0.870 \\ 
		& (0.001) & (0.000) & (0.002) & (0.001) & (0.000) & (0.001)  \\ \hline 
		\texttt{StaHawkes}    & 0.000 & 0.848 & 0.023 & 0.002 & 0.824 & 0.024 \\ 
		 & (0.000) & (0.001)  & (0.000)  & (0.000) & (0.001)  & (0.000)\\ \hline
		\texttt{BinGLM}    & 0.379 & 0.121 & 0.090 & 0.787 & 0.060 & 0.059  \\ 
		  & (0.003) & (0.000)  & (0.000)  & (0.004) & (0.000)  & (0.001) \\ \hline
	\end{tabular}}}
	\caption{Comparison of the three methods with varying network parameters in Simulation 2. \texttt{NStaHawkes} refers to the proposed method, \texttt{StaHawkes} refers to \cite{chen2017multivariate} and \texttt{BinGLM} refers to \cite{zhang2016statistical}. Standard errors are shown in parentheses.}
	\label{tab2}
\end{table}

In this simulation, we evaluate the edge selection accuracy of our proposed method. We consider two types of networks.
The first type of networks are assumed to follow an Erdos-Renyi network model with edge probability $p_e$ \citep{erdos1959}. In an Erdos-Renyi network model, edges are generated independently from a Bernoulli distribution with probability $p_e$. 
The second type of networks follow a scale-free network model, with the degrees of nodes generated from a power-law distribution with parameter $\alpha$; such networks have a skewed degree distributions and a larger $\alpha$ indicates a higher degree heterogeneity \citep{clauset2009power}.
We set $p=100$, $p_e=0.025,0.05$ and $\alpha=1,2$. The generated networks are shown in Figure~\ref{sim2}. 
Based on the generated networks, we simulate data using the following setting.
\begin{description}
\item \textbf{Setting 2:}
\begin{eqnarray*}
&&\nu_{j} (t)=\alpha_j+\alpha_j\times\sin(2\pi f t/T),\quad j=1,\ldots,100,
\end{eqnarray*}
\end{description}
where $\alpha_j$ is generated independently from $N(100,5^2)$ for each node. The transfer functions are the same as in Setting 1.2. 
We simulate events in $[0,T]$ with intensity function \eqref{intensity} with $f=5$ and $T=20$. 
To estimate the background intensities, we use cubic B-splines with equally spaced knots and to estimate the transfer functions, we use step functions with equally spaced knots, as considered in \cite{hansen2015lasso,chen2017multivariate}. 
The numbers of basis functions $m_0$ and $m_1$ are selected following the same procedure as in Simulation 1.
Table~\ref{tab2} compares the false negative rate, false positive rate and F$_1$ score of the three methods over 100 data replications. 
It is seen that \texttt{NStaHawkes} achieves the best edge selection accuracy, in terms of F$_1$ scores, across all settings; \texttt{StaHawkes} shows a large false positive rate and this is likely due to the biased estimation of the background intensity functions; \texttt{BinGLM} shows a large false negative rate and this is possibly due to the loss of information in the binning approach. 
We have also considered $p=200$, where \texttt{NStaHawkes} continues to achieve a satisfactory edge selection accuracy (see Section \ref{sec:simadd}).}

{
\subsubsection*{Simulation 3} 
\begin{figure}[!t]
\centering
\includegraphics[scale=0.55,trim=0 15mm 0 0]{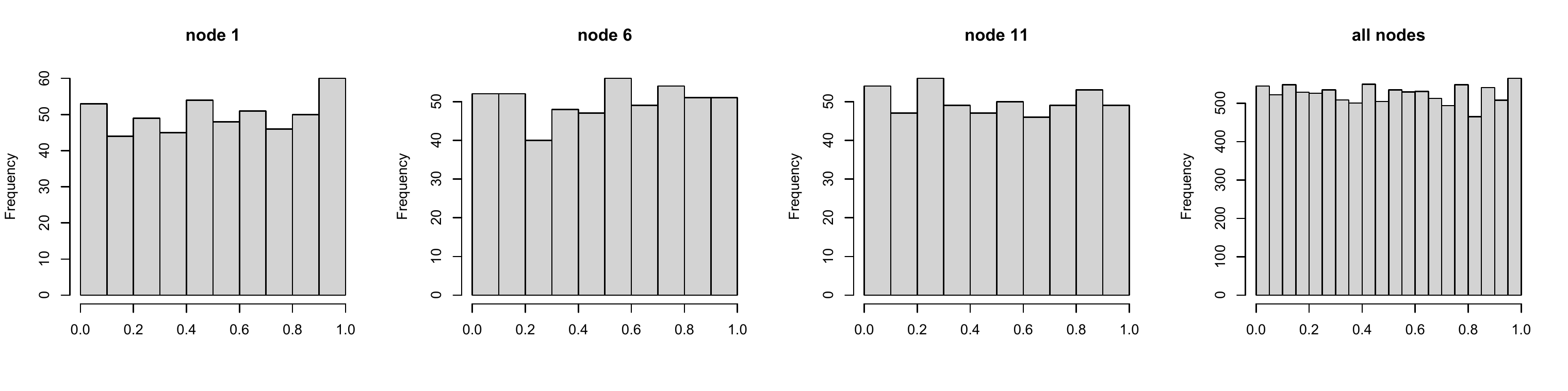}
\caption{Distribution of $p$-values under the null from 500 data replicates for nodes 1, 6, 11 and all nodes combined.}
\label{type1}
\end{figure}

In this simulation study, we evaluate the size and power of the proposed test in Section \ref{sec:test} under network (b) in Simulation 1. 
To evaluate the test size, we consider 
\begin{description}
\item \textbf{Setting 3.1:} $\nu_{j} (t)=\alpha_{j},\quad j=1,\ldots,21$,
\end{description}
where $\alpha_j$ is generated from $N(50,5^2)$. The transfer functions are the same as in Setting 1.2. We simulate events in $[0,T]$ with $T=20$. 
The B-spline bases used to estimate the background intensities and transfer functions are selected following the same procedure as in Simulation 1.
Figure~\ref{type1} shows the $p$-values from nodes 1, 6, 11 and all nodes combined, from 500 data replicates. 
It is seen that, under the null, the $p$-values are approximately uniformly distributed, suggesting that the size of our proposed test is well controlled.

\begin{table}[!t]
	\centering	\setlength{\tabcolsep}{4pt}
{\renewcommand{\arraystretch}{0.85}
	{\begin{tabular}{c|ccccccc} \hline
		& node 1  & node 2  & node 3  & node 4  & node 5  & node 6  & node 7\\ \hline
		$\rho=0.25$ & 0.816 & 0.998 & 1.000 & 0.998 & 1.000 & 0.994 & 1.000 \\
		$\rho=1$ & 0.998 & 1.000 & 1.000 & 1.000 & 1.000 & 1.000 & 1.000 \\ \hline 
		& node 8  & node 9  & node 10  & node 11  & node 12  & node 13  & node 14\\ \hline
		$\rho=0.25$ & 1.000 & 1.000 & 0.998 & 1.000 & 1.000 & 1.000 & 0.994\\
		$\rho=1$ & 1.000 & 1.000 & 1.000 & 1.000 & 1.000 & 1.000 & 1.000 \\ \hline 
		& node 15  & node 16  & node 17  & node 18  & node 19  & node 20  & node 21\\ \hline
		$\rho=0.25$ & 1.000 & 1.000 & 1.000 & 1.000 & 0.998 & 1.000 & 0.998\\
		$\rho=1$ & 1.000 & 1.000 & 1.000 & 1.000 & 1.000 & 1.000 & 1.000 \\ \hline 
	\end{tabular}}}
	\caption{{Proportions of rejection for tests on all nodes in 500 data replicates of Setting 3.2. The significance level is set to 0.05.}}
	\label{power}
\end{table}

Next, to examine the power of the proposed test, we consider 
\begin{description}
\item \textbf{Setting 3.2:} $\nu_{j} (t)=\alpha_{j}+\rho\alpha_{j}\times\sin(2\pi t/T),\quad j=1,\ldots,21$,
\end{description}
where $\alpha_j$ is generated from $N(50,5^2)$ and $\rho=0.25,1$. 
With a larger $\rho$, the alternative hypothesis departs more from the null, i.e., constant background intensity. 
The transfer functions are the same as in Setting 1.2. 
We simulate events in $[0,T]$ with $T=20$. 
The B-spline bases used to estimate the background intensities and transfer functions are selected following the same procedure as in Simulation 1.
We perform the test of hypothesis with significance level set to 0.05, and the results from 500 data replications are summarized in Table \ref{power}. 
It is seen that the proportions of rejection are close to 1 for all nodes, even when $\rho=0.25$. 
This suggests that our test is powerful against alternatives. 
}

\section{Application to Neurophysiological Data}
\label{sec::spike}

In this section, we apply our proposed method to a neuron spike train data set and estimate the functional connectivity network of neurons in the rat prefrontal cortex. 
The data were obtained from adult male Sprague-Dawley rats performing a T-maze based delayed-alternation task of working memory \citep{devilbiss2004effects}. 
In the experiment, the animal was trained to navigate down the T-maze and choose one of two arms (opposite to the one previously visited) for food rewards. 
In each trial, the animal was released after being placed in a start box for a fixed length of delay.
On a correct trial (i.e., the arm with food was chosen), the animal was rewarded and returned to the start box.
On an incorrect trial (i.e., the arm without food was chosen), the animal was returned to the start box without being rewarded. 
In the study, the animal remained in a training period until it reached 90\%-100\% accuracy on 40 trials. 
After the training period, a recording session was performed.
The spike train recording consisted of 73 neurons in an experiment of 40 trials. 
Each trials took about 36 seconds and the total recording had 1434.22 seconds.
See \cite{zhang2016statistical} for more information about data collection and processing.

We applied our proposed method to this dataset. 
{To estimate the background intensities, we used cubic B-splines with equally spaced knots and to estimate the transfer functions, we used step functions with equally spaced knots, as considered in \cite{hansen2015lasso,chen2017multivariate}. 
The numbers of basis functions $m_0$ and $m_1$ were selected using the proposed procedure in Section \ref{sec:m0m1}.
The GIC in \eqref{bic} was used to select the tuning parameters with $\alpha_T=(\log p)^2\log T/2$.}
The range of the transfer functions were set to $[0,2]$. We have also considered a larger range, and the results remain very similar. 
First, we performed the proposed test of hypothesis for each neuron to assess the if the background intensity is constant in time. 
Based on the $p$-values from the tests, 38 neurons had time-varying background intensity functions (significance level was set to 0.05).
Next, we move to estimate the neuronal connectivity network using the proposed \texttt{NStaHawkes}. 
When estimating the network structure, we also considered \texttt{StaHawkes} and \texttt{BinGLM}.
The tuning parameters in \texttt{StaHawkes} and \texttt{BinGLM} were selected using their recommended BIC functions, respectively.
\begin{figure}[!t]
\centering
\includegraphics[scale=0.825, trim=0 0 0 0]{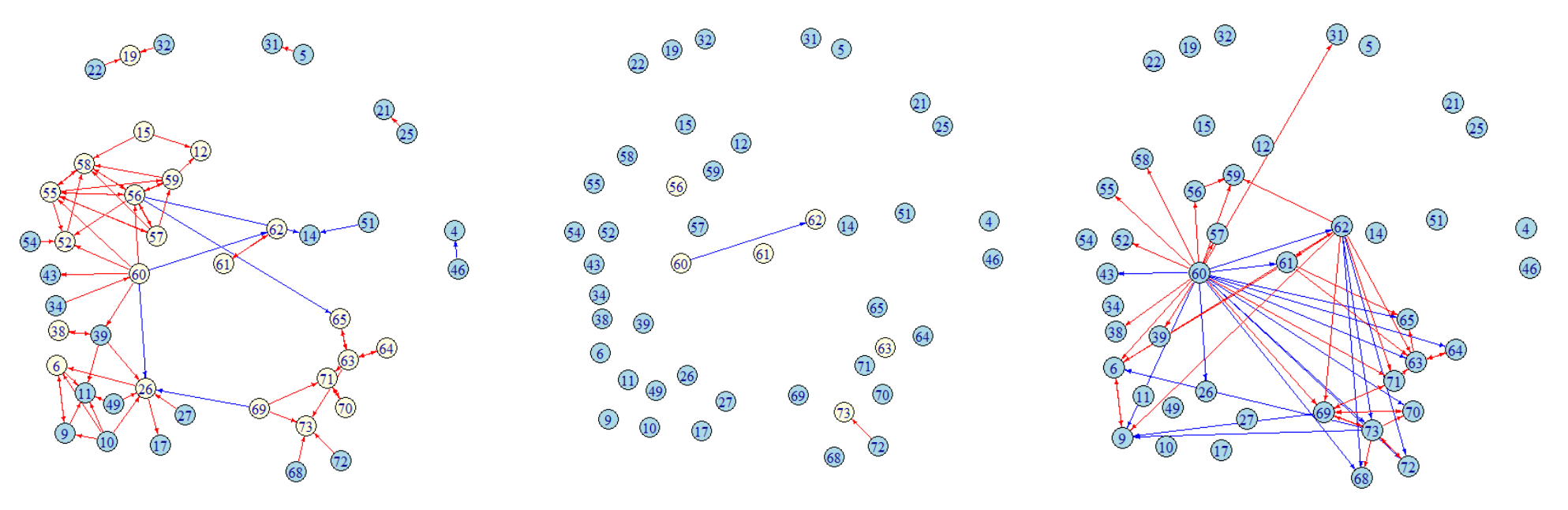}

\caption{Estimated neuronal networks using \texttt{NStaHawkes} (left), \texttt{StaHawkes} (middle) and \texttt{BinGLM} (right).
The red and blue arrows represent excitatory and inhibitory effects, respectively. Light colored nodes represent neurons that have self-exciting effects.}
\label{network}
\end{figure}
Figure~\ref{network} shows the estimated neuronal networks from the three different methods. 
We can see all three estimated networks are sparse, with both excitatory and inhibitory relationships. 
However, their structures are quite different.
The network estimated from our method is highly clustered and has a power-law degree distribution, which are two unique features of real world networks \citep{barabasi1999emergence}. 
Also interestingly, about 70\% of the identified edges in our estimated network are within the right prefrontal cortex, which agrees with existing findings that the right prefrontal cortex is highly related to the episodic memory retrieval \citep{henson1999right}. The biological significance of the identified edges requires further investigation.

Compared to our estimated network, \texttt{StaHawkes} identified a very sparse network. 
This difference is likely due to the bias in estimating the background intensity function from their method. 
\texttt{BinGLM} also identified a very different network structure. 
This network has two hub (or densely connected) nodes, namely, neurons 60 and 62, and a small clustering coefficient.
We find that neurons 60 and 62 are the two most frequently fired neurons in the ensemble. 
Specifically, neurons 60 and 62 have 14,433 and 8,191 firing events, respectively, while other neurons have on average 501 firing events during the experiment.
The regularized generalized linear model framework in \texttt{BinGLM} penalizes the frequently and infrequently firing neurons equally when encouraging sparsity. 
This can potentially lead to over selection for the frequently firing neurons, and under selection for the infrequently firing neurons.

\section{Discussion}
\label{sec:dis}
{
We conclude the paper with a brief discussion on some potential future directions.
The limiting distribution result in Theorem \ref{thm5} requires conditions for establishing selection consistency (i.e., irrepresentable condition in Assumption \ref{irr} and beta-min condition in Assumption \ref{betamin}). 
To derive a valid inferential procedure that does not reply on such conditions, we could consider the de-correlated score testing procedure in \citet{neykov2018unified,wang2020statistical} or the double selection procedure in \citet{bach2020uniform}.
In our work, we investigated empirically the reliance of the testing procedure on model selection accuracy. 
In Simulation 3.1, we showed that the size of the proposed test is well controlled; for this setting, the average false negative rate is 0, false positive rate is 0.015 and F$_1$ score is 0.764.
In Simulation 3.2, we showed that the proposed test is powerful against alternatives; for this setting and $\rho=0.25$, the average false negative rate is 0, false positive rate is 0.016 and F$_1$ score is 0.753. 
It is seen that the proposed testing procedure is not overly sensitive to errors in model selection. }

\bibliographystyle{asa}
\begingroup
\baselineskip=15pt
\bibliography{hawkes}
\endgroup

\newpage
\renewcommand{\thesection}{A}
\renewcommand{\thesubsection}{A\arabic{subsection}}
\renewcommand{\theequation}{S\arabic{equation}}
\renewcommand{\thelemma}{S\arabic{lemma}}
\setcounter{equation}{0}
\setcounter{table}{0}
\setcounter{section}{0}
\setcounter{subsection}{0}  
\setcounter{page}{1}
\def\eop
{\hfill $\Box$
}

\begin{center}
{\large\bf Supplementary Materials for ``Learning Latent Network Structure from High Dimensional Multivariate Point Processes"} \\
\medskip
\end{center}

\section{Proofs}
\subsection{Technical lemmas} 
\label{sec:lemmas}

We first state a number of technical lemmas. The proof of Lemmas \ref{lemma:dom}, \ref{couple}, \ref{exptail}, \ref{poisson2}, \ref{lemma:block} and \ref{lemma:re} are delayed to Sections \ref{sec::dom}, \ref{sec::couple}, \ref{sec::exptail}, \ref{exp::poisson2}, \ref{sec:block} and \ref{sec:re}, respectively.

\begin{lemma}[Proposition 2.4.2 in \cite{Neveu1965finitelimit}]\label{finitelimit} In order that a sequence $\{X_{n},n\geq 1\}$ of a.s. finite random variables converge a.s., it is sufficient that there exist a summable sequence $\{\epsilon_{n},n\geq 1\}$ of positive numbers such that
\[
\sum_{n=1}^{\infty}\mathbb{P}(|X_{n+1}-X_{n}|\geq\epsilon_{n})< \infty;
\]
the limit is then a.s. finite.
\end{lemma}

\begin{lemma}[Lemma 3 in \cite{bremaud1996stability}]\label{thinexp} Let $\overline{N}$ be a Poisson process of intensity $1$ on $\mathbb{R}^2$. Let $\mathcal{F}_{t}$ be a history of $\overline{N}$ and $\{\mathcal{F}_{t}^{\overline{N}}\}_{t\in\mathbb{R}}$ be the internal history of $\overline{N}$ (i.e., $\mathcal{F}_{t}^{\overline{N}}\subset\mathcal{F}_{t}$, $t\in \mathbb{R}$). Let $\{\lambda(t)\}_{t\in \mathbb{R}}$ be a nonnegative $\mathcal{F}_{t}^{\overline{N}}$-predictable process and define the point process $N$ by 
\[
N(C)=\int_{C\times \mathbb{R}}1_{[0,\lambda(t)]}(z)\overline{N}(\dd t\times \dd z)
\]
for all $C\in\mathcal{B}(\mathbb{R})$. Then $N$ admits the $\mathcal{F}_{t}^{\overline{N}}$-intensity $\{\lambda(t)\}_{t\in \mathbb{R}}$.
\end{lemma}

\begin{lemma}[Theorem 3.1 in \cite{van1995exponential}]\label{martingale}
Suppose that there exists $\lambda_{\max}$ such that $\lambda_{j}(t)\leq \lambda_{\max}$ for all $t$ and $1\leq j\leq p$. Let $H(t)$ be a bounded function that is $\mathcal{H}_{t}$-predictable. Then, for any $\epsilon>0$, the inequality 
$$
\frac{1}{T}\int_{0}^T H(t)\left\{\lambda_{j}(t)\dd t-\dd N_j(t)\right\}\leq 4\epsilon^{1/2}\left\{\frac{\lambda_{\max}}{2T}\int_{0}^T H^{2}(t)\dd t \right\}^{1/2}
$$
holds with probability at least $1-C_3\exp(-\epsilon T)$ for some $C_3>0$. 
\end{lemma}

\begin{lemma}[Lemma 3 in \cite{dedecker2004coupling}]\label{randombound}
Let $X$ be an integrable random variable and $\mathcal{M}$ a $\sigma$-algebra defined on the same probability space. If the random variable $Y$ has the same distribution as $X$, and is independent of $\mathcal{M}$, then 
\begin{equation*}
\tau(\mathcal{M},X)\equiv\mathbb{E}\left[\sup_{h}\left\{\left|\int f(x)\mathbb{P}_{X|\mathcal{M}}(\dd x)-\int f(x)\mathbb{P}_{X}(\dd x)\right| \right\}\right]\leq \mathbb{E}|X-Y|
\end{equation*} 
where $f:\mathbb{R}\rightarrow\mathbb{R}$ can be any $1$-Lipschitz function and $\mathbb{P}_{X|\mathcal{M}}$ denotes the probability measure of $X$ conditional on $\mathcal{M}$.
\end{lemma}

\begin{lemma}[Chernoff bound]\label{poisson}
Suppose that $x$ is a Poisson random variable with mean $u$. Then for any $k>0$,
$$
\mathbb{P}(x\geq k)\le \exp\left\{-u-k\log\,(k/u)+k\right\}.
$$
\end{lemma}

{
\begin{lemma}\label{lemma:dom}
Consider $p$-dimensional Hawkes processes $(N_j)_{j\in[p]}$ and $(N_j^*)_{j\in[p]}$ on $[0,T]$ with intensities defined as 
\begin{equation*}
\begin{split}
&\lambda_{j}(t)=h\left\{\nu_j(t)+\sum_{k=1}^p\int_{0}^t\omega_{j,k}(t-u)\dd N_{k}(u)\right\},\,\, j\in[p],\\
&\lambda_{j}^*(t)=\nu^*+\sum_{k=1}^p\int_{0}^t|\omega_{j,k}(t-u)|\dd N_{k}^*(u),\,\, j\in[p],
\end{split}
\end{equation*}
where $h(\cdot)$ is a $\theta$-Lipschitz link function with $\theta\le 1$, $\nu_{j}(\cdot):\mathbb{R}^+\rightarrow \mathbb{R}^+$, $\omega_{j,k}(\cdot):\mathbb{R}^{+}\rightarrow \mathbb{R}$ and $h\{\nu_j(t)\}\leq \nu^*$ for any $t$ and $j$. 
Let $\bar\lambda_{j}(t)=\mathbb{E}\{\dd N_j(t)\}/\dd t$ and $\bar\lambda_{j}^*(t)=\mathbb{E}\{\dd N^*_j(t)\}/\dd t$.
Assuming Assumption \ref{ass1},  it then holds that $\bar\lambda_{j}(t)\leq \bar\lambda_{j}^*(t)$, $j\in[p]$.
\end{lemma}
}

\begin{lemma}\label{couple}
Let $\N$ be a Hawkes process with intensity in \eqref{intensity} satisfying Assumptions \ref{ass1}-\ref{ass3}. 
For any $z>0$, there exists a point process $\widetilde\N$ such that $\widetilde{\N}$ has the same distribution as $\N$ and is independent of the history of $\N$ up to time $z$. 
Moreover, for any $u\geq 0$ and $n=\lfloor u/b+1\rfloor$,
\begin{eqnarray*}
&&\mathbb{E}\left|\dd\widetilde{\N}(z+u)-\dd\N(z+u)\right|/\dd u\preceq 2\lambda_{\max}\bOmega^{n}\mathbf{1}_{p} \\
&&\mathbb{E}\left|\dd\widetilde{\N}(t')\dd\widetilde{\N}^\top (z+u)-\dd\N(t')\dd\N^\top(z+u)\right|/(\dd u\dd t') \preceq 2c_{0}^{2}\sum_{i=1}^{n+1}\bOmega^{i-1}\mathbf{1}_{p\times p}(\bOmega^\top)^{n-i+1}
\end{eqnarray*}
where $\preceq$ denotes the element-wise inequality, $b$ is as defined in Assumption \ref{ass2} and $c_{0}=\max\{\lambda_{\max},\nu+\max_{j,k}\|\omega_{j,k}\|_{\infty} \}$. 
\end{lemma}

\begin{lemma}\label{exptail}
Let $\N$ be a Hawkes process with intensity in \eqref{intensity} satisfying Assumptions \ref{ass1}-\ref{ass3}.
For a given $z>0$, consider the coupling process $\widetilde{\N}$ constructed as in Lemma~\ref{couple}. For $j\in[p]$, we have
\begin{equation*}
\mathbb{E}\left|\dd\widetilde{N}_{j}(z+u)-\dd N_{j}(z+u)\right|\dd u\leq a_{1}\exp(-a_{2}u);
\end{equation*}
for any $j,k$ and $t'>z+u$, we have
\begin{equation*}
\mathbb{E}\left|\dd\widetilde{N}_{j}(t')\dd\widetilde{N}_{k}(z+u)-\dd N_{j}(t')\dd N_{k}(z+u)\right|/(\dd t'\dd u)\leq a_{3}\exp(-a_{4}u),
\end{equation*}
where $a_1$, $a_2$, $a_3$ and $a_4$ are positive constants that depend on $\lambda_{\max}$, $\rho_{\bOmega}$ and $b$ is as defined in Assumption \ref{ass2}.
\end{lemma}

\begin{lemma}\label{poisson2}
Suppose that $Z_{1}$ and $Z_{2}$ satisfy, for all $n\geq 0$,  
$$
\mathbb{P}(|Z_{i}|>n)\leq \exp(1-n/K_{i}), \quad i=1,2.
$$
Then, for any $n\geq 0$ and $K'=K_{1}K_{2}(\log (2) +1)$,we have
$$
\mathbb{P}(|Z_{1}Z_{2}|>n)\leq 
\exp(1-(n/K')^{1/2}).
$$
\end{lemma}

{

\begin{lemma}\label{lemma:block}
Consider a Hawkes process on $[0,T]$ with intensity as defined in \eqref{intensity} satisfying Assumptions \ref{ass1}-\ref{ass4} and the matrix $\G^{(k,k)}$, $k\in[p]$ as defined in \eqref{Gmatrix} with normalized B-spline basis $\bphi_1(t)$ of dimension $m_1$. 
There exist constants $\gamma_{\min}, \gamma_{\max}>0$ such that for $k\in[p]$
\begin{equation*}
\frac{\gamma_{\min}}{2m_1}\leq \sigma_{\min}(\G^{(k,k)})\leq 
\sigma_{\max}(\G^{(k,k)})\leq \frac{3\gamma_{\max}}{2m_1},
\end{equation*} 
with probability at least $1-c_{2}T\exp(-c_{3}T^{1/5})-c_{2}'T\exp(-c_{3}'T^{1/5})$, where $c_2$, $c_2'$, $c_3$ and $c_3'$ are as defined in Theorem \ref{thm2}.
\end{lemma}

\begin{lemma}\label{lemma:re}
Consider a Hawkes process on $[0,T]$ with intensity as defined in \eqref{intensity} satisfying Assumptions \ref{ass1}-\ref{ass4} and the matrix $\G$ as defined in \eqref{Gmatrix} with normalized B-spline bases $\bphi_0(t)$ of dimension $m_0$ and $\bphi_1(t)$ of dimension $m_1$. Let $\Delta=(\Delta_0, \Delta_1,\ldots,\Delta_p)\in\mathbb{R}^{m_0+pm_1}$, such that $\Delta_0\in\mathbb{R}^{m_0}$, $\Delta_k\in\mathbb{R}^{m_1}$, $k\in[p]$.
Given $\mE_j\subset[p]$, write $\Delta_{\mE_j}=(\Delta_k)_{k\in\mE_j}$. 
Assume that $m_1/m_0=\mathcal{O}(1)$, $s=o(T^{2/5})$ and $sm_1=\mathcal{O}(T^{4/5})$, where $s=\vert\mE_j\vert$.
For any constant $c>0$, it holds with probability at least $1-2c_{2}pT\exp(-c_{3}T^{1/5})-c_{2}'p^2T\exp(-c_{3}'T^{1/5})$ that
\begin{equation*}
\min_{\Delta\neq\bm{0}}\left\{\frac{\Delta^\top\G\Delta}{\|\Delta_{\mE_j}\|^2_2}: \sum_{k\notin \mE_j}\|\Delta_k\|_2\leq c\sum_{k\in \mE_j}\|\Delta_k\|_2, \,1\le j\le p\right\}\geq \frac{\gamma_{\min}}{4m_1},
\end{equation*}
where $c_2$, $c_2'$, $c_3$ and $c_3'$ are defined in Theorem~\ref{thm2}.
\end{lemma}

}

\subsection{Proof of Theorem~\ref{thm1}}
We first show the convergence of $\blambda^{(n)}(t)$ to $\blambda(t)$, and then show the limiting process of $\{\N^{(n)}\}_{n=1}^{\infty}$ is uniquely $\N$ with intensity \eqref{intensity}. Recall $h\{\nu_j(t)\}\leq \nu^*$ for any $t$ and $j$. We have that, for $j\in[p]$,
\begin{align*}
\begin{split}
&\quad \mathbb{E}\left|\lambda_{j}^{(n+1)}(t)-\lambda_{j}^{(n)}(t)\right|\\
&\leq \theta \mathbb{E}\left|\psi_{j}^{(n+1)}(t)-\psi_{j}^{(n)}(t)\right|\\
&=\theta \mathbb{E}\left|\int_{0}^{t}\sum_{k_{1}=1}^{p}\omega_{j,k_{1}}(t-s_{n-1})\left\{\dd N_{k_{1}}^{(n)}(s_{n-1})-\dd N_{k_{1}}^{(n-1)}(s_{n-1})\right\}\right|\\
&=\theta \int_{0}^{t}\sum_{k_{1}=1}^{p}|\omega_{j,k_{1}}(t-s_{n-1})|\times\mathbb{E}{\left|\overline{N}_{k_{1}}\left([0,\lambda_{k_{1}}^{(n)}(s_{n-1})]\times \dd s_{n-1}\right)-\overline{N}_{k_{1}}\left([0,\lambda_{k_{1}}^{(n-1)}(s_{n-1})]\times \dd s_{n-1}\right)\right|.}\\
&\leq\theta\int_{0}^{t}\sum_{k_{1}=1}^{p}|\omega_{j,k_{1}}(t-s_{n-1})|\times\mathbb{E}\left|\lambda_{k_{1}}^{(n)}(s_{n-1})-\lambda_{k_{1}}^{(n-1)}(s_{n-1})\right|\dd s_{n-1}\\
&\leq\theta^{2}\int_{0}^{t}\sum_{k_{1}=1}^{p}|\omega_{j,k_{1}}(t-s_{n-1})|\int_{0}^{s_{n-1}}\sum_{k_{2}=1}^{p}|\omega_{k_{1},k_{2}}(s_{n-1}-s_{n-2})|\\
&\hspace{2.5in}\times\mathbb{E}\left|\lambda_{k_{2}}^{(n-1)}(s_{n-2})-\lambda_{k_{2}}^{(n-2)}(s_{n-2})\right|\dd s_{n-2}\dd s_{n-1}\\
&\leq \theta^{n}\int_{0}^{t}\sum_{k_{1}=1}^{p}|\omega_{j,k_{1}}(t-s_{n-1})|\int_{0}^{s_{n-1}}\cdots\int_{0}^{s_{1}}\sum_{k_{n}=1}^{p}|\omega_{k_{n-1},k_{n}}(s_{1}-s_{0})|\times\mathbb{E}\left|\lambda_{k_{n}}^{(1)}(s_{1})\right|\dd s_{0}\cdots \dd s_{n-1}\\
&\leq {\nu^\ast} \theta^{n}\int_{0}^{t}\sum_{k_{1}=1}^{p}|\omega_{j,k_{1}}(t-s_{n-1})|\int_{0}^{s_{n-1}}\cdots\int_{0}^{s_{2}}\sum_{k_{n-1}=1}^{p}|\omega_{k_{n-2},k_{n-1}}(s_{2}-s_{1})|\left\{\bOmega_{k_{n-1},\cdot}^\top\mathbf{1}_{p}\right\}\dd s_{1}\cdots \dd s_{n-1}\\
&\leq {\nu^\ast} \theta^{n}\int_{0}^{t}\sum_{k_{1}=1}^{p}|\omega_{j,k_{1}}(t-s_{n-1})|\int_{0}^{s_{n-1}}\cdots\int_{0}^{s_{3}}\sum_{k_{n-2}=1}^{p}|\omega_{k_{n-3},k_{n-2}}(s_{3}-s_{2})|\left\{\bOmega_{k_{n-2},\cdot}^\top \bOmega\mathbf{1}_{p}\right\}\dd s_{2}\cdots \dd s_{n-1}\\
&\leq {\nu^\ast}\bOmega_{j,.}^\top \bOmega^{n-1}\mathbf{1}_{p},
\end{split}
\end{align*}
where the first inequality is due to the $\theta$-Lipschitz property of $h(\cdot)$, and the second equality is due to the construction in \eqref{thin}. 
Consequently, we have, in the vector form,
$$
\mathbb{E}\left|\blambda^{(n+1)}(t)-\blambda^{(n)}(t)\right|\preceq {\nu^\ast}\bOmega^{n}\mathbf{1}_{p}.
$$
Under Assumption~\ref{ass1}, we have that
\begin{equation}\label{recbound}
\sum_{j=1}^{p}\mathbb{E}\left|\lambda_{j}^{(n+1)}(t)-\lambda_{j}^{(n)}(t)\right|
\leq \nu \mathbf{1}_{p}^\top \bOmega^{n}\mathbf{1}_{p}\leq {\nu^\ast} p \sigma_{\bOmega}^{n}.
\end{equation}
By Chebyshev's inequality, it holds that
$$
\mathbb{P}\left\{\left|\lambda_{j}^{(n+1)}(t)-\lambda_{j}^{(n)}(t)\right|\geq\sigma_{\bOmega}^{n/2}\right\}\leq \frac{\mathbb{E}\left|\lambda_{j}^{(n+1)}(t)-\lambda_{j}^{(n)}(t)\right|}{\sigma_{\bOmega}^{n/2}}\leq {\nu^\ast} p\sigma_{\bOmega}^{n/2}.
$$
Therefore, we have
$$
\sum_{n=1}^{\infty}\mathbb{P}\left\{\left|\lambda_{j}^{(n+1)}(t)-\lambda_{j}^{(n)}(t)\right|\geq\sigma_{\bOmega}^{n/2}\right\}\leq{\nu^\ast} p\frac{\sigma_{\bOmega}^{1/2}}{1-\sigma_{\bOmega}^{1/2}}<\infty.
$$ 
By Lemma~\ref{finitelimit}, $\lambda_{j}^{(n)}(t)$ converges to a limit $\lambda_{j}(t)$ as $n\rightarrow\infty$.

Next, we show that the limiting process of $N_{j}^{(n)}$ exists.
First, for any bounded set $C\in \mathcal{B}(R)$, we have that
\begin{eqnarray*}
&&\sum_{n\geq 1}\mathbb{P}\left(\int_{C}\left|\dd N_{j}^{(n+1)}(t)-\dd N_{j}^{(n)}(t)\right|\neq 0\right)\\
&=& \sum_{n\geq 1}\mathbb{P}\left(\int_{C}\left|\dd N_{j}^{(n+1)}(t)-\dd N_{j}^{(n)}(t)\right|\geq 1\right)\\
&\leq& \sum_{n\geq 1}\mathbb{E}\left(\int_{C}\left|\dd N_{j}^{(n+1)}(t)-\dd N_{j}^{(n)}(t)\right|\right)\\
&=& \sum_{n\geq 1}\int_{C}\mathbb{E}\left|\lambda_{j}^{(n+1)}(t)-\lambda_{j}^{(n)}(t)\right|\dd t\\
&\leq& \left(\int_{C}\dd t \right){\nu^\ast} p\frac{\sigma_{\bOmega}}{1-\sigma_{\bOmega}}<\infty,
\end{eqnarray*}
where the first inequality is due to the Chebyshev inequality, the second equality is due to the construction in \eqref{thin}, and the last inequality is due to \eqref{recbound}. 
By the Borel-Cantelli lemma, the process $N_{j}^{(n)}$ remains eventually constant on any bounded set as $n$ increases. Therefore, $N_{j}^{(n)}$ converges to a limiting point process $N_{j}$ as $n\rightarrow \infty$.

We then show that the limiting process $N_{j}$ counts the points in $\overline{N}_{j}\left([0,\lambda_j(s)]\times\dd s\right)$. By Fatou's lemma, we have that, for any bounded set $C\in \mathcal{B}(R)$,
\begin{equation*}
\begin{split}
&\quad \mathbb{E}\int_{C}\left|\dd N_{j}(s)-\overline{N}_{j}\left([0,\lambda_j(s)]\times\dd s]\right)\right|\\
&\leq \lim_{n\rightarrow\infty}\mathbb{E}\int_{C}\left|\overline{N}_{j}\left([0,\lambda_j^{(n)}(s)]\times\dd s]\right)-\overline{N}_{j}\left([0,\lambda_j(s)]\times\dd s]\right)\right|\\
&=\left(\int_{C}\dd s\right)\lim_{n\rightarrow\infty}\mathbb{E}\left|\lambda_{j}^{(n)}(t)-\lambda_{j}(t)\right|=0.
\end{split}
\end{equation*}
Hence, the limiting process $N_{j}$ counts the points in $\overline{N}_{j}\left([0,\lambda_j(s)]\times\dd s\right)$.

At last, we verify that the limit $\lambda_{j}(t)=h(\psi_j(t))$, which is the intensity function of the limiting process $N_{j}$. By the Lipschitz property of $h(\cdot)$, we have
\begin{equation*}
\begin{split}
&\quad \mathbb{E}\left|\lambda_{j}(t)-h(\psi_j(t))\right|\\
&\leq \mathbb{E}\left|\lambda_{j}(t)-\lambda_{j}^{(n)}(t)\right|+\mathbb{E}\int_{0}^{t}\sum_{k=1}^{p}|\omega_{j,k}(t-s)|\times|\dd N_{j}(s)-\dd N_{j}^{(n-1)}(s)|\\
&=\mathbb{E}\left|\lambda_{j}(t)-\lambda_{j}^{(n)}(t)\right|+\int_{0}^{t}\sum_{k=1}^{p}|\omega_{j,k}(t-s)|\mathbb{E}\left|\lambda_{j}(s)-\lambda_{j}^{(n-1)}(s)\right|\dd s,
\end{split}
\end{equation*}
and let $n$ tends to $\infty$ in the above expression to conclude.

The uniqueness of the limit $N_{j}$ is guaranteed by Theorem 1 of \cite{massoulie1998stability}. 
{To apply Theorem 1 of \cite{massoulie1998stability}, several conditions need to be verified under our setting. 
The first condition requires that $\lambda_j(t)$ is upper bounded by a positive constant and this is satisfied under Assumption \ref{ass2}.
The second condition requires that the function $\lambda_j(t)$ satisfies a Lipschitz property, i.e., there exists $g_{j,k}$'s such that
\begin{equation}\label{eqn:lip}
\left|\lambda_j(t;N)-\lambda_j(t;N')\right|\leq \sum_{k=1}^p\int_0^t g_{j,k}(t-s)|N(\dd s)-N'(\dd s)|,
\end{equation}
where in our setting $\lambda_{j}(t;N)=h\left\{\nu_{j}(t)+\sum_{k=1}^{p}\int_{0}^{t}\omega_{j,k}(t-u)\dd N_{k}(u)\right\}$ 
and $\lambda_{j}(t;N')=h\left\{\nu_{j}(t)+\sum_{k=1}^{p}\int_{0}^{t}\omega_{j,k}(t-u)\dd N'_{k}(u)\right\}$; it is seen that \eqref{eqn:lip} holds by taking $g_{j,k}(t)=|\omega_{j,k}(t)|$ and by noting that $h$ is a $\theta$-Lipschitz function with $\theta\leq 1$.
The third condition requires that the $g_{j,k}$'s in \eqref{eqn:lip} satisfies $\int_{0}^{\infty}|g_{j,k}(t)|\dd t<\infty$. When $g_{j,k}(t)=|\omega_{j,k}(t)|$, this third condition is satisfied under Assumption \ref{ass2}.
This last condition requires that the initial condition satisfies 
\begin{equation*}
\lim_{t\rightarrow \infty}\sum_{k=1}^p\int_t^\infty \dd t'\int_{-\infty}^0 g_{j,k}(t'-s) N_k(\dd s)=0\quad a.s..
\end{equation*}
This holds when $g_{j,k}(t)=|\omega_{j,k}(t)|$, as $\omega_{j,k}$'s have a bounded support $[0,b]$ and $\int_{-\infty}^0 g_{j,k}(t'-s) N_k(\dd s)=0$ for $t'>b$. Hence, Theorem 1 of \cite{massoulie1998stability} is applicable and the uniqueness of the limit is ensured.
}

Putting together the above steps, we have shown that $\blambda^{(n)}(t)$ converges to $\blambda(t)$ almost surely for any $t$, and the limiting process of the sequence $\left\{\N^{(n)}\right\}_{n=1}^{\infty}$ is uniquely $\N$ with intensity \eqref{intensity}. 

\medskip
\noindent
\textbf{Remark:} When $p$ is diverging, we additionally need Assumption~\ref{ass3} for Theorem \ref{thm1} to hold. Specifically, for \eqref{recbound}, under Assumption~\ref{ass3}, we may bound it as
\begin{equation*}
\mathbb{E}\left|\lambda_{j}^{(n+1)}(t)-\lambda_{j}^{(n)}(t)\right|
\leq {\nu^\ast}\rho_{\bOmega}^n.
\end{equation*}
By Chebyshev's inequality, we have 
$$
\mathbb{P}\left\{\left|\lambda_{j}^{(n+1)}(t)-\lambda_{j}^{(n)}(t)\right|\geq\rho_{\bOmega}^{n/2}\right\}\leq \frac{\mathbb{E}\left|\lambda_{j}^{(n+1)}(t)-\lambda_{j}^{(n)}(t)\right|}{\rho_{\bOmega}^{n/2}}\leq {\nu^\ast} \rho_{\bOmega}^{n/2}.
$$
Consequently, it holds that
$$
\sum_{n=1}^{\infty}\mathbb{P}\left\{\left|\lambda_{j}^{(n+1)}(t)-\lambda_{j}^{(n)}(t)\right|\geq\rho_{\bOmega}^{n/2}\right\}\leq {\nu^\ast} \frac{\rho_{\bOmega}^{1/2}}{1-\rho_{\bOmega}^{1/2}}<\infty.
$$ 
Thus, by Lemma~\ref{finitelimit}, $\lambda_{j}^{(n)}(t)$ converges to a limit $\lambda_{j}(t)$ as $n\rightarrow\infty$.
\eop

\subsection{Proof of Theorem~\ref{thm2}}
\noindent
This proof is divided into two parts that show the concentration inequalities of the first and second order statistics, respectively.
 
\noindent
\textbf{Part I:} we show the concentration inequality of the first order statistic of the proposed Hawkes process. Here we employ a similar technique as in \cite{chen2017multivariate}. For a small constant $\epsilon$ such that $T/(2\epsilon)$ is an integer, define
$$
y_{k,i}=\frac{1}{2\epsilon}\int_{2\epsilon(i-1)}^{2\epsilon i} f_{1}(t)\dd N_{k}(t).
$$ 
We can regard $y_{k}$ as the average of $\{y_{k,i} \}_{i=1}^{T/(2\epsilon)}$, that is, $y_{k}=\frac{2\epsilon}{T}\sum_{i=1}^{T/(2\epsilon)}y_{k,i}$. Following \cite{dedecker2004coupling}, the temporal dependence of this sequence can be measured by
$$
\tau_{y}(l)=\sup_{u}\tau(\mathcal{H}_{u}^{y},y_{k,u+l})
$$
where $\tau(\cdot,\cdot)$ is defined as in Lemma~\ref{randombound}, $\mathcal{H}_{u}^{y}$ is a $\sigma$-field determined by $\{y_{k,i}\}_{i=1}^{u}$, $l$ is any positive gap and $u$ can be any positive integer. 
If we can show that $\{y_{k,i}\}_{i=1}^{T/(2\epsilon)}$ is a weakly dependent sequence, we can use the Bernstein type inequality in \cite{merlevede2011bernstein} to bound $\mathbb{P}(|y_{k}-\mathbb{E}y_{k}|\geq T\epsilon_{1})$. 
To do this, we need to verify two conditions in Theorem 1 of \cite{merlevede2011bernstein}.

First, we show that $y_{k,i}$ has an exponential tail of order $1$, that is,
\begin{equation}\label{seqtail1}
\sup_{i>0}\mathbb{P}(|y_{k,i}|\geq x)\leq a_5\exp(-a_6x),
\end{equation}
for some positive constants $a_{5}$ and $a_6$. Since $f_{1}$ is a bounded function, i.e., $\|f_1\|_{\infty}\le f_{\max}$, we know that
\begin{equation*}
y_{k,i}\leq \frac{f_{\max}}{2\epsilon}N_{k}\left([2\epsilon (i-1),2\epsilon]\right).
\end{equation*}
We can construct a dominating process $\hat{\bm{N}}$ as
\begin{equation*}
\dd \hat{N}_k(t)=\bar{N}_j([0,\lambda_{\max}]\times\dd t),\quad k\in[p].
\end{equation*}
Since $\lambda_k(t)\leq \lambda_{\max}$ as assumed in Assumption \ref{ass2}, by the construction in \eqref{thin}, it holds that $\dd \hat{N}_k(t)\geq \dd N_k^{(n)}(t)$ for any $n$ and we have
\begin{equation*}
\dd \hat{N}_k(t)\geq \lim_{n\rightarrow \infty}\dd N_k^{(n)}(t)=\dd N_j(t).
\end{equation*}
As such, we can conclude that for any $n$,
\begin{equation}\label{compos1}
\mathbb{P}\left\{N_{k}([2\epsilon (i-1),2\epsilon])\geq n\right\}\leq \mathbb{P}\left\{\hat{N}_{k}([2\epsilon (i-1),2\epsilon])\geq n\right\}.
\end{equation}
By definition, $\hat{N}_k([2\epsilon (i-1),2\epsilon])$ is a Poisson random variable. Then, by Lemma~\ref{poisson}, the tail probability in \eqref{compos1} decreases exponentially fast with $n$.


Second, we show that for any positive integer $l$, 
\begin{equation}
\label{tautail1}
\tau_{y}(l)\leq a_{7}\exp(-a_{8}l),
\end{equation}
for some positive constants $a_{7}$, $a_{8}$.
From Lemma~\ref{couple}, for any $z$, let $\widetilde{\N}$ be the process such that it has the same distribution as $\N$, and is independent of $\mathcal{H}_{2\epsilon z}$. 
We define $\widetilde{y}_{k,i}$ as 
\begin{equation}
\widetilde{y}_{k,i}\equiv \frac{1}{2\epsilon}\int_{2\epsilon(i-1)}^{2\epsilon i} f_{1}(t)\dd \widetilde{N}_{k}(t),
\end{equation}
From the definition of $\widetilde{\N}$, we have that $\{\widetilde{y}_{k,i}\}_{i=1}^{T/(2\epsilon)}$ has the same distribution as $\{y_{k,i}\}_{i=1}^{T/(2\epsilon)}$, and is independent of $\{y_{k,i}\}_{i=1}^{z}$. We have that,
\begin{eqnarray*}
	\mathbb{E}\left|\widetilde{y}_{k,z+l}-y_{k,z+l}\right|&=&\frac{1}{2\epsilon}\mathbb{E}\left|\int_{2\epsilon(z+l-1)}^{2\epsilon(z+l)}f_{1}(t)\left\{\dd\widetilde{N}_{k}(t)-\dd N_{k}(t)\right\}\right|\\
	&\leq& \frac{1}{2\epsilon}\int_{2\epsilon(z+l-1)}^{2\epsilon(z+l)}|f_{1}(t)|\times\mathbb{E}\left|\left\{\dd\widetilde{N}_{k}(t)-\dd N_{k}(t)\right\}\right|\\
	& \leq &a_{1}f_{\max}\exp\left\{-2a_2 \epsilon (z+l-1)\right\},
\end{eqnarray*}
where the last inequality follows from Lemma~\ref{exptail}. Let $a_7=a_1f_{\max}\exp(-2a_2\epsilon (z-1))$ and $a_8=2a_2\epsilon$, we have
\begin{equation*}
\mathbb{E}|\widetilde{y}_{k,z+l}-y_{k,z+l}|\leq a_{7}\exp(-a_{8}l).
\end{equation*}
By Lemma~\ref{randombound}, we have that
\begin{equation*}
\tau_{y}(l)\leq\mathbb{E}|\widetilde{y}_{k,z+l}-y_{k,z+l}|\leq a_{7}\exp(-a_{8}l).
\end{equation*}
With \eqref{seqtail1}, \eqref{tautail1} and Theorem 1 in \cite{merlevede2011bernstein}, we have that
\begin{eqnarray}\label{conineq1}
&&\mathbb{P}\left(\left|\sum_{i=1}^{T/(2\epsilon)}y_{k,i}-\frac{T}{2\epsilon}\mathbb{E}y_{k,i}\right|\geq T\epsilon_{1}\right)\\\nonumber
&&\leq \frac{T}{2\epsilon}\exp\left\{-\frac{(\epsilon_{1}T)^{1/2}}{C_{1}}\right\}+\exp\left\{-\frac{\epsilon_{1}^{2}T^{2}}{C_{2}(1+Tv_{y}/(2\epsilon))}\right\}+\exp\left[-\frac{\epsilon_{1}^{2}T^{2}}{C_{3}T/(2\epsilon)}\exp\left\{-\frac{(\epsilon_{1}T)^{1/4}}{C_{4}(\log(\epsilon_{1}T))^{1/3}}\right\}\right],
\end{eqnarray}
where $\epsilon_{1}$ is to be specified later, and $v_{y}$ is defined as
\begin{equation*}
v_{y}=\sup_{i>0}\left\{\mathbb{E}\left\{(y_{k,i}-\mathbb{E}y_{k,i})^{2}\right\}+2\sum_{l\geq 1}\mathbb{E}\left[(y_{k,i}-\mathbb{E}y_{k,i})(y_{k,i+l}-\mathbb{E}y_{k,i+l})\right] \right\}.
\end{equation*}
Furthermore, we have that
\begin{eqnarray*}
	v_{y}&=&\sup_{i>0}\left\{\mathbb{E}\left\{(y_{k,i}-\mathbb{E}y_{k,i})^{2}\right\}+2\sum_{l\geq 1}\mathbb{E}\{(y_{k,i}-\mathbb{E}y_{k,i})\mathbb{E}(y_{k,i+l}-\mathbb{E}y_{k,i+l}|y_{k,i})\} \right\}\\
	&\leq&  \sup_{i>0}\left\{\mathbb{E}\left\{(y_{k,i}-\mathbb{E}y_{k,i})^{2}\right\}+2\sum_{l\geq 1}a_{7}\exp(-a_{8}l)\mathbb{E}(|y_{k,i}-\mathbb{E}y_{k,i}|) \right\}.
\end{eqnarray*}
where the inequality follows from properties of $\tau$-dependence. 
From the above expression, we can see that $v_{y}$ is bounded. Let $\epsilon_{1}=a_{9}T^{-3/5}/(2\epsilon)$ and note that $\epsilon$ is fixed, we have
\begin{equation}\label{conineq11}
\mathbb{P}\left(\frac{1}{T}\left|\sum_{i=1}^{T/(2\epsilon)}y_{j,k,i}-\frac{T}{2\epsilon}\mathbb{E}y_{j,k,i}\right|\geq a_{9}T^{-3/5}/(2\epsilon)\right)\leq a_{10}T \exp(-a_{11}T^{1/5}).
\end{equation}
Reorganizing the terms in \eqref{conineq11}, we have that
\begin{equation*}
\mathbb{P}\left(|y_{j,k}-\mathbb{E}y_{j,k}|\geq c_{1}T^{-3/5}\right)\leq c_{2} T \exp(-c_{3}T^{1/5}),
\end{equation*}
for some positive constants $c_1$, $c_2$ and $c_3$.

\medskip
\noindent
\textbf{Part II:} we establish the concentration inequality of the second order statistic $y_{j,k}$. 
Similarly, for a small constant $\epsilon$ such that $T/(2\epsilon)$ is an integer, define
$$
y_{j,k,i}=\frac{1}{2\epsilon}\int_{2\epsilon(i-1)}^{2\epsilon i}\int_0^T f_{2}(t-t')\dd N_{k}(t')\dd N_{j}(t).
$$ 
$y_{j,k}$ can be regarded as the average of $\{y_{j,k,i} \}_{i=1}^{T/(2\epsilon)}$, that is, $y_{j,k}=\frac{2\epsilon}{T}\sum_{i=1}^{T/(2\epsilon)}y_{j,k,i}$. Following \cite{dedecker2004coupling}, the temporal dependence of this sequence can be measured by
$$
\tau'_{y}(l)=\sup_{u}\tau'(\mathcal{H}_{u}^{y},y_{j,k,u+l})
$$
where $\tau'(\cdot,\cdot)$ is defined as in Lemma~\ref{randombound}, $\mathcal{H}_{u}^{y}$ is a $\sigma$-field determined by $\{y_{j,k,i}\}_{i=1}^{u}$, $l$ is any positive gap and $u$ can be any positive integer. 
If we can show that $\{y_{j,k,i}\}_{i=1}^{T/(2\epsilon)}$ is a weakly dependent sequence, we can use the Bernstein type inequality in \cite{merlevede2011bernstein} to bound $\mathbb{P}(|y_{j,k}-\mathbb{E}y_{j,k}|\geq T\epsilon_{1})$. 
To this end, we need to verify two conditions in Theorem 1 of \cite{merlevede2011bernstein}.

First, we show that $y_{j,k,i}$ has an exponential tail of order $1/2$, that is,
\begin{equation}\label{seqtail}
\sup_{i>0}\mathbb{P}(|y_{j,k,i}|\geq x)\leq \exp(1-a_{12}x^{1/2}),
\end{equation}
for some constant $a_{12}$. Since $f_{2}$ is a bounded function, we have $\|f_2\|_{\infty}\le f_{\max}$ for some constant $f_{\max}>0$. As $f_{2}$ also has a bounded support, we assume, without loss of generality, that $\text{supp}(f)\subset [-b,0]$ for positive constant $b$. This can be generalized to $f$ with an arbitrary bounded support. To see this, consider three cases for $\text{supp}(f)=[b_1,b_2]$. The first case is $b_1<b_2<0$. It is then easy to see that $[b_1,b_2]\subset[-b,0]$ with $b=-b_1$. The second case is $0<b_1<b_2$. We can define $g(x)=f(-x)$ and then the proof can be applied to $g(x)$. The last case is $b_1<0<b_2$. In this case, $f$ can be written as $f=f_{+}+f_{-}$ where $f_{+}=f(x)\1_{[x>0]}$ and $f_{-}=f(x)\1_{[x\leq 0]}$. Let $g(x)=f_{+}(-x)$, then the proof is applicable for both $g(x)$ and $f_{-}(x)$ and thus to $f$. 
Next, we have that
\begin{equation*}
y_{j,k,i}\leq \frac{f_{\max}}{2\epsilon}N_{j}\left([2\epsilon (i-1),2\epsilon]\right)N_{k}\left([2\epsilon (i-1)-b,2\epsilon]\right).
\end{equation*}
{Given \eqref{compos1} and Lemma~\ref{poisson}, it holds that the tail probability of $N_{j}\left([2\epsilon (i-1),2\epsilon]\right)$ decreases exponentially fast with $n$. The same argument holds for $N_{k}\left([2\epsilon (i-1)-b,2\epsilon]\right)$. Thus, applying Lemma~\ref{poisson2}, we have that $y_{j,k,i}$ has an exponential tail of order $1/2$.
}

Second, we show that for any positive integer $l$, 
\begin{equation}
\label{tautail}
\tau'_{y}(l)\leq a_{13}\exp(-a_{14}l),
\end{equation}
for some constants $a_{13}$, $a_{14}$.
From Lemma~\ref{couple}, for any $z$, let $\widetilde{\N}$ be the process such that it has the same distribution as $\N$, and is independent of $\mathcal{H}_{2\epsilon z}$. 
We define $\widetilde{y}_{j,k,i}$ as 
\begin{equation}
\widetilde{y}_{j,k,i}\equiv \frac{1}{2\epsilon}\int_{2\epsilon(i-1)}^{2\epsilon i}\int_{0}^T f_{2}(t-t')\dd \widetilde{N}_{k}(t')\dd \widetilde{N}_{j}(t),
\end{equation}
From the definition of $\widetilde{\N}$, we have that $\{\widetilde{y}_{j,k,i}\}_{i=1}^{T/(2\epsilon)}$ has the same distribution as $\{y_{j,k,i}\}_{i=1}^{T/(2\epsilon)}$, and is independent of $\{y_{j,k,i}\}_{i=1}^{z}$. We have that,
\begin{eqnarray*}
\mathbb{E}\left|\widetilde{y}_{j,k,z+l}-y_{j,k,z+l}\right|&=&\frac{1}{2\epsilon}\mathbb{E}\left|\int_{2\epsilon(z+l-1)}^{2\epsilon(z+l)}\int_{t{-b}}^{t}f_{2}(t-t')\left\{\dd\widetilde{N}_{k}(t')\dd\widetilde{N}_{j}(t)-\dd N_{k}(t')\dd N_j(t)\right\}\right|\\
&\leq& \frac{1}{2\epsilon}\int_{2\epsilon(z+l-1)}^{2\epsilon(z+l)}\int_{t{-b}}^{t}|f_{2}(t-t')|\times\mathbb{E}\left|\left\{\dd\widetilde{N}_{k}(t')\dd\widetilde{N}_{j}(t)-\dd N_{k}(t')\dd N_j(t)\right\}\right|\\
& \leq &\frac{f_{\max}}{2\epsilon}\int_{2\epsilon(z+l-1)}^{2\epsilon(z+l)}\int_{t-b}^{t}\mathbb{E}\left|\left\{\dd\widetilde{N}_{k}(t')\dd\widetilde{N}_{j}(t)-\dd N_{k}(t')\dd N_j(t)\right\}\right|\\
& \leq &\frac{f_{\max}}{2\epsilon}\int_{2\epsilon(z+l-1)}^{2\epsilon(z+l)}\int_{t-b}^{t}a_{3}\exp\left\{-a_{4}(\min(t',t)-2\epsilon z)\right\}\dd t'\dd t,
\end{eqnarray*}
where the last inequality follows from Lemma~\ref{exptail}. Given that $\min(t',t)-2\epsilon z=2\epsilon (z+l-1)-b-2\epsilon z=2\epsilon(l-1)-b$, we have
\begin{equation*}
\mathbb{E}|\widetilde{y}_{j,k,z+l}-y_{j,k,z+l}|\leq a_{15}\exp(-a_{16}l),
\end{equation*}
where $a_{15}=a_{3}f_{\max}b\exp(-b-2\epsilon)$ and $a_{16}=2  a_{4}\epsilon$. By Lemma~\ref{randombound}, we have that
\begin{equation*}
\tau_{y}(l)\leq\mathbb{E}|\widetilde{y}_{j,k,z+l}-y_{j,k,z+l}|\leq a_{15}\exp(-a_{16}l).
\end{equation*}
With \eqref{seqtail}, \eqref{tautail} and Theorem 1 in \cite{merlevede2011bernstein}, we have that
\begin{eqnarray}\label{conineq2}
&&\mathbb{P}\left(\left|\sum_{i=1}^{T/(2\epsilon)}y_{j,k,i}-\frac{T}{2\epsilon}\mathbb{E}y_{j,k,i}\right|\geq T\epsilon_{1}\right)\\\nonumber
&&\leq \frac{T}{2\epsilon}\exp\left\{-\frac{(\epsilon_{1}T)^{1/3}}{C'_{1}}\right\}+\exp\left\{-\frac{\epsilon_{1}^{2}T^{2}}{C'_{2}(1+Tv_{y}/(2\epsilon))}\right\}+\exp\left[-\frac{\epsilon_{1}^{2}T^{2}}{C'_{3}T/(2\epsilon)}\exp\left\{-\frac{(\epsilon_{1}T)^{2/9}}{C'_{4}(\log(\epsilon_{1}T))^{1/3}}\right\}\right],
\end{eqnarray}
where $\epsilon_{1}$ is to be specified later, and $v_{y}$ is defined as
\begin{equation*}
v_{y}'=\sup_{i>0}\left\{\mathbb{E}\left\{(y_{j,k,i}-\mathbb{E}y_{j,k,i})^{2}\right\}+2\sum_{l\geq 1}\mathbb{E}\left[(y_{j,k,i}-\mathbb{E}y_{j,k,i})(y_{j,k,i+l}-\mathbb{E}y_{j,k,i+l})\right] \right\}.
\end{equation*}
Furthermore, we have that
\begin{eqnarray*}
v_{y}'&=&\sup_{i>0}\left\{\mathbb{E}\left\{(y_{j,k,i}-\mathbb{E}y_{j,k,i})^{2}\right\}+2\sum_{l\geq 1}\mathbb{E}\{(y_{j,k,i}-\mathbb{E}y_{j,k,i})\mathbb{E}(y_{j,k,i+l}-\mathbb{E}y_{j,k,i+l}|y_{j,k,i})\} \right\}\\
&\leq& \sup_{i>0}\left\{\mathbb{E}\left\{(y_{j,k,i}-\mathbb{E}y_{j,k,i})^{2}\right\}+2\sum_{l\geq 1}a_{15}\exp(-a_{16}l)\mathbb{E}(|y_{j,k,i}-\mathbb{E}y_{j,k,i}|) \right\}.
\end{eqnarray*}
where the inequality follows from properties of $\tau$-dependence. 
From the above expression, we can see that $v_{y}'$ is bounded. Let $\epsilon_{1}=a_{17}T^{-2/5}/(2\epsilon)$ and note that $\epsilon$ is fixed, we have
\begin{equation}\label{conineq21}
\mathbb{P}\left(\frac{1}{T}\left|\sum_{i=1}^{T/(2\epsilon)}y_{j,k,i}-\frac{T}{2\epsilon}\mathbb{E}y_{j,k,i}\right|\geq a_{17}T^{-2/5}/(2\epsilon)\right)\leq a_{18}T \exp(-a_{19}T^{1/5}).
\end{equation}
Reorganizing the terms in \eqref{conineq21}, we have that
\begin{equation*}
\mathbb{P}\left(|y_{j,k}-\mathbb{E}y_{j,k}|\geq c_{1}'T^{-2/5}\right)\leq c_{2}' T \exp(-c_{3}'T^{1/5}),
\end{equation*}
for some positive constants $c'_1$, $c'_2$ and $c'_3$.
\eop

\subsection{Proof of Corollary~\ref{cor1}}
We consider the following two types of entries in $\G$.
\begin{itemize}
\item[(\romannumeral1)] The entries in $\G^{(0,k)}$, $k\in[p]$, can be written as
\[
\frac{1}{T}\int_{0}^T\phi_{0,l_{0}}(t)\left\{\int_{0}^{t}\phi_{1,l_{k}}(t-s)\dd N_{k}(s)\right\}\dd t.
\]
\item[(\romannumeral2)] The entries in $\G^{(k_1,k_2)}=\frac{1}{T}\int_{0}^T\bPsi_{k_1}(t)\bPsi_{k_2}(t)\dd t$, $k_1,k_2\in[p]$, can be written as 
\[
\frac{1}{T}\int_{0}^T\left\{\int_{0}^{t}\phi_{1,l_{1}}(t-s_{1})\dd N_{k_1}(s_{1})\right\} \left\{\int_{0}^{t}\phi_{1,l_{2}}(t-s_{2})\dd N_{k_2}(s_{2})\right\}\dd t.
\]
\end{itemize} 

For $\G^{(0,k)}$, $k\in[p]$, we have that
\begin{equation}\label{G0k}
\G^{(0,k)}_{l_1l_2}=\frac{1}{T}\int_{0}^T\int_{s}^T\phi_{0,l_{0}}(t)\phi_{1,l_k}(t-s)\dd t \dd N_{k}(s), 
\end{equation}
where $l_1\in[m_0]$ and $l_2\in[m_1]$.
This is a direct result from the Fubini's Theorem. By Cauchy-Schwarz inequality, we have
\begin{equation*}
\int_{s}^T\phi_{0,l_{0}}(t)\phi_{1,l_{k}}(t-s)\dd t\le \sqrt{\int_{s}^{s+b}\phi_{0,l_0}^2(t)\dd t}\sqrt{\int_{s}^{s+b}\phi_{1,l_k}^2(t-s)\dd t}
\end{equation*}
Since $\phi_{0,l_{0}}$ and $\phi_{1,l_{k}}$ are bounded functions, \eqref{firstorder} in Theorem~\ref{thm2} is applicable, we have that 
$$
\mathbb{P}\left(|\G^{(0,k)}_{l_1l_2}-\mathbb{E}(\G^{(0,k)}_{l_1l_2})|\ge c_1T^{-3/5}\right)\le c_2T\exp(-c_3T^{1/5}).
$$
For $\G^{(k_1,k_2)}$, $k_1,k_2\in[p]$, we have that 
\begin{eqnarray*}
\G^{(k_1,k_2)}_{l_1l_2}&=&\frac{1}{T}\int_{0}^T\left\{\int_{0}^{t}\phi_{1,l_{1}}(t-s_{1})\dd N_{k_1}(s_{1})\right\} \left\{\int_{0}^{t}\phi_{1,l_{2}}(t-s_{2})\dd N_{k_2}(s_{2})\right\}\dd t\\
&=&\frac{1}{T}\int_{0}^T\dd t\int_{0}^{t}\dd N_{k_1}(s_{1})\int_{0}^{t}\dd N_{k_2}(s_{2})\phi_{1,l_{1}}(t-s_{1})\phi_{1,l_{2}}(t-s_{2})\\
&=&\frac{1}{T}\int_{0}^T\dd N_{k_1}(s_{1})\int_{s_{1}}^T\int_{0}^{t}\dd N_{k_2}(s_{2})\phi_{1,l_{1}}(t-s_{1})\phi_{1,l_{2}}(t-s_{2})\dd t\\
&=&\frac{1}{T}\int_{0}^T\dd N_{k_1}(s_{1})\int_{0}^T\dd N_{k_2}(s_{2})\int_{\max\{s_{1},s_{2}\}}^{{b}+\min\{s_{1},s_{2}\}}\phi_{1,l_{1}}(t-s_{1})\phi_{1,l_{2}}(t-s_{2})\dd t\\
&=&\frac{1}{T}\int_{0}^T\dd N_{k_1}(s_{1})\int_{0}^T\dd N_{k_2}(s_{2})\int_{\max\{0,s_{2}-s_{1}\}}^{{b}+\min\{0,s_{2}-s_{1}\}}\phi_{1,l_{1}}(s)\phi_{1,l_{2}}(s+s_{1}-s_{2})\dd s,
\end{eqnarray*}
where the second and third equalities are direct consequences of the Fubini's Theorem, and the fourth equality is due to that $\phi_{1,l}(\cdot)$ has a bounded support. By Cauchy-Schwarz inequality, we have
\begin{equation*}
\int_{\max\{0,s_{2}-s_{1}\}}^{{b}+\min\{0,s_{2}-s_{1}\}}\phi_{1,l_{j}}(s)\phi_{1,l_{k}}(s+s_{1}-s_{2})\dd s\leq\sqrt{\int_0^b\phi_{1,l_1}^2(s)\dd s}\sqrt{\int_0^b\phi_{1,l_2}^2(s)\dd s}
\end{equation*}
Since $\phi_{1,l_k}$ are bounded functions with bounded support, $\int_{\max\{0,s_{2}-s_{1}\}}^{{b}+\min\{0,s_{2}-s_{1}\}}\phi_{1,l_{1}}(s)\phi_{1,l_{2}}(s+s_{1}-s_{2})\dd s$ is a bounded function with a bounded support.
Applying \eqref{secondorder} in Theorem~\ref{thm2}, we have that 
\begin{equation}\label{Gk1k2}
\mathbb{P}\left(|\G^{(k_1,k_2)}_{l_1l_2}-\mathbb{E}(\G^{(k_1,k_2)}_{l_1l_2})|\ge c'_1T^{-2/5}\right)\le c'_2T\exp(-c'_3T^{1/5}).
\end{equation}
Putting \eqref{G0k} and \eqref{Gk1k2} together, it is sraightforward to get that
\begin{equation*}
\mathbb{P}\left[\bigcap_{i\neq j}\left\{\left|\G_{ij}-\mathbb{E}(\G_{ij})\right|\le c_{4}T^{-2/5}\right\}\right]\ge 1-c_{5}(p+1)^{2}T \exp(-c_{6}T^{1/5}).
\end{equation*}
for some positive constants $c_{4}$, $c_{5}$ and $c_{6}$.
\eop

\subsection{Proof of Theorem~\ref{thm3}}
Since $\hat{\bbeta}_{j}$ is the minimizer of \eqref{obj}, we have that
\begin{equation}\label{hattilde}
\begin{split}
&-\int_{0}^T \hat{\lambda}_{j}(t)\dd N_j(t)+\frac{1}{2}\int_{0}^T \hat{\lambda}_{j}(t)^{2}\dd t+\eta_j\sqrt{T}\sum_{k=1}^p\left\{\int_{0}^T \left(\bPsi_{k}^\top(t)\hat{\bbeta}_{j,k}\right)^{2}\dd t \right\}^{1/2} \\
&\quad \leq -\int_{0}^T \widetilde{\lambda}_{j}(t)\dd N_j(t)+\frac{1}{2}\int_{0}^T \widetilde{\lambda}_{j}(t)^{2}\dd t+\eta_j\sqrt{T}\sum_{k=1}^p\left\{\int_{0}^T \left(\bPsi_{k}^\top(t)\widetilde{\bbeta}_{j,k}\right)^{2}\dd t \right\}^{1/2}, 
\end{split}
\end{equation}
where $\widetilde{\lambda}_{j}(t)=\bPsi^\top(t)\widetilde{\bbeta}_{j}$ is as defined in Assumption~\ref{ass6}.
By adding $\frac{1}{2}\int_{0}^T \lambda_{j}^{2}(t)\dd t-\int_{0}^T \hat{\lambda}_{j}(t)\lambda_{j}(t)\dd t$ to both sides of \eqref{hattilde}, we have
\begin{eqnarray}\label{orgbound}
&&\frac{1}{2}\int_{0}^T \left\{\hat{\lambda}_{j}(t)-\lambda_{j}(t)\right\}^{2}\dd t \\\nonumber
&\leq&\underbrace{\frac{1}{2}\int_{0}^T \left\{\widetilde{\lambda}_{j}(t) -\lambda_{j}(t)\right\}^{2}\dd t}_{I_1}+\underbrace{\int_{0}^T \left\{\widetilde{\lambda}_{j}(t)- \hat{\lambda}_{j}(t)\right\}\left\{\lambda_{j}(t)\dd t-\dd N_j(t)\right\}}_{I_2} \\\nonumber
&& \qquad+\eta_{j}\sqrt{T}\underbrace{\sum_{k=1}^p\left\{\left\{\int_{0}^T \left(\bPsi_{k}^\top(t)\widetilde{\bbeta}_{j,k}\right)^{2}\dd t  \right \}^{1/2} -\left\{\int_{0}^T \left(\bPsi_{k}^\top(t)\hat{\bbeta}_{j,k}\right)^{2}\dd t \right\}^{1/2}   \right\}}_{I_3}.
\end{eqnarray}
For term $I_3$, we consider two different cases with $I_3>0$ and $I_3\le 0$, respectively. First, by Cauchy-Schwarz inequality, we have that
\begin{eqnarray}\label{penaltybound}
&&\sum_{k=1}^{p}\left\{\int_{0}^T (\bPsi_{k}^\top(t)\widetilde{\bbeta}_{j,k})^{2}\dd t \right\}^{1/2}\\\nonumber
&\leq&\sum_{k\in \mathcal{E}_{j}}\left[\int_{0}^T \{\bPsi_{k}^\top(t)(\hat{\bbeta}_{j,k}-\widetilde{\bbeta}_{j,k})\}^{2}\dd t \right]^{1/2}+\sum_{k\in \mathcal{E}_{j}}\left\{\int_{0}^T (\bPsi_{k}^\top(t)\hat{\bbeta}_{j,k})^{2}\dd t \right\}^{1/2},
\end{eqnarray}
where we used the fact that $\widetilde{\bbeta}_{j,k}=\mathbf{0}$ when $k\in \mathcal{E}_{j}^c$.
Additionally, we have that
\begin{eqnarray}\label{penaltydec}
&&\sum_{k=1}^{p}\left\{\int_{0}^T (\bPsi_{k}^\top(t)\hat{\bbeta}_{j,k})^{2}\dd t \right\}^{1/2}\\\nonumber
&=&\sum_{k\in \mathcal{E}_{j}}\left\{\int_{0}^T (\bPsi_{k}^\top(t)\hat{\bbeta}_{j,k})^{2}\dd t\right\}^{1/2}+\sum_{k\in \mathcal{E}_{j}^c}\left[\int_{0}^T (\bPsi_{k}^\top(t)\hat{\bbeta}_{j,k})^{2}\dd t  \right]^{1/2}\\\nonumber
&=&\sum_{k\in \mathcal{E}_{j}}\left\{\int_{0}^T (\bPsi_{k}^\top(t)\hat{\bbeta}_{j,k})^{2}\dd t\right\}^{1/2}+\sum_{k\in \mathcal{E}_{j}^c}\left[\int_{0}^T \{\bPsi_{k}^\top(t)(\hat{\bbeta}_{j,k}-\widetilde{\bbeta}_{j,k})\}^{2}\dd t  \right]^{1/2},
\end{eqnarray}
where we again used the fact that $\widetilde{\bbeta}_{j,k}=\mathbf{0}$ when $k\in \mathcal{E}_{j}^c$.
If $I_3>0$, combining \eqref{penaltybound} and \eqref{penaltydec}, we have that
\begin{equation}\label{edgemore}
\sum_{k\in \mathcal{E}_{j}}\left[\int_{0}^T \left\{\bPsi_{k}^\top(t)(\hat{\bbeta}_{j,k}-\widetilde{\bbeta}_{j,k})\right\}^{2}\dd t \right]^{1/2}\geq\sum_{{k\in \mathcal{E}_{j}^c}}\left[\int_{0}^T \left\{\bPsi_{k}^\top(t)(\hat{\bbeta}_{j,k}-\widetilde{\bbeta}_{j,k})\right\}^{2}\dd t  \right]^{1/2}.
\end{equation}
Write $\Delta=\hat{\bbeta}_j-\widetilde{\bbeta}_j$. {With \eqref{edgemore} and Lemma~\ref{lemma:block}, we have
\begin{equation*}
\begin{aligned}
\sqrt{\frac{3\gamma_{\max}}{2m_1}} \sum_{k\in \mathcal{E}_{j}}\|\Delta_k\|_2&\geq \sum_{k\in \mathcal{E}_{j}}\left[\frac{1}{T}\int_{0}^T \left\{\bPsi_{k}^\top(t)(\hat{\bbeta}_{j,k}-\widetilde{\bbeta}_{j,k})\right\}^{2}\dd t \right]^{1/2}\\
&\geq\sum_{{k\in \mathcal{E}_{j}^c}}\left[\frac{1}{T}\int_{0}^T \left\{\bPsi_{k}^\top(t)(\hat{\bbeta}_{j,k}-\widetilde{\bbeta}_{j,k})\right\}^{2}\dd t  \right]^{1/2}\geq\sqrt{\frac{\gamma_{\min}}{2m_1}}\sum_{k\in\mathcal{E}_j^c}\|\Delta_k\|_2
\end{aligned}
\end{equation*}	
with probability at least $1-c_{2}pT\exp(-c_{3}T^{1/5})-c_{2}'pT\exp(-c_{3}'T^{1/5})$. This implies that 
\begin{equation*}
\sum_{k\in\mathcal{E}_j^c}\|\Delta_k\|_2\leq\sqrt{\frac{3\gamma_{\max}}{\gamma_{\min}}}\sum_{k\in \mathcal{E}_{j}}\|\Delta_k\|_2,
\end{equation*}
with probability at least $1-c_{2}pT\exp(-c_{3}T^{1/5})-c_{2}'pT\exp(-c_{3}'T^{1/5})$. Apply Lemma~\ref{lemma:re} with $c=\sqrt{\frac{3\gamma_{\max}}{\gamma_{\min}}}$ gives that
\begin{equation}\label{alltrue}
\frac{1}{T}\int_{0}^T \left\{\hat{\lambda}_{j}(t)-\widetilde{\lambda}_{j}(t)\right\}^{2}\dd t=\Delta^\top\G\Delta\geq \frac{\gamma_{\min}}{4m_1}\sum_{k\in\mathcal{E}_j}\|\Delta_k\|_2^2,
\end{equation} 
with probability at least {$1-2c_2pT\exp(-c_{3}T^{1/5})-c_2'p^2T\exp(-c_{3}'T^{1/5})$}.
By \eqref{penaltybound} and Lemma~\ref{lemma:block}, it holds with probability at least $1-c_{2}pT\exp(-c_{3}T^{1/5})-c_{2}'pT\exp(-c_{3}'T^{1/5})$,
\begin{equation}\label{pentrue}
 \begin{aligned}
&\sum_{k=1}^{p}\left[\left\{\frac{1}{T}\int_{0}^T\left(\bPsi_{k}^\top(t)\widetilde{\bbeta}_{j,k}\right)^{2}\dd t \right\}^{1/2} -\left\{\frac{1}{T}\int_{0}^T \left(\bPsi_{k}^\top(t)\hat{\bbeta}_{j,k}\right)^{2}\dd t \right\}^{1/2}   \right]\\
\le& \sum_{k\in\mathcal{E}_j}\left[\frac{1}{T}\int_{0}^T \left\{\bPsi_{k}^\top(t)(\hat{\bbeta}_{j,k}-\widetilde{\bbeta}_{j,k})\right\}^{2}\dd t \right]^{1/2} 
\leq \sqrt{\frac{3s\gamma_{\max}}{2m_1}}\left(\sum_{k\in\mathcal{E}_j}\|\Delta_{k}\|_2^2\right)^{1/2},
\end{aligned}
\end{equation} 
where we used the fact that $(\sum_{k\in\mathcal{E}_j}\|\Delta_{k}\|_2)^2\le s\sum_{k\in\mathcal{E}_j}\|\Delta_{k}\|_2^2$.
Plugging \eqref{pentrue} into \eqref{alltrue}, we have that
\begin{eqnarray}\label{penbound}
&&\sum_{k=1}^{p}\left[\left\{\frac{1}{T}\int_{0}^T \left(\bPsi_{k}^\top(t)\widetilde{\bbeta}_{j,k}\right)^{2}\dd t \right\}^{1/2} -\left\{\frac{1}{T}\int_{0}^T \left(\bPsi_{k}^\top(t)\hat{\bbeta}_{j,k}\right)^{2}\dd t \right\}^{1/2} \right]\\\nonumber
&\leq&\sqrt{\frac{6s\gamma_{\max}}{\gamma_{\min}}}\left[\frac{1}{T}\int_{0}^T \left\{\hat{\lambda}_{j}(t)-\widetilde{\lambda}_{j}(t)\right\}^{2}\dd t \right]^{1/2},
\end{eqnarray}
holds with probability at least {$1-3c_{2}pT\exp(-c_{3}T^{1/5})-c_{2}'p(p+1)T\exp(-c_{3}'T^{1/5})$}.}
Combining \eqref{orgbound} and \eqref{penbound}, we have that
\begin{equation}\label{orgbound1}
\begin{aligned}
\frac{1}{2}\int_{0}^T \left\{\hat{\lambda}_{j}(t)-\lambda_{j}(t)\right\}^{2}\dd t 
\leq &\frac{1}{2}\int_{0}^T \left\{\widetilde{\lambda}_{j}(t) -\lambda_{j}(t)\right\}^{2}\dd t+\int_{0}^T \left\{\widetilde{\lambda}_{j}(t)- \hat{\lambda}_{j}(t)\right\}\left\{\lambda_{j}(t)\dd t-\dd N_j(t)\right\} \\
&+\eta_{j}\sqrt{T}\sqrt{\frac{6s\gamma_{\max}}{\gamma_{\min}}}\left[\int_{0}^T \left\{\hat{\lambda}_{j}(t)-\widetilde{\lambda}_{j}(t)\right\}^{2}\dd t \right]^{1/2},
\end{aligned}
\end{equation}
with probability at least {$1-3c_{2}pT\exp(-c_{3}T^{1/5})-c_{2}'p(p+1)T\exp(-c_{3}'T^{1/5})$}. If $I_3\le0$, it is straightforward that \eqref{orgbound1} still holds. 

Next, by Lemma~\ref{martingale} and setting $\epsilon=\frac{2\log p}{T}$, we have that 
\begin{equation}\label{stoerror}
I_2\leq4\sqrt{\lambda_{\max}\log p}\left[\int_{0}^T \left\{\widetilde{\lambda}_{j}(t)-\hat{\lambda}_{j}(t)\right\}^{2}\dd t \right]^{1/2},
\end{equation}
holds with probability at least $1-C_{3}p^{-2}$. Plugging \eqref{stoerror} into \eqref{orgbound1}, we have that
\begin{equation*}
\begin{aligned}
\frac{1}{2}\int_{0}^T \left\{\hat{\lambda}_{j}(t)-\lambda_{j}(t)\right\}^{2}\dd t\leq& \frac{1}{2}\int_{0}^T \left\{\widetilde{\lambda}_{j}(t) -\lambda_{j}(t)\right\}^{2}\dd t\\
&+\sqrt{T}\left(\eta_{j}\sqrt{\frac{6s\gamma_{\max}}{\gamma_{\min}}}+4\sqrt{\frac{\lambda_{\max}\log p}{T}}\right)\left[\int_{0}^T \left\{\hat{\lambda}_{j}(t)-\widetilde{\lambda}_{j}(t)\right\}^{2}\dd t \right]^{1/2}
\end{aligned}
\end{equation*}
holds with the probability at least {$1-C_{3}p^{-2}-3c_{2}pT\exp(-c_{3}T^{1/5})-c_{2}'p(p+1)T\exp(-c_{3}'T^{1/5})$}. Setting $\eta_{j}=\left\{C_2\log p/T\right\}^{1/2}$ with $C_2\geq \frac{\gamma_{\min}\lambda_{\max}}{6\gamma_{\max}}$, we have that
\begin{eqnarray}\label{orgbound2}
&&\frac{1}{2}\int_{0}^T \left\{\hat{\lambda}_{j}(t)-\lambda_{j}(t)\right\}^{2}\dd t\\\nonumber
&\leq& \frac{1}{2}\int_{0}^T \left\{\widetilde{\lambda}_{j}(t) -\lambda_{j}(t)\right\}^{2}\dd t+6\sqrt{s\lambda_{\max}\log p}\left[\int_{0}^T \left\{\hat{\lambda}_{j}(t)-\widetilde{\lambda}_{j}(t)\right\}^{2}\dd t \right]^{1/2},
\end{eqnarray}
with the probability at least $1-C_{3}p^{-2}-3c_{2}pT\exp(-c_{3}T^{1/5})-c_{2}'p(p+1)T\exp(-c_{3}'T^{1/5})$.
Next, we consider three different cases:
\begin{eqnarray*}
&&\text{i.}\int_{0}^T \left\{\hat{\lambda}_{j}(t)-\lambda_{j} (t)\right\}^{2}\dd t\geq \max\left\{\int_{0}^T \left\{\widetilde{\lambda}_{j}(t)-\lambda_{j} (t)\right\}^{2}\dd t,\int_{0}^T \left\{\hat{\lambda}_{j}(t)-\widetilde{\lambda}_{j} (t)\right\}^{2}\dd t\right\},\\
&&\text{ii.}\int_{0}^T \left\{\hat{\lambda}_{j}(t)-\lambda_{j} (t)\right\}^{2}\dd t< \int_{0}^T \left\{\widetilde{\lambda}_{j}(t)-\lambda_{j} (t)\right\}^{2}\dd t,\\
&&\text{iii.}\int_{0}^T \left\{\hat{\lambda}_{j}(t)-\lambda_{j} (t)\right\}^{2}\dd t< \int_{0}^T \left\{\hat{\lambda}_{j}(t)-\widetilde{\lambda}_{j} (t)\right\}^{2}\dd t.
\end{eqnarray*}
Under Case (i), \eqref{orgbound2} can be written as
\begin{eqnarray}\label{case1}
&&\frac{1}{2}\int_{0}^T \left\{\hat{\lambda}_{j}(t)-\lambda_{j}(t)\right\}^{2}\dd t\leq \frac{1}{2}\left[\int_{0}^T \left\{\widetilde{\lambda}_{j}(t)-\lambda_{j} (t)\right\}^{2}\dd t\int_{0}^T \left\{\hat{\lambda}_{j}(t)-\lambda_{j} (t)\right\}^{2}\dd t\right]^{1/2}\\\nonumber
&&\hspace{2.5in}+6\left\{s\lambda_{\max}\log p\right\}^{1/2}\left[\int_{0}^T \left\{\hat{\lambda}_{j}(t)-\lambda_{j} (t)\right\}^{2}\dd t\right]^{1/2}.\nonumber
\end{eqnarray}
By Assumption~\ref{ass6}, we have that 
\begin{equation}\label{approerror1}
\frac{1}{T}\int_{0}^T \left\{\widetilde{\lambda}_{j}(t)-\lambda_{j}(t)\right\}^{2}\dd t\leq {C_1(s+1)^2m_1^{-2d}},
\end{equation}
with probability at least {$1-C_{2}'pT\exp(-C_{3}'T^{1/5})$}. 
Plugging \eqref{approerror1} into \eqref{case1}, we have {with the probability at least $1-C_{3}p^{-2}-3c_{2}pT\exp(-c_{3}T^{1/5})-c_{2}'p(p+1)T\exp(-c_{3}'T^{1/5})-C_{2}'pT\exp(-C_{3}'T^{1/5})$,}
$$
\frac{1}{T}\int_{0}^T \left\{\hat{\lambda}_{j}(t)-\lambda_{j} (t)\right\}^{2}\dd t\leq {2C_1(s+1)^2m_1^{-2d}}+288s\lambda_{\max}\frac{\log p}{T}.
$$
Under Case (ii), we directly have
\begin{equation*}
\frac{1}{T}\int_{0}^T \left\{\hat{\lambda}_{j}(t)-\lambda_{j} (t)\right\}^{2}\dd t\leq {C_1(s+1)^2m_1^{-2d}}.
\end{equation*}
with probability at least {$1-C_{2}'pT\exp(-C_{3}'T^{1/5})$}.
Next, we consider Case (iii).
Similar to \eqref{orgbound}, we can get that
\begin{eqnarray*}
&&\frac{1}{2}\int_{0}^T \left\{\hat{\lambda}_{j}(t)-\widetilde{\lambda}_{j}(t)\right\}^{2}\dd t\\
&\leq& \int_{0}^T \left\{\widetilde{\lambda}_{j}(t)-\hat{\lambda}_{j}(t)\right\}\left\{\lambda_{j} (t)\dd t-\dd N(t)\right\}+\int_{0}^T \left\{\widetilde{\lambda}_{j}(t)-\hat{\lambda}_{j}(t)\right\}\left\{\widetilde{\lambda}_{j}(t)-\lambda_{j} (t)\right\}\dd t\\
&&\qquad +\eta_{j}\sqrt{T}\sum_{k=1}^{p}\left[\left\{\int_{0}^T \left(\bPsi_{k}^\top(t)\widetilde{\bbeta}_{j,k}\right)^{2}\dd t \right\}^{1/2}-\left\{\int_{0}^T \left(\bPsi_{k}^\top(t)\hat{\bbeta}_{j,k}\right)^{2}\dd t \right\}^{1/2}  \right].
\end{eqnarray*}
Similar to \eqref{orgbound1}, we have that
\begin{eqnarray}\label{case31}
&&\frac{1}{2}\int_{0}^T \left\{\hat{\lambda}_{j}(t)-\widetilde{\lambda}_{j}(t)\right\}^{2}\dd t\\\nonumber
&\leq& \underbrace{\int_{0}^T \left\{\widetilde{\lambda}_{j}(t)-\hat{\lambda}_{j}(t)\right\}\left\{\lambda_{j} (t)\dd t-\dd N(t)\right\}}_{A_1}+\underbrace{\int_{0}^T \left\{\widetilde{\lambda}_{j}(t)-\hat{\lambda}_{j}(t)\right\}\left\{\widetilde{\lambda}_{j}(t)-\lambda_{j} (t)\right\}\dd t}_{A_2}\\\nonumber
&&\qquad +\left\{s\lambda_{\max}\log p\right\}^{1/2}\left[\int_{0}^T \left\{\widetilde{\lambda}_{j}(t)-\hat{\lambda}_{j}(t)\right\}^{2}\dd t \right]^{1/2},
\end{eqnarray}
with probability at least {$1-3c_{2}pT\exp(-c_{3}T^{1/5})-c_{2}'p(p+1)T\exp(-c_{3}'T^{1/5})$}.
By Lemma~\ref{martingale}, we have that 
\begin{equation}\label{case32}
A_1\leq4\left[\frac{\lambda_{\max}}{T}\int_{0}^T \left\{\widetilde{\lambda}_{j}(t)-\hat{\lambda}_{j}(t)\right\}^{2}\dd t \right]^{1/2}\sqrt{\frac{\log p}{T}},
\end{equation}
with probability at least $1-C_3p^{-2}$. Additionally, we have that 
\begin{eqnarray}\label{case33}
A_2&\leq& 2\left[\int_{0}^T \left\{\widetilde{\lambda}_{j}(t)-\hat{\lambda}_{j}(t)\right\}^{2}\dd t\right]^{1/2}\left[\int_{0}^T \left\{\widetilde{\lambda}_{j}(t)-\lambda_{j} (t)\right\}^{2}\dd t\right]^{1/2}\\\nonumber
&\leq& 2(s+1)\sqrt{T}{C_1^{1/2}m_1^{-d}}\left[\int_{0}^T \left\{\widetilde{\lambda}_{j}(t)-\hat{\lambda}_{j} (t)\right\}^{2}\dd t\right]^{1/2},
\end{eqnarray}
with probability at least {$1-C_2'pT\exp(-C_3'T^{1/5})$}.

Plugging \eqref{case32} and \eqref{case33} into \eqref{case31}, we have 
\begin{eqnarray*}
&&\frac{1}{2T}\int_{0}^T \left\{\hat{\lambda}_{j}(t)-\widetilde{\lambda}_{j}(t)\right\}^{2}\dd t\\
&\leq& \left\{{2^{3/2}(s+1)C_{1}^{1/2}m_1^{-d}}+6\sqrt{2}(s\lambda_{\max})^{1/2}\sqrt{\frac{\log p}{T}}\right\}\left[\frac{1}{2T}\int_{0}^T \left\{\widetilde{\lambda}_{j}(t)-\hat{\lambda}_{j} (t)\right\}^{2}\dd t\right]^{1/2},
\end{eqnarray*}
with probability at least {$1-C_3p^{-2}-3c_{2}pT\exp(-c_{3}T^{1/5})-c_{2}'p(p+1)T\exp(-c_{3}'T^{1/5})-C_2'pT\exp(-C_3'T^{1/5})$}.
Thus we have that with probability at least {$1-C_3p^{-2}-3c_{2}pT\exp(-c_{3}T^{1/5})-c_{2}'p(p+1)T\exp(-c_{3}'T^{1/5})-C_2'pT\exp(-C_3'T^{1/5})$}
\begin{eqnarray*}
\frac{1}{2T}\int_{0}^T \left\{\hat{\lambda}_{j}(t)-\widetilde{\lambda}_{j}(t)\right\}^{2}\dd t\le {16C_1(s+1)^2m_1^{-2d}}+144s\lambda_{\max}\frac{\log p}{T}.
\end{eqnarray*}
Hence, under Case (iii), we have
$$
\frac{1}{T}\int_{0}^T \left\{\hat{\lambda}_{j}(t)-\lambda_{j}(t)\right\}^{2}\dd t\leq 32\left\{{C_1(s+1)^2m_1^{-2d}}+9s\lambda_{\max}\frac{\log p}{T}\right\},
$$
with probability at least {$1-C_3p^{-2}-3c_{2}pT\exp(-c_{3}T^{1/5})-c_{2}'p(p+1)T\exp(-c_{3}'T^{1/5})-C_2'pT\exp(-C_3'T^{1/5})$}. Let $C_4=3c_2+2c_2'+C_2'$ and $C_5=\min\{c_3,c_3',C_3'\}$. Considering the above three cases together, we have \eqref{bound} holds for any $j$ with probability at least {$1-C_{3}p^{-2}-C_{4}p^{2}T\exp(-C_{5}T^{1/5})$}.
\eop

\subsection{Proof of Theorem~\ref{thm4}}
The proof is divided into two steps, and loosely follows the primal-dual witness arguments in \cite{ravikumar2009sparse}. 
The first step is to show that $\mathcal{E}_j\subset\hat{\mathcal{E}_j}$, and the second step is to show that $\mathcal{E}_j^c\bigcap\hat{\mathcal{E}_j}=\emptyset$. Together these two steps will complete the argument for selection consistency. 

We aim to construct an estimator $(\hat{\bbeta}_{j,k})_{k\in[p]}$ that satisfies the KKT conditions for optimizing \eqref{obj}, i.e., 
\begin{equation}\label{firstder}
-\frac{1}{T}\int_{0}^T\bPsi_{k}(t)\dd N_j(t)+\frac{1}{T}\int_{0}^T\bPsi_{k}(t)\left(\sum_{k=0}^{p}\bPsi_{k}^\top(t)\hat{\bbeta}_{j,k}\right)\dd t+\eta_j\hat{\g}_{j,k}=0
\end{equation}
where
\begin{equation}\label{pender1}
\hat{\g}_{j,k}=\frac{\frac{1}{T}\int_{0}^T\bPsi_{k}(t)\bPsi_{k}^\top(t)\dd t\hat{\bbeta}_{j,k}}{\left(\frac{1}{T}\hat{\bbeta}_{j,k}^\top\int_{0}^{T}\bPsi_{k}(t)\bPsi_{k}^\top(t)\dd t\hat{\bbeta}_{j,k}\right)^{1/2}},\quad \mathrm{if}\,\,\hat{\bbeta}_{j,k}\ne 0
\end{equation}
\begin{equation}
\label{pender2}
\hat{\g}_{j,k}^\top\left(\frac{1}{T}\int_{0}^T \bPsi_{k}(t)\bPsi_{k}^\top(t)\dd t\right)^{-1}\hat{\g}_{j,k}<1,\quad \mathrm{if}\,\,\hat{\bbeta}_{j,k}= 0.
\end{equation}
We construct the primal-dual pair $(\hat{\bbeta}_{j,k},\hat{\g}_{j,k})_{j\in[p]}$ as follows.
For $k\in\mathcal{E}_j$, we let 
\begin{eqnarray}\label{hatbeta}
&&\hat{\bbeta}_{j,k}=\arg\min_{\bbeta_{j,k}\in\mathbb{R}^{m_1}}-\frac{1}{T}\int_{0}^{T}\left(\sum_{k\in\bar\mE_j}\bPsi_{k}^\top(t)\bbeta_{j,k}\right)\dd N_j(t)\\\nonumber
&&\hspace{1in}+\frac{1}{2T}\int_{0}^T \left(\sum_{k\in\bar\mE_j}\bPsi_{k}^\top(t)\bbeta_{j,k}\right)^2\dd t+\eta_j\sum_{k=1}^{p}\left\{\bbeta_{j,k}^\top\left(\frac{1}{T}\int_{0}^T\bPsi_{k}(t)\bPsi_{k}^\top(t)\dd t\right)\bbeta_{j,k}\right\}^{1/2}
\end{eqnarray}
where $\bar\mE_j=\mathcal{E}_j\bigcup\{0\}$; for $k\in \mathcal{E}_j^c$, we let $\hat{\bbeta}_{j,k}=\0$. Define $\hat{\g}_{j,k}$ for $k\in\mathcal{E}_j$ as in \eqref{pender1}, and define $\hat{\g}_{j,k}$ for $k\in\mathcal{E}_j^c$ by solving \eqref{firstder}.

{
\bigskip
\noindent\textbf{Step 1.}
To establish $\mathcal{E}_j\subset\hat{\mathcal{E}_j}$, we first show that 
\begin{equation}\label{betabound}
\sum_{k\in {\mathcal{E}_{j}}}\|\hat{\bbeta}_j-\widetilde{\bbeta}_j\|_2^2=o(1),
\end{equation}
with probability at least {$1-C_{3}p^{-2}-2C_{4}p^2 T\exp(-C_{5}T^{1/5})$}.
If this holds, then $\max_{k\in\mathcal{E}_j}\|\hat{\bbeta}_{j,k}-\tilde{\bbeta}_{j,k}\|_2^2\leq \frac{\gamma_{\min}\beta_{\min}}{6\gamma_{\max}}$ with probability at least {$1-C_{3}p^{-2}-2C_{4}p^2 T\exp(-C_{5}T^{1/5})$}. Combined with Lemma \ref{lemma:block}, this leads to
\begin{equation*}
\max_{k\in\mathcal{E}_j} \frac{1}{T}\int_{0}^T \left\{\bPsi_{k}^\top(t)(\hat{\bbeta}_{j,k}-\tilde{\bbeta}_{j,k})\right\}^2\dd t \leq \frac{3\gamma_{\max}}{2m_1}\max_{k\in\mathcal{E}_j}\|\hat{\bbeta}_{j,k}-\tilde{\bbeta}_{j,k}\|_2^2\leq \frac{\gamma_{\min}\beta_{\min}}{4m_1},
\end{equation*}
with probability at least {$1-C_{3}p^{-2}-3C_{4}p^2 T\exp(-C_{5}T^{1/5})$}. By Assumption~\ref{betamin} and Lemma \ref{lemma:block}, it holds for $k\in \mathcal{E}_j$ that
\begin{equation*}
\begin{aligned}
\left\{\frac{1}{T}\int_{0}^T \left(\bPsi_{k}^\top(t)\hat{\bbeta}_{j,k}\right)^2\dd t \right\}^{1/2}
&\geq \left\{\frac{1}{T}\int_{0}^T \left(\bPsi_{k}^\top(t)\tilde{\bbeta}_{j,k}\right)^2\dd t \right\}^{1/2}-\left\{\frac{1}{T}\int_{0}^T \bPsi_{k}^\top(t)\left(\hat{\bbeta}_{j,k}-\tilde{\bbeta}_{j,k}\right)^2\dd t \right\}^{1/2}\\
&\geq\sqrt{\frac{\gamma_{\min}\beta_{\min}}{2m_1}}-\sqrt{\frac{\gamma_{\min}\beta_{\min}}{4m_1}}>0,
\end{aligned}
\end{equation*}
which implies that $\hat{\bbeta}_{j,k}\neq \0$ for $k\in \mathcal{E}_j$. Consequently, we have that $\mathcal{E}_j\subset \hat{\mathcal{E}_j}$ holds with probability at least {$1-C_{3}p^{-2}-3C_{4}p^2 T\exp(-C_{5}T^{1/5})$}. 
Therefore, what remains to complete Step 1 is to verify \eqref{betabound}. To ease notation, we denote $\Delta_k=\hat{\bbeta}_{j,k}-\tilde{\bbeta}_{j,k}$.

By Cauchy-Schwarz inequality, Theorem \ref{thm3} and Assumption \ref{ass6}, it holds with probability at least {$1-C_{3}p^{-2}-C_{4}p^2 T\exp(-C_{5}T^{1/5})$} that
\begin{equation}\label{hattilde1}
\begin{aligned}
\frac{1}{T}\int_{0}^T \left\{\hat{\lambda}_{j}(t)-\widetilde{\lambda}_{j}(t)\right\}^{2}\dd t &\leq \frac{2}{T}\int_{0}^T\left\{\hat{\lambda}_j(t)-\lambda_j(t)\right\}^2\dd t+\frac{2}{T}\int_{0}^T\left\{\tilde{\lambda}_j(t)-\lambda_j(t)\right\}^2\dd t\\
&\leq 64\left\{C_1(s+1)^2m_1^{-2d}+9s\lambda_{\max}\frac{\log p}{T}\right\}+2C_{1}(s+1)^2m_1^{-2d}\\\nonumber
&= 66 C_1(s+1)^2m_1^{-2d}+576 s\lambda_{\max}\frac{\log p}{T}.
\end{aligned}
\end{equation}  
Consequently, it holds with probability at least {$1-C_{3}p^{-2}-C_{4}p^2 T\exp(-C_{5}T^{1/5})$} that
\begin{equation}\label{case1bound}
\begin{aligned}
&\frac{1}{T}\int_{0}^T \left\{\widetilde{\lambda}_{j}(t)- \hat{\lambda}_{j}(t)\right\}\left\{\widetilde{\lambda}_{j}(t)\dd t-\dd N_j(t)\right\}\\
=&\frac{1}{T}\int_{0}^T \left\{\widetilde{\lambda}_{j}(t)- \hat{\lambda}_{j}(t)\right\}\left\{\widetilde{\lambda}_{j}(t)-\lambda_j(t)\right\}\dd t+\frac{1}{T}\int_{0}^T \left\{\widetilde{\lambda}_{j}(t)- \hat{\lambda}_{j}(t)\right\}\left\{\lambda_{j}(t)\dd t-\dd N_j(t)\right\}\\
\leq& \left\{\frac{1}{T}\int_{0}^T \left\{\widetilde{\lambda}_{j}(t)- \hat{\lambda}_{j}(t)\right\}^2\dd t\right\}^{1/2}\left[\left\{\frac{1}{T}\int_{0}^T \left\{\widetilde{\lambda}_{j}(t)- \lambda_{j}(t)\right\}^2\dd t\right\}^{1/2}+4\sqrt{\frac{\lambda_{\max}\log p}{T}}\right]\\
\leq& \left\{C_1(s+1)^2m_1^{-2d}+576 s\lambda_{\max}\frac{\log p}{T}\right\}^{1/2}\left\{\sqrt{C_1}(s+1)m_1^{-d}+4\sqrt{\frac{\lambda_{\max}\log p}{T}}\right\},
\end{aligned}
\end{equation}
where the first inequality holds due to the Cauchy-Schwarz inequality and \eqref{stoerror}, and the last inequality holds due to Assumption~\ref{ass6}.
To verify \eqref{betabound}, we consider two different cases:
\begin{eqnarray*}
&(i)&\frac{1}{T}\int_{0}^T \left\{\widetilde{\lambda}_{j}(t)- \hat{\lambda}_{j}(t)\right\}\left\{\widetilde{\lambda}_{j}(t)\dd t-\dd N_j(t)\right\}\geq \frac{\eta_j}{2}\sum_{k=1}^p\left\{\frac{1}{T}\int_{0}^T \left(\bPsi_{k}^\top(t)(\widetilde\bbeta_{j,k}-\hat\bbeta_{j,k})\right)^{2}\dd t\right\}^{1/2}\\
&(ii)&\frac{1}{T}\int_{0}^T \left\{\widetilde{\lambda}_{j}(t)- \hat{\lambda}_{j}(t)\right\}\left\{\widetilde{\lambda}_{j}(t)\dd t-\dd N_j(t)\right\}< \frac{\eta_j}{2}\sum_{k=1}^p\left\{\frac{1}{T}\int_{0}^T \left(\bPsi_{k}^\top(t)(\widetilde\bbeta_{j,k}-\hat\bbeta_{j,k})\right)^{2}\dd t\right\}^{1/2}.	
\end{eqnarray*}	
Under case (i), by Lemma~\ref{lemma:block}, it holds with probability at least {$1-c_2pT\exp(-c_3T^{1/5})+c_2'pT\exp(-c_3'T^{1/5})$},
\begin{equation*}
\begin{aligned}
&4\eta_j^{-2}\left[\frac{1}{T}\int_{0}^T \left\{\widetilde{\lambda}_{j}(t)- \hat{\lambda}_{j}(t)\right\}\left\{\widetilde{\lambda}_{j}(t)\dd t-\dd N_j(t)\right\}\right]^2\\
\geq& \left(\sum_{k=1}^p\sqrt{\frac{\gamma_{\min}}{2m_1}}\|\Delta_k\|_2\right)^2\geq  \frac{\gamma_{\min}}{2m_1}\sum_{k\in\mE_j}\|\Delta_k\|_2^2,
\end{aligned}
\end{equation*} 
where the last inequality follows the fact that $(\sum_i a_i)^2\geq \sum_i a_i^2$ when $a_i\geq 0$. Combining with $\eta_j^2=O\left(\frac{\log p}{T}\right)$ and \eqref{case1bound}, it holds that 
$$
\sum_{k\in\mE_j}\|\Delta_k\|_2^2= \mathcal{O}\left\{(m_1T/\log p)(s^4m_1^{-4d}+s^2\log p^2T^{-2})\right\}=o(1),
$$ 
where we used the fact that $s^2m_1=o(T^{4/5})$, $\log p=\mathcal{O}(T^{1/5})$, $s^2m_1^{-2d}=\mathcal{O}(\log p/T)=o(1)$.
Next we consider case (ii). A similar argument as in \eqref{orgbound} leads to
\begin{equation}\label{dishattilde}
\begin{aligned}
\frac{1}{2T}\int_{0}^T \left\{\hat{\lambda}_{j}(t)-\widetilde{\lambda}_{j}(t)\right\}^{2}\dd t 
&\leq\frac{1}{T}\int_{0}^T \left\{\widetilde{\lambda}_{j}(t)- \hat{\lambda}_{j}(t)\right\}\left\{\widetilde{\lambda}_{j}(t)\dd t-\dd N_j(t)\right\}\\ &+\eta_{j}\underbrace{\sum_{k=1}^p\left\{\left\{\frac{1}{T}\int_{0}^T \left(\bPsi_{k}^\top(t)\widetilde\bbeta_{j,k}\right)^{2}\dd t  \right \}^{1/2} -\left\{\frac{1}{T}\int_{0}^T \left(\bPsi_{k}^\top(t)\hat{\bbeta}_{j,k}\right)^{2}\dd t \right\}^{1/2}   \right\}}_{\uppercase\expandafter{\romannumeral1}}.
\end{aligned}
\end{equation}
Noting $\widetilde{\bbeta}_{j,k}=\0$ for $k\in \mathcal{E}_j^c$, it holds that 
\begin{equation*}
\begin{aligned}
&\sum_{k=1}^p\left\{\frac{1}{T}\int_{0}^T \left(\bPsi_{k}^\top(t)\widetilde\bbeta_{j,k}\right)^{2}\dd t  \right \}^{1/2}=\sum_{k\in \mathcal{E}_j}\left\{\frac{1}{T}\int_{0}^T \left(\bPsi_{k}^\top(t)(\widetilde\bbeta_{j,k}-\hat\bbeta_{j,k}+\hat\bbeta_{j,k})\right)^{2}\dd t  \right \}^{1/2}\\
\leq &\sum_{k\in \mathcal{E}_j}\left\{\frac{1}{T}\int_{0}^T \left(\bPsi_{k}^\top(t)(\widetilde\bbeta_{j,k}-\hat\bbeta_{j,k})\right)^{2}\dd t  \right \}^{1/2}+\sum_{k\in \mathcal{E}_j}\left\{\frac{1}{T}\int_{0}^T \left(\bPsi_{k}^\top(t)\hat\bbeta_{j,k}\right)^{2}\dd t  \right \}^{1/2}.
\end{aligned}
\end{equation*}
Plugging this into $\uppercase\expandafter{\romannumeral1}$, it is straightforward to obtain that 
\begin{equation}\label{pencon}
\uppercase\expandafter{\romannumeral1}\leq \sum_{k\in \mathcal{E}_j}\left\{\frac{1}{T}\int_{0}^T \left(\bPsi_{k}^\top(t)(\widetilde\bbeta_{j,k}-\hat\bbeta_{j,k})\right)^{2}\dd t  \right \}^{1/2}-\sum_{k\in \mathcal{E}_j^c}\left\{\frac{1}{T}\int_{0}^T \left(\bPsi_{k}^\top(t)\hat\bbeta_{j,k}\right)^{2}\dd t  \right \}^{1/2}.
\end{equation}
Together with \eqref{dishattilde}, \eqref{pencon} and case (ii), we have that
\begin{equation*}
\begin{aligned}
0&\leq\frac{1}{2T}\int_{0}^T \left\{\hat{\lambda}_{j}(t)-\widetilde{\lambda}_{j}(t)\right\}^{2}\dd t \\
&\leq\frac{3}{2}\eta_j\sum_{k\in \mathcal{E}_j}\left\{\frac{1}{T}\int_{0}^T \left(\bPsi_{k}^\top(t)(\widetilde\bbeta_{j,k}-\hat\bbeta_{j,k})\right)^{2}\dd t  \right \}^{1/2}-\frac{1}{2}\eta_j\sum_{k\in \mathcal{E}_j^c}\left\{\frac{1}{T}\int_{0}^T \left(\bPsi_{k}^\top(t)(\widetilde\bbeta_{j,k}-\hat\bbeta_{j,k})\right)^{2}\dd t  \right \}^{1/2}.
\end{aligned}
\end{equation*}
Combined with Lemma~\ref{lemma:block}, it holds with probability at least {$1-c_2pT\exp(-c_3T^{1/5})+c_2'pT\exp(-c_3'T^{1/5})$,}
\begin{equation}\label{restrictedeigencon}
\sum_{k\in {\mathcal{E}_{j}^c}}\|\Delta_k\|\leq 3\sqrt{\frac{3\gamma_{\max}}{\gamma_{\min}}}\sum_{k\in {\mathcal{E}_{j}}}\|\Delta_k\|.
\end{equation}
As such, Lemma~\ref{lemma:re} is applicable and we have
\begin{equation*}
\frac{\gamma_{\min}}{4m_1}\sum_{k\in {\mathcal{E}_{j}}}\|\Delta_k\|^2\leq\Delta^\top\G\Delta=\frac{1}{T}\int_{0}^T \left\{\hat{\lambda}_{j}(t)-\widetilde{\lambda}_{j}(t)\right\}^{2}\dd t\leq 66 C_1(s+1)^2m_1^{-2d}+576 s\lambda_{\max}\frac{\log p}{T}
\end{equation*}
with probability at least $1-C_3p^{-2}-2C_{4}p^2 T\exp(-C_{5}T^{1/5})$, where $C_4$ and $C_5$ are as defined in Theorem \ref{thm3}. Since $s^2m_1^{-2d+1}=\mathcal{O}(m_1\log p/T)=o(1)$ and $sm_1\log p/T=o(1)$ (due to $\log p=\mathcal{O}(T^{1/5})$ and $s^2m_1=o(T^{1/5})$), \eqref{betabound} holds.}

\bigskip
\noindent\textbf{Step 2.}
To show that $\hat{\mathcal{E}_j}\bigcap\mathcal{E}_j^{c}=\emptyset$, we verify the strict dual feasibility \eqref{pender2} for $k\notin\mathcal{E}_j$. For $k\notin\mathcal{E}_j$, we calculate $\hat{\g}_{j,k}$ from the subgradient condition in \eqref{firstder}.
Adding and removing $\frac{1}{T}\int_{0}^T\bPsi^\top_{k}(t)\tilde{\lambda}_j(t)\dd t$ to the left side of \eqref{firstder}, we have 
\begin{eqnarray}\label{nonedge0}
-\eta_j\hat{\g}_{j,k}&=&\frac{1}{T}\int_{0}^T\bPsi_{k}(t)\left\{\hat{\lambda}_j(t)-\tilde{\lambda}_j(t)\right\}\dd t+\frac{1}{T}\int_{0}^T\bPsi_{k}(t)\left\{\tilde{\lambda}_j(t)\dd t-\dd N_j(t)\right\}\\\nonumber
&=&\frac{1}{T}\int_{0}^T\bPsi_{k}(t)\bPsi_{\bar\mE_j}^\top(t)\dd t(\hat{\bbeta}_{\bar\mE_j}-\tilde{\bbeta}_{\bar\mE_j})+\frac{1}{T}\int_{0}^T\bPsi_{k}(t)\left\{\tilde{\lambda}_j(t)\dd t-\dd N_j(t)\right\}.
\end{eqnarray}
Jointly for $k\in\bar\mE_j$, it holds that
\begin{equation*}
\frac{1}{T}\int_{0}^T\bPsi_{\bar\mE_j}(t)\bPsi_{\bar\mE_j}^\top(t)\dd t(\hat{\bbeta}_{\bar\mE_j}-\tilde{\bbeta}_{\bar\mE_j})+\frac{1}{T}\int_{0}^T\bPsi_{\bar\mE_j}(t)\left\{\tilde{\lambda}_j(t)\dd t-\dd N_j(t)\right\}+\eta_j\hat{\g}_{\bar\mE_j}=0,
\end{equation*}
where $\hat{\g}_{\bar\mE_j}^\top=\{\0_{m_0}^\top,(\hat{\g}_{j,k}^\top)_{k\in\mathcal{E}_j}\}$. Rearranging the terms gives that
\begin{equation}\label{betadis}
\hat{\bbeta}_{\bar\mE_j}-\tilde{\bbeta}_{\bar\mE_j}=-\left\{\frac{1}{T}\int_{0}^T\bPsi_{\bar\mE_j}(t)\bPsi_{\bar\mE_j}^\top(t)\dd t\right\}^{-1}\left[\frac{1}{T}\int_{0}^T\bPsi_{\bar\mE_j}(t)\left\{\tilde{\lambda}_j(t)\dd t-\dd N_j(t)\right\}+\eta\hat{\g}_{\bar\mE_j} \right].
\end{equation}
Note that $\hat{\g}_{j,0}=\0_{m_0}$, this implies that $\frac{1}{T}\int_{0}^T\bPsi_{0}(t)\left\{\hat{\lambda}_j(t)\dd t-\dd N_j(t)\right\}=\0_{m_0}$ from \eqref{firstder}.
Plugging \eqref{betadis} into \eqref{nonedge0}, we get that
\begin{eqnarray}\label{nonedge1}
&&-\eta_j\hat{\g}_{j,k}=\frac{1}{T}\int_{0}^T\bPsi_{k}(t)\bPsi_{\bar\mE_j}^\top(t)\dd t(\hat{\bbeta}_{\bar\mE_j}-\tilde{\bbeta}_{\bar\mE_j})+\frac{1}{T}\int_{0}^T\bPsi_{k}(t)\left\{\tilde{\lambda}_j(t)\dd t-\dd N_j(t)\right\}\\\nonumber
&&=\underbrace{-\left(\frac{1}{T}\int_{0}^T\bPsi_{k}(t)\bPsi_{\bar\mE_j}^\top(t)\dd t\right)\left(\frac{1}{T}\int_{0}^T\bPsi_{\bar\mE_j}(t)\bPsi_{\bar\mE_j}^\top(t)\dd t\right)^{-1}\frac{1}{T}\int_{0}^T\bPsi_{\bar\mE_j}(t)\left\{\tilde{\lambda}_j(t)\dd t-\dd N_j(t)\right\}}_{I_1}\\\nonumber
&&-\underbrace{\left(\frac{1}{T}\int_{0}^T\bPsi_{k}(t)\bPsi_{\bar\mE_j}^\top(t)\dd t\right)\left(\frac{1}{T}\int_{0}^T\bPsi_{\bar\mE_j}(t)\bPsi_{\bar\mE_j}^\top(t)\dd t\right)^{-1}\eta_j\hat{\g}_{\bar\mE_j}}_{I_2}+\underbrace{\frac{1}{T}\int_{0}^T\bPsi_{k}(t)\left\{\tilde\lambda_j(t)\dd t-\dd N_j(t)\right\}}_{I_3}.\\\nonumber
\end{eqnarray}
To show $\|\hat{\g}_{j,k}\|_2^2<1$, we move to bound terms $\|I_1\|_2^2$, $\|I_2\|_2^2$ and $\|I_1\|_3^2$, respectively.

{
\medskip
\noindent
Consider the term $I_1$. First, we have 
\begin{equation*}
\begin{aligned}
\frac{1}{T}\int_{0}^T\bPsi_{0}(t)\left\{\tilde{\lambda}_j(t)\dd t-\dd N_j(t)\right\}&=\frac{1}{T}\int_{0}^T\bPsi_{0}(t)\left\{\tilde{\lambda}_j(t)\dd t-\hat{\lambda}_j(t)\dd t+\hat{\lambda}_j(t)\dd t-\dd N_j(t)\right\}\\
&=\frac{1}{T}\int_{0}^T\bPsi_{0}(t)\left\{\tilde{\lambda}_j(t)-\hat{\lambda}_j(t)\right\}\dd t,
\end{aligned}
\end{equation*}
where we used the fact that $\frac{1}{T}\int_{0}^T\bPsi_{0}(t)\left\{\hat{\lambda}_j(t)\dd t-\dd N_j(t)\right\}=\0_{m_0}$ (because $\hat{\g}_{j,0}=\0_{m_0}$). Recall that we defined $\bPsi_{0}(t)=\bphi_{0}(t)$ and therefore,
$$
\frac{1}{T}\int_{0}^T\bPsi_{0,l}(t)\left\{\tilde{\lambda}_j(t)\dd t-\dd N_j(t)\right\}=\frac{1}{T}\int_{0}^T\phi_{0,l}(t)\left\{\tilde{\lambda}_j(t)-\hat{\lambda}_j(t)\right\}\dd t.
$$
Let $\text{supp}(\cdot)$ be the support $\{t\in\mX: f(t)\neq 0\}$ for a continuous function $f$ defined on $\mX\in\mathbb{R}$.
It holds by H\"{o}lder's inequality that
\begin{equation*}
\begin{aligned}
&\left[\frac{1}{T}\int_{0}^T\phi_{0,l}(t)\left\{\tilde{\lambda}_j(t)-\hat{\lambda}_j(t)\right\}\dd t\right]^2=\left[\frac{1}{T}\int_{\text{supp}(\phi_{0,l})}\phi_{0,l}(t)\left\{\tilde{\lambda}_j(t)-\hat{\lambda}_j(t)\right\}\dd t\right]^2\\
\leq&\frac{1}{T}\int_{\text{supp}(\phi_{0,l})}\phi_{0,l}^2(t)\dd t\cdot\frac{1}{T}\int_{\text{supp}(\phi_{0,l})}\left\{\tilde{\lambda}_j(t)-\hat{\lambda}_j(t)\right\}^2\dd t.
\end{aligned}
\end{equation*}	
By the local property of B-spline basis \citep{de1972calculating}, there exist some constants ${c_4'}>0$ and ${c_5'}>0$ such that $\frac{1}{T}\int_{0}^T\phi_{0,l}^2(t)\dd t\le {c_4'}/m_0$ and $\sum_{l\in[m_0]}\int_{\text{supp}(\phi_{0,l})}f(t)^2\dd t<{c_5'}\int_0^T f(t)^2\dd t$ for some function $f$. 
By Theorem \ref{thm3}, we have that
\begin{equation}\label{error0}
\begin{aligned}
&\left\|\frac{1}{T}\int_{0}^T\bPsi_{0}(t)\left\{\tilde{\lambda}_j(t)-\hat{\lambda}_j(t)\right\}\dd t\right\|_2^2\\
\leq& \max_{l\in[m_0]}\left\{\frac{1}{T}\int_{\text{supp}(\phi_{0,l})}\phi_{0,l}^2(t)\dd t\right\}\sum_{l=1}^{m_0}\frac{1}{T}\int_{\text{supp}(\phi_{0,l})}\left\{\tilde{\lambda}_j(t)-\hat{\lambda}_j(t)\right\}^2\dd t\\
\leq& \frac{32{c_4'c_5'}}{m_0}\left\{C_1(s+1)^2m_1^{-2d}+9s\lambda_{\max}\frac{\log p}{T}\right\} 
\end{aligned}
\end{equation}
with probability at least {$1-C_{3}p^{-2}-C_{4}p^2 T\exp(-C_{5}T^{1/5})$}. 
Next, for any $k\in[p]$, it holds that
\begin{equation*}
\begin{aligned}
&\left|\frac{1}{T}\int_{0}^T\bPsi_{k,l}(t)\left\{\tilde{\lambda}_j(t)\dd t-\dd N_j(t)\right\}\right|\\
\leq& \left|\frac{1}{T}\int_{0}^T\bPsi_{k,l}(t)\left\{\tilde{\lambda}_j(t)-\lambda_j(t)\right\}\dd t\right|+\left|\frac{1}{T}\int_{0}^T\bPsi_{k,l}(t)\left\{\lambda_j(t)\dd t-\dd N_j(t)\right\}\right|\\
\leq& \underbrace{\left|\frac{1}{T}\int_0^TH_{1}(s)\dd N_k(s)\right|}_{B_1}+\underbrace{\left|\frac{1}{T}\int_{0}^T\bPsi_{k,l}(t)\left\{\lambda_j(t)\dd t-\dd N_j(t)\right\}\right|}_{B_2},
\end{aligned}
\end{equation*}
where $H_{1}(s)=\int_0^{T-s}\phi_{1,l}(t)\left\{\tilde{\lambda}_j(t+s)-\lambda_j(t+s)\right\}\dd t$. For term $B_1$, it holds that 
$$
B_1<\left|\frac{1}{T}\int_0^TH_{1}(s)\{\lambda_k(s)\dd s-\dd N_k(s)\}\right|+\left|\frac{1}{T}\int_0^TH_{1}(s)\lambda_k(s)\dd s\right|.
$$ 
Again by the local property of B-spline basis \citep{de1972calculating}, there exist some constant ${c_6'}>0$ such that $\int_{0}^b\phi_{1,l}^2(t)\dd t\le {c_6'}/m_1$.
Consequently, we have
$$
\{H_{1}(s)\}^2\leq \int_0^b\phi_{1,l}^2(t)\dd t\int_s^{s+b}\left\{\tilde{\lambda}_j(t)-\lambda_j(t)\right\}^2\dd t\leq\frac{{c_6'}}{m_1}\int_s^{s+b}\left\{\tilde{\lambda}_j(t)-\lambda_j(t)\right\}^2\dd t,
$$ 
where we used the fact that the support of $\phi_{1,l}$ is upper bounded by $b$.
By Lemma~\ref{martingale} with $\epsilon=\frac{2\log p}{T}$, it holds with probability at least {$1-C_3p^{-2}-C_2'pT\exp(-C_3'T^{1/5})$} that
\begin{equation*}
\begin{aligned}
\left[\frac{1}{T}\int_0^TH_{1}(s)\{\dd N_k(s)-\lambda_k(s)\dd s\}\right]^2&\leq \frac{8{c_6'}\log p}{m_1T}\frac{\lambda_{\max}}{T}\int_0^T\int_s^{s+b}\left\{\tilde{\lambda}_j(t)-\lambda_j(t)\right\}^2 \dd t\dd s\\
&\leq \frac{8{c_6'}C_1b\lambda_{\max}(s+1)^2\log p}{m_1^{2d+1}T},
\end{aligned}
\end{equation*}
where the last equality holds due to
\begin{equation*}
\begin{aligned}
\frac{1}{T}\int_0^T\int_s^{s+b}\left\{\tilde{\lambda}_j(t)-\lambda_j(t)\right\}^2\dd t\dd s&\leq \frac{b}{T}\int_{0}^T\left\{\tilde{\lambda}_j(t)-\lambda_j(t)\right\}^2\dd t.
\end{aligned}
\end{equation*} 
By Assumption~\ref{ass6}, we have with probability at least {$1-C_2'pT\exp(-C_3'T^{1/5})$},
\begin{equation*}
\begin{aligned}
&\left\{\frac{1}{T}\int_0^TH_{1}(s)\lambda_k(s)\dd s\right\}^2=\left[\frac{1}{T}\int_0^T\left\{\int_0^t\phi_{1,l}(t-s)\lambda_k(s)\dd s\right\}\left\{\tilde{\lambda}_j(t)-\lambda_j(t)\right\}\dd t\right]^2\\
\leq&\frac{1}{T}\int_0^T\left\{\int_0^t\phi_{1,l}(t-s)\lambda_k(s)\dd s\right\}^2\dd t\cdot\frac{1}{T}\int_0^T\left\{\tilde{\lambda}_j(t)-\lambda_j(t)\right\}^2\dd t\\
\leq& \frac{{{c_6'}}^2C_1\lambda_{\max}^2(s+1)^2}{m_1^{2d+2}},
\end{aligned}
\end{equation*}
where the first inequality holds due to H\"{o}lder's inequality, and the second inequality holds due to $\lambda_k(t)\le\lambda_{\max}$ and $\int_{0}^b\phi_{1,l}^2(t)\dd t\le {c_6'}/m_1$.
Since $\frac{m_1\log p}{T}=o(1)$, it holds that $\frac{(s+1)^2\log p}{m_1^{2d+1}T}=o\left\{\frac{(s+1)^2}{m_1^{2d+2}}\right\}$. Consequently, it holds that
\begin{equation}\label{errork1}
B_1^2=\left|\frac{1}{T}\int_{0}^T\bPsi_{k,l}(t)\left\{\tilde{\lambda}_j(t)-\lambda_j(t)\right\}\dd t\right|^2\leq \frac{2{{c_6'}}^2C_1\lambda_{\max}^2(s+1)^2}{m_1^{2d+2}}
\end{equation}
with probability at least {$1-C_3p^{-2}-C_2'pT\exp(-C_3'T^{1/5})$}. 

Next, we consider term $B_2$. Following the same argument used to bound $B_1$, we can get that 
$$
\left\vert\int_0^t\phi_{1,l}(t-s)\dd N_k(s)\right\vert\leq 2 {c_6'}\lambda_{\max}/m_1
$$
with with probability at least {$1-c_2T\exp(-c_3T^{1/5})-c_2'T\exp(-c_3'T^{1/5})$}. 
By Lemma~\ref{martingale} with $\epsilon=\frac{2\log p}{T}$, it holds that
\begin{equation}\label{errork2}
\begin{aligned}
B_2^2&=\left[\frac{1}{T}\int_0^T\bPsi_{k,l}(t)\left\{\lambda_j(t)\dd t-\dd N_j(t)\right\}\right]^2\\
&\leq 16\frac{\log p}{T}\frac{\lambda_{\max}}{2T}\int_0^T\left\{\int_0^t\phi_{1,l}(t-s)\dd N_k(s)\right\}^2\dd t\leq\frac{16{c_6'}^2\lambda_{\max}^3\log p}{m_1^2T},
\end{aligned}
\end{equation} 
with probability at least {$1-C_3p^{-2}-c_2T\exp(-c_3T^{1/5})-c_2'T\exp(-c_3'T^{1/5})$}. 
Together with \eqref{errork1}, \eqref{errork2} and $\frac{s^2}{m_1^{2d}}= O\left(\frac{\log p}{T}\right)$, there exists one positive constant ${c_{7}}$ such that
\begin{equation}\label{errork}
\left|\frac{1}{T}\int_{0}^T\bPsi_{k,l}(t)\left\{\tilde{\lambda}_j(t)\dd t-\dd N_j(t)\right\}\right|^2\leq \frac{{c_{7}}\log p}{m_1^2T}, 
\end{equation}
with probability at least {$1-C_3p^{-2}-C_2'pT\exp(-C_3'T^{1/5})-c_2T\exp(-c_3T^{1/5})-c_2'T\exp(-c_3'T^{1/5})$}.
Combining \eqref{error0}, \eqref{errork}, Assumption~\ref{irr}, Corollary \ref{cor1} and definition of $C_4,C_5$, it holds with probability at least {$1-C_3p^{-2}-C_4p^2T\exp(-C_5T^{1/5})$},
\begin{equation}\label{I1bound}
\begin{aligned}
&\|I_{1}\|_2^2 \leq\frac{\gamma_{\min}^2}{36s\gamma_{\max}^2}\left[sm_1\frac{{c_{7}}\log p}{m_1^2T}+\frac{32{c_{4}'c_5'}}{m_0}\left\{C_1(s+1)^2m_1^{-2d}+9s\lambda_{\max}\frac{\log p}{T}\right\} \right]\\
&=\frac{{c_{8}}\log p}{m_1T},
\end{aligned}
\end{equation}
for some constant ${c_{8}}>0$, where we used $s^2T/\log p=\mathcal{O}(m_1^{2d})$ and $m_1/m_0=\mathcal{O}(1)$.}

\medskip
\noindent
Consider the term $I_2$. By Lemma \ref{lemma:block}, it holds with probability at least {$1-c_2pT\exp(-c_3T^{1/5})-c_2'pT\exp(-c_3'T^{1/5})$,}
$$
\|\hat{g}_{j,k}\|_2^2\leq \frac{9\gamma_{\max}^2\|\hat{\bbeta}_{j,k}\|_2^2/(4m_1^2)}{\gamma_{\min}\|\hat{\bbeta}_{j,k}\|_2^2/(2m_1)}=\frac{9\gamma_{\max}^2}{2m_1\gamma_{\min}}.
$$ 
By Assumption~\ref{irr} and Corollary \ref{cor1}, we have with probability at least {$1-c_5(p+1)^2T\exp(-c_6T^{1/5})$,}
\begin{equation}\label{I2bound}
\| I_{2}\|_{2}^2\leq \eta_j^2\frac{\gamma_{\min}^2}{36\gamma_{\max}^2s}\|\hat{g}_{\bar\mE_j}\|_2^2\leq \eta_j^2\frac{\gamma_{\min}^2}{36\gamma_{\max}^2}\max_{k\in\mathcal{E}_j}\|\hat{g}_{j,k}\|_2^2\leq \frac{\eta_j^2\gamma_{\min}}{8m_1}.
\end{equation}

\medskip
\noindent
Consider the term $I_3$. By \eqref{errork}, we have with probability at least {$1-C_3p^{-2}-C_4pT\exp(-C_5T^{1/5})$,}
\begin{equation}\label{I3bound}
\| I_{3}\|_{2}^2\leq \sum_{l=1}^{m_1}\left|\frac{1}{T}\int_{0}^T\bPsi_{k,l}(t)\left\{\tilde{\lambda}_j(t)\dd t-\dd N_j(t)\right\}\right|^2=\frac{{c_{9}}\log p}{m_1T},\end{equation}
for some constant ${c_{9}}>0$.

Combining \eqref{I1bound}, \eqref{I2bound} and \eqref{I3bound}, we have that
$$
\eta_j^{2}\| \hat{\g}_{j,k}\|_{2}^{2}< \frac{3c_{8}\log p}{m_1T}+\frac{3\eta_j^2\gamma_{\min}}{8m_1}+\frac{3{c_{9}}\log p}{m_1T},
$$
with probability at least {$1-2C_3p^{-2}-2C_4p^2T\exp(-C_5T^{1/5})$}. As $\eta_j=(C_2\log p/T)^{1/2}$, when $C_2$ is sufficiently large, we have that
$$
\hat{\g}_{j,k}^\top\left(\frac{1}{T}\int_0^T\bPsi_{k}(t)\bPsi_{k}^\top(t)\dd t\right)^{-1}\hat{\g}_{j,k}\le\frac{2m_1}{\gamma_{\min}}\|\hat{\g}_{j,k}\|_{2}^2< \frac{2m_1}{\gamma_{\min}}\cdot\frac{\gamma_{\min}}{2m_1}=1,
$$
with probability at least {$1-2C_3p^{-2}-2C_4p^2T\exp(-C_5T^{1/5})$}.
This implies that \eqref{pender2} is satisfied for $k\in\mathcal{E}_j^c$. Consequently, we have that $\mathcal{E}_j^c\bigcap\hat{\mathcal{E}_j}=\emptyset$ {with probability at least $1-2C_3p^{-2}-3C_4p^2T\exp(-C_5T^{1/5})$}, which completes the proof.
\eop

\subsection{Proof of Theorem~\ref{thm4.2}}
\label{sec:gic}
{
Let $\hat\bbeta_j^{\eta_j}$ denote the estimator obtained from \eqref{obj} with $\eta_j$.
To ease notation, we write $\hat\bbeta^{\eta_j}$ without emphasizing its dependence on $j$, when there is no ambiguity. To facilitate our investigation of the asymptotic properties of $\text{GIC}(\eta_j)$, we introduce a proxy function $\text{GIC}^-(\cdot)$. 
For a set $\mathcal{S}_j\subset [p]$ with $|\mS_j|=o(T)$, the function $\text{GIC}^-(\mS_j)$ is defined as
\begin{equation}\label{ebicproxy}
\text{GIC}^-(\mS_j)={\ell_j(\hat{\bbeta}^0_{\bar{\mS}_j})\cdot\kappa_j}+(\alpha_T/T)\cdot|\mS_j|,
\end{equation}
where $\hat\bbeta^0_{\bar{\mS}_j}$ is the unpenalized maximum least squares estimator restricted to the set $\bar{\mS}_j=\mS_j\bigcup\{0\}$, that is,
\begin{equation}\label{unbiasedest}
\hat\bbeta^0_{\bar{\mS}_j}=\arg\min_{\text{supp}_1(\bbeta_j)=\mS_j}-2\bbeta_j^\top\balpha_j+\bbeta_j^\top\G\bbeta_j.
\end{equation}
where $\text{supp}_1(\bbeta_j)=\{k\in [p]:\bbeta_{j,k}\neq \0\}$.
Recall that $\eta^*_j$ is the tuning parameter that identifies the true model $\mE_j$.
The proof of Theorem \ref{thm4.2} is divided into the following three steps.\\
\textbf{Step 1:} We show that $\text{GIC}(\eta_j^\ast)$ and $\text{GIC}^-(\mE_j)$, defined as in \eqref{bic} and \eqref{ebicproxy}, respectively, are close, that is,
\begin{equation}\label{unbiaserror}
\text{GIC}(\eta_j^\ast)-\text{GIC}^-(\mE_j) \leq \frac{128}{\lambda_{\min}}\left\{C_1(s+1)^2m_1^{-2d}+9\lambda_{\max}\frac{s\log p}{T}\right\}
\end{equation}
with probability at least {$1-2C_3p^{-2}-2C_4p^2T\exp(-C_5T^{1/5})$}.\\
\textbf{Step 2:} We consider the under-fitted case and show that
\begin{equation*}
\inf_{\mS_j\not\supset \mE_j,|\mS_j|\leq s_0}\text{GIC}^-(\mS_j)-\text{GIC}^-(\mE_j)>\frac{\gamma_{\min}\beta_{\min}}{32\lambda_{\min}m_1},
\end{equation*}
with probability at least {$1-3C_3p^{-2}-5C_4p^2T\exp(-C_5T^{1/6})$}.\\
\textbf{Step 3:} We consider the over-fitted case and show that 
\begin{equation*}
\inf_{\mS_j\supset \mE_j}\text{GIC}^-(\mS_j)-\text{GIC}^-(\mE_j)>\alpha_T/(2T),
\end{equation*}
with probability at least {$1-5C_3p^{-2}-4C_4p^2T\exp(-C_5T^{1/5})$}.

By the definition of GIC$(\cdot)$ and GIC$^-(\cdot)$, for any $\eta_j>0$ and $\mS_j$ such that $\mS_j=\text{supp}_1(\hat{\bbeta}^{\eta_j})$, it holds that $\text{GIC}(\eta_j)\geq\text{GIC}^-(\mS_j)$. 
This is true because $\text{GIC}^-(\mS_j)$ is calculated using the unpenalized estimator $\hat\bbeta^0_{\bar{\mS}_j}$ and GIC$(\eta_j)$ is calculated using the penalized estimator $\hat\bbeta^{\eta_j}$ while these two estimators share the same support (i.e., $\mS_j=\text{supp}_1(\hat{\bbeta}^{\eta_j})$). 
Thus, for any $\eta_j\in\Gamma_{-}\bigcup \Gamma_{+}$, it holds with probability at least {$1-C'_3p^{-1}-C'_4p^2T\exp(-C_5T^{1/6})$} for some $C_3',C_4'>0$ that
\begin{equation*}
\begin{aligned}
&\text{GIC}(\eta_j)-\text{GIC}(\eta_j^\ast)\\
\geq &\text{GIC}^-(\mS_j)-\text{GIC}^-(\mE_j)+\text{GIC}^-(\mE_j)-\text{GIC}(\eta_j^\ast)\\
\geq &\text{GIC}^-(\mS_j)-\text{GIC}^-(\mE_j)-\frac{{128}}{\lambda_{\min}}\left\{C_1(s+1)^2m_1^{-2d}+9\lambda_{\max}\frac{s\log p}{T}\right\}\\
\geq& \min\left\{\frac{\gamma_{\min}\beta_{\min}}{32\lambda_{\min}m_1},\alpha_T/(2T)\right\}-\frac{{128}}{\lambda_{\min}}\left\{C_1(s+1)^2m_1^{-2d}+9\lambda_{\max}\frac{s\log p}{T}\right\}>0,
\end{aligned}
\end{equation*}
where the second inequality is true from Step 1 and the third inequality is true from Steps 2 and 3, where we used the fact that 
$s^2m_1^{-2d}=\mathcal{O}(s\log p/T)$, $s\log p/T=o(1/m_1)$, $s\log p=o(\alpha_T)$. 
Thus, to finish our proof, it only remains to show results in Steps 1-3.

\medskip
\noindent
\textbf{Proof of Step 1:} 
Recall $\tilde{\bbeta}_j$ in \eqref{berror} and $\text{supp}_1(\tilde\bbeta_j)=\mE_j$ as assumed in Assumption~\ref{ass6}. 
By definition, $\hat{\bbeta}_{\bar{\mE}_j}^0$ minimizes $\ell(\bbeta_{j})$ subject to $\text{supp}_1(\bbeta_j)=\mE_j$. 
Therefore, it holds that $\ell(\hat{\bbeta}_{\bar{\mE}_j}^0)\leq \ell(\tilde{\bbeta}_{j})$. 
Adding $-\frac{2}{T}\int_0^T\lambda_j(t)\dd N_j(t)+\frac{1}{T}\int_0^T\lambda_j^2(t)\dd t$ to both sides  
and using a similar argument as in \eqref{orgbound}, gives
\begin{equation}\label{unbiasest0}
\frac{1}{T}\int_0^T\{\hat\lambda_{\mE_j}^0(t)-\lambda_{j}(t)\}^2\dd t
\leq \frac{1}{T}\int_0^T\{\tilde{\lambda}_{j}(t)-\lambda_j(t)\}^2\dd t+\frac{2}{T}\int_0^T\{\tilde{\lambda}_j(t)-\hat\lambda_{\mE_j}^0(t)\}\{\lambda_j(t)\dd t-\dd N_j(t)\},
\end{equation}
where $\hat\lambda_{\mE_j}^0(t)=\bPsi_{\bar{\mE}_j}^\top(t)\hat{\bbeta}_{\bar{\mE}_j}^0$. By Lemma~\ref{martingale} with $\epsilon=\frac{2\log p}{T}$, we have that
\begin{equation}\label{unbiasest}
\frac{2}{T}\int_0^T\left\{\tilde{\lambda}_j(t)-\hat\lambda_{\mE_j}^0(t)\right\}\left\{\lambda_j(t)\dd t-\dd N_j(t)\right\}\leq 8\sqrt{\frac{2\log p}{T}}\left[\frac{\lambda_{\max}}{T}\int_0^T\left\{\tilde{\lambda}_{j}(t)-\hat\lambda_{\mE_j}^0(t)\right\}^2\dd t\right]^{1/2},
\end{equation}
with probability at least $1-C_3p^{-2}$.
Plugging \eqref{unbiasest} into \eqref{unbiasest0}, we get
\begin{equation*}
\begin{aligned}
&\frac{1}{T}\int_0^T\left\{\hat\lambda_{\mE_j}^0(t)-\lambda_{j}(t)\right\}^2\dd t\\
\leq& \frac{1}{T}\int_0^T\left\{\tilde{\lambda}_j(t)-\lambda_j(t)\right\}^2\dd t+8\sqrt{\frac{2\log p}{T}}\left[\frac{\lambda_{\max}}{T}\int_0^T\left\{\tilde{\lambda}_j(t)-\hat\lambda_{\mE_j}^0(t)\right\}^2\dd t\right]^{1/2},
\end{aligned}
\end{equation*}
{with probability at least $1-C_3p^{-2}$.}
Using a similar argument as in \eqref{orgbound}, we have
\begin{equation}\label{error1}
\frac{1}{T}\int_0^T\left\{\hat\lambda_{\mE_j}^0(t)-\lambda_j(t)\right\}^2\dd t\leq 32\left\{C_1(s+1)^2m_1^{-2d}+8\lambda_{\max}\frac{\log p}{T}\right\}
\end{equation}
with probability at least {$1-C_3p^{-2}-C_4p^2T\exp(-C_5T^{1/5})$}.
By the definitions of $\text{GIC}^-(\mE_j)$ and $\text{GIC}(\eta_j^\ast)$, it holds that
\begin{equation}\label{bicdis}
\begin{aligned}
&\text{GIC}^-(\mE_j)-\text{GIC}(\eta_j^\ast)\\
=&{\kappa_j}\left\{\ell(\hat{\bbeta}_{\bar{\mE}_j}^0)-\ell(\hat{\bbeta}^{\eta_j^\ast}) \right\}+\{(\alpha_T/T)\cdot |\mE_j|-(\alpha_T/T)\cdot |\mE_j|\}\\
=&{\kappa_j}\left[\left\{-\frac{2}{T}\int_0^T\hat{\lambda}_{\mE_j}^0(t)\dd N_j(t)+\frac{1}{T}\int_0^T\{\hat{\lambda}_{\mE_j}^0(t)\}^2\dd t\right\}-\left\{-\frac{2}{T}\int_0^T\hat{\lambda}_j(t)\dd N_j(t)+\frac{1}{T}\int_0^T\hat{\lambda}_j^2(t)\dd t\right\}\right]\\
=&{\kappa_j}\left[\frac{1}{T}\int_0^T\left\{\hat\lambda_{\mE_j}^0(t)-\lambda_j(t)\right\}^2\dd t-\frac{1}{T}\int_0^T\left\{\hat\lambda_{j}(t)-\lambda_j(t)\right\}^2\dd t\right]\\
&\qquad\qquad\qquad\qquad\qquad\qquad+\frac{{2\kappa_j}}{T}\int_0^T\left\{\hat\lambda_j(t)-\hat\lambda_{\mE_j}^0(t)\right\}\left\{\dd N_j(t)-\lambda_j(t)\dd t\right\}.
\end{aligned}
\end{equation}
It follows from Theorem~\ref{thm3} and \eqref{error1} that
\begin{equation}\label{error11}
\frac{1}{T}\left\vert\int_0^T\left\{\hat\lambda_{\mE_j}^0(t)-\lambda_j(t)\right\}^2\dd t-\int_0^T\left\{\hat\lambda_j(t)-\lambda_j(t)\right\}^2\dd t\right\vert\leq 64\left\{C_1(s+1)^2m_1^{-2d}+9\lambda_{\max}\frac{s\log p}{T}\right\}
\end{equation}
with probability at least {$1-C_3p^{-2}-C_4p^2T\exp(-C_5T^{1/5})$}. Again, by Lemma~\ref{martingale} with $\epsilon=\frac{2\log p}{T}$, it holds with probability at least {$1-2C_3p^{-2}-C_4p^2T\exp(-C_5T^{1/5})$},
\begin{equation}\label{error12}
\begin{aligned}
&\frac{2}{T}\int_0^T\left\{\hat\lambda_j(t)-\hat\lambda_{\mE_j}^0(t)\right\}\left\{\lambda_j(t)\dd t-\dd N_j(t)\right\}\leq  8\sqrt{\frac{2\log p}{T}}\left[\frac{\lambda_{\max}}{T}\int_0^T\left\{\hat\lambda_j(t)-\hat\lambda_{\mE_j}^0(t)\right\}^2\dd t\right]^{1/2}\\
\leq &8\sqrt{\frac{2\lambda_{\max}\log p}{T}}\left[\frac{1}{T}\int_0^T\left\{\hat\lambda_j(t)-\lambda_j(t)\right\}^2+\left\{\lambda_j(t)-\hat\lambda_{\mE_j}^0(t)\right\}^2\dd t\right]^{1/2}\\
\leq & 64\sqrt{\frac{2\lambda_{\max}\log p}{T}}\left\{C_1(s+1)^2m_1^{-2d}+9\lambda_{\max}\frac{s\log p}{T}\right\}^{1/2}
\end{aligned}
\end{equation}
where the last inequality is true due to Theorem~\ref{thm3} and \eqref{error1}. 
Since
$1/\kappa_j=\frac{1}{T}\int_0^T\dd N_j(t)$ and $\mathbb{E}\left\{\dd N_j(t)\right\}/\dd t\in[\lambda_{\min},\lambda_{\max}]$, by Theorem \ref{thm2}, we have $\kappa_j\leq \frac{1}{\lambda_{\min}}+c_1T^{-3/5}$ with probability at least $1-c_2T\exp(-c_3T^{1/5})$.
Combining \eqref{error11} and \eqref{error12} leads to the desired result in \eqref{unbiaserror}, that is,
\begin{equation*}
\left\vert\text{GIC}(\eta_j^\ast)-\text{GIC}^-(\mE_j)\right\vert\leq \frac{128}{\lambda_{\min}}\left\{C_1(s+1)^2m_1^{-2d}+9\lambda_{\max}\frac{s\log p}{T}\right\}
\end{equation*}
with probability at least {$1-2C_3p^{-1}-2C_4p^2T\exp(-C_5T^{1/5})$}.

\medskip
\noindent
\textbf{Proof of Step 2:} 
We first state a useful result that is similar to Lemma \ref{lemma:re}. Its proof is delay to Section \ref{sec:re2}.
\begingroup
\setcounter{lemma}{11} 
\renewcommand\thelemma{\arabic{lemma}b}
\begin{lemma}\label{lemma:re2}
Consider a Hawkes process on $[0,T]$ with intensity as defined in \eqref{intensity} satisfying Assumptions \ref{ass1}-\ref{ass4} and $\G$ is as defined in \eqref{Gmatrix} with normalized B-spline bases $\bphi_0(t)$ of dimension $m_0$ and $\bphi_1(t)$ of dimension $m_1$. Let $\Delta=(\Delta_0, \Delta_1,\ldots,\Delta_p)\in\mathbb{R}^{m_0+pm_1}$, and given $\mS_j\subset[p]$, write $\Delta_{\mS_j}=(\Delta_k)_{k\in\mS_j}$. 
Assume that $m_1/m_0=\mathcal{O}(1)$, $\vert\mS_j\vert=o(T^{1/2})$ and $\vert\mS_j\vert m_1=\mathcal{O}(T^{5/6})$.
For any constant $c>0$, it holds that
\begin{equation*}
\min_{\Delta\neq\bm{0}}\left\{\frac{\Delta^\top\G\Delta}{\|\Delta_{\mS_j}\|^2_2}: \sum_{k\notin \mS_j}\|\Delta_k\|_2\leq c\sum_{k\in \mS_j}\|\Delta_k\|_2, \,1\le j\le p\right\}\geq \frac{\gamma_{\min}}{4m_1},
\end{equation*}
with probability at least $1-2c_{2}pT\exp(-c_{3}T^{1/6})-c_{2}'p^2T\exp(-c_{3}'T^{1/6})$, where $c_2$, $c_2'$, $c_3$ and $c_3'$ are defined in Theorem~\ref{thm2}.
\end{lemma}
\endgroup
By \eqref{ebicproxy}, we have, for any $\mS_j\not\supset \mE_j$ and $|\mS_j|\leq s_0$,
\begin{equation}
\begin{aligned}
&\text{GIC}^-(\mS_j)-\text{GIC}^-(\mE_j)
=&\kappa_j\left\{\ell(\hat{\bbeta}_{\bar{\mS}_j}^0)-\ell(\hat{\bbeta}_{\bar{\mE}_j}^0)\right\}+(|\mS_j|-|\mE_j|)\frac{\alpha_T}{T}.\\
\end{aligned}
\end{equation}
The term $\ell(\hat{\bbeta}_{\bar{\mS}_j}^0)-\ell(\hat{\bbeta}_{\bar{\mE}_j}^0)$ can be expanded as
\begin{equation}\label{likedis}
\begin{aligned}
&\ell(\hat{\bbeta}_{\bar{\mS}_j}^0)-\ell(\hat{\bbeta}_{\bar{\mE}_j}^0)\\
=&\left\{-2\balpha_j^\top\hat{\bbeta}_{\bar{\mS}_j}^0+(\hat{\bbeta}_{\bar{\mS}_j}^0)^\top\G\hat{\bbeta}_{\bar{\mS}_j}^0\right\}-\left\{-2\balpha_j^\top\hat{\bbeta}_{\bar{\mE}_j}^0+(\hat{\bbeta}_{\bar{\mE}_j}^0)^\top\G\hat{\bbeta}_{\bar{\mE}_j}^0\right\}\\
=&\left[-\frac{2}{T}\int_0^T\hat{\lambda}_{\mS_j}^0(t)\dd N_j(t)+\frac{1}{T}\int_0^T\left\{\hat{\lambda}_{\mS_j}^0(t)\right\}^2\dd t \right]-\left[-\frac{2}{T}\int_0^T\hat{\lambda}_{\mE_j}^0(t)\dd N_j(t)+\frac{1}{T}\int_0^T\left\{\hat{\lambda}_{\mE_j}^0(t)\right\}^2\dd t \right]\\
=&\underbrace{\frac{1}{T}\int_0^T\left\{\hat{\lambda}_{\mS_j}^0(t)-\hat{\lambda}_{\mE_j}^0(t) \right\}^2\dd t}_{B_1}+\underbrace{\frac{2}{T}\int_0^T\left\{\hat{\lambda}_{\mE_j}^0(t)-\hat{\lambda}_{\mS_j}^0(t)\right\}\left\{\dd N_j(t)-\hat{\lambda}_{\mE_j}^0(t)\right\}}_{B_2}.
\end{aligned}
\end{equation}
Let $\check\mS_j=\mS_j\bigcup\mE_j$, then $|\check\mS_j|\leq s_0+s$. Since it is assumed that $s_0T^{-1/2}=o(1)$ and $s_0m_1T^{-5/6}=o(1)$ (as $m_1=\mathcal{O}(T^{1/3})$, by Lemma~\ref{lemma:re2}, it holds with probability at least {$1-2c_{2}pT\exp(-c_{3}T^{1/6})-c_{2}'p^2T\exp(-c_{3}'T^{1/6})$},
\begin{equation*}
\begin{aligned}
B_1&=\left(\hat{\bbeta}_{\bar{\mS}_j}^0-\hat{\bbeta}_{\bar{\mE}_j}^0\right)^\top\G\left(\hat{\bbeta}_{\bar{\mS}_j}^0-\hat{\bbeta}_{\bar{\mE}_j}^0\right)\geq \frac{\gamma_{\min}}{4m_1}\sum_{k\in\mE_j/\mS_j}\left\Vert\hat{\bbeta}_{\bar{\mE}_j,k}^0\right\Vert_2^2,
\end{aligned}
\end{equation*}
where $\mE_j/\mS_j$ collects indices that are in $\mE_j$ but not in $\mS_j$ and consequently, $\hat{\bbeta}_{\bar\mS_j,k}^0=\0$ for $k\in\mE_j/\mS_j$.
By \eqref{error1} and Assumption~\ref{ass6}, we have
\begin{equation}\label{eq:bound08}
\begin{aligned}
\frac{1}{T}\int_0^T\left\{\hat\lambda_{\mE_j}^0(t)-\tilde{\lambda}_j(t)\right\}^2\dd t
&\leq \frac{2}{T}\int_0^T\left\{\hat\lambda_{\mE_j}^0(t)-\lambda_j(t)\right\}^2\dd t+ \frac{2}{T}\int_0^T\left\{\lambda_j(t)-\tilde{\lambda}_j(t)\right\}^2\dd t\\
&\leq 66C_1(s+1)^2m_1^{-2d}+512\lambda_{\max}\frac{\log p}{T},
\end{aligned}
\end{equation}
with probability at least {$1-C_3p^{-2}-C_4p^2T\exp(-C_5T^{1/5})$}. 
By Lemma~\ref{lemma:re} and $\hat\bbeta^0_{\mE_j}-\tilde\bbeta_j$ is $\0$ on $\mE_j^c$, it follows that
\begin{equation*}
\frac{1}{T}\int_0^T\left\{\hat\lambda_{\mE_j}^0(t)-\tilde{\lambda}_j(t)\right\}^2\dd t\geq\frac{\gamma_{\min}}{4m_1}\sum_{k\in\mE_j}\left\Vert\hat{\bbeta}_{\mE_j,k}^0-\tilde{\bbeta}_{\mE_j,k}\right\Vert_2^2,
\end{equation*}
with probability at least {$1-2c_{2}pT\exp(-c_{3}T^{1/5})-c_{2}'p^2T\exp(-c_{3}'T^{1/5})$}.
Putting together Assumption~\ref{betamin} and \eqref{eq:bound08}, it holds for $k\in\mE_j$,
\begin{equation*}
\begin{aligned}
B_1\ge\frac{\gamma_{\min}}{4m_1}\left\Vert\hat{\bbeta}_{\bar{\mE}_j,k}^0\right\Vert_2^2&\geq \frac{\gamma_{\min}}{4m_1}\left\Vert\tilde{\bbeta}_{\bar{\mE}_j,k}\right\Vert_2^2-\frac{\gamma_{\min}}{4m_1}\left\Vert\hat{\bbeta}_{\bar{\mE}_j,k}^0-\tilde{\bbeta}_{\bar{\mE}_j,k}\right\Vert_2^2\\
&\geq \frac{\gamma_{\min}\beta_{\min}}{4m_1}-66C_1(s+1)^2m_1^{-(2d+1)}-512\lambda_{\max}\frac{\log p}{m_1T}
\end{aligned}
\end{equation*}
holds with probability at least {$1-C_3p^{-2}-3C_4p^2T\exp(-C_5T^{1/6})$}. Therefore, $B_1\geq \frac{\gamma_{\min}\beta_{\min}}{8m_1}$ holds with probability at least {$1-C_3p^{-2}-3C_4p^2T\exp(-C_5T^{1/6})$}.

Next, for term $B_2$, we have, with probability at least {$1-2C_3p^{-2}-C_4p^2\exp(-C_5T^{1/5})$},
\begin{equation*}
\begin{aligned}
B_2&=\frac{2}{T}\int_0^T\left\{\hat{\lambda}_{\mE_j}^0(t)-\hat{\lambda}_{\mS_j}^0(t)\right\}\left\{\dd N_j(t)-\lambda_j(t)+\lambda_j(t)-\hat{\lambda}_{\mE_j}^0(t)\right\}\\
&\leq 2\left[\frac{1}{T}\int_0^T\left\{\hat{\lambda}_{\mE_j}^0(t)-\hat{\lambda}_{\mS_j}^0(t)\right\}^2\dd t\right]^{1/2}\left(\sqrt{\frac{32\lambda_{\max}\log p}{T}}+\left[\frac{1}{T}\int_0^T\left\{\lambda_j(t)-\hat{\lambda}_{\mE_j}^0(t)\right\}^2\dd t\right]^{1/2}\right)\\
&\leq 2\left[\frac{1}{T}\int_0^T\left\{\hat{\lambda}_{\mE_j}^0(t)-\hat{\lambda}_{\mS_j}^0(t)\right\}^2\dd t\right]^{1/2}\left(\sqrt{\frac{32\lambda_{\max}\log p}{T}}+\sqrt{32C_1(s+1)^2m_1^{-2d}+256\lambda_{\max}\frac{\log p}{T}}\right),
\end{aligned}
\end{equation*}
where the first inequality is due to H\"{o}lder's inequality and Lemma~\ref{martingale} by setting $\epsilon=\frac{2\log p}{T}$, and the last inequality follows \eqref{error1}.
Since $B_1\geq \frac{\gamma_{\min}\beta_{\min}}{8m_1}$, $m_1\log p/T=o(1)$ and $s^2m_1^{-2d+1}=o(1)$, we have
\begin{equation*}
\begin{aligned}
&B_1+B_2\ge\left[\frac{1}{T}\int_0^T\left\{\hat{\lambda}_{\mE_j}^0(t)-\hat{\lambda}_{\mS_j}^0(t)\right\}^2\dd t\right]^{1/2}\times\\
&\left(\left[\frac{1}{T}\int_0^T\left\{\hat{\lambda}_{\mE_j}^0(t)-\hat{\lambda}_{\mS_j}^0(t)\right\}^2\dd t\right]^{1/2}-2\sqrt{\frac{32\lambda_{\max}\log p}{T}}-2\sqrt{32C_1(s+1)^2m_1^{-2d}+256\lambda_{\max}\frac{\log p}{T}}\right)\\
\geq &\sqrt{\frac{\gamma_{\min}\beta_{\min}}{8m_1}}\times\frac{1}{2}\sqrt{\frac{\gamma_{\min}\beta_{\min}}{m_1}},
\end{aligned}
\end{equation*}
when $T$ is sufficiently large. Plugging this into \eqref{likedis}, we can get that
\begin{equation*}
\ell(\hat{\bbeta}_{\bar{\mS}_j}^0)-\ell(\hat{\bbeta}_{\bar{\mE}_j}^0)\geq \frac{\gamma_{\min}\beta_{\min}}{16m_1}
\end{equation*}
with probability at least {$1-3C_3p^{-2}-4C_4p^2\exp(-C_5T^{1/5})$}.
When $\vert\mS_j\vert\le\vert\mE_j\vert$, given $sm_1\alpha_T/T=o(1)$, we get that $(|\mS_j|-|\mE_j|)\frac{\alpha_T}{T}=o(m_1^{-1})$. Hence, for sufficiently large $T$,
\begin{equation*}
\begin{aligned}
&\text{GIC}^-(\mS_j)-\text{GIC}^-(\mE_j)
\geq \frac{\gamma_{\min}\beta_{\min}}{16m_1\lambda_{\min}}+(|\mS_j|-|\mE_j|)\frac{\alpha_T}{T}
\geq\frac{\gamma_{\min}\beta_{\min}}{32\lambda_{\min}m_1}
\end{aligned}
\end{equation*}
with probability at least {$1-3C_3p^{-2}-5C_4p^2T\exp(-C_5T^{1/6})$}.

\medskip
\noindent
\textbf{Proof of Step 3:} By \eqref{ebicproxy}, we have, for any $\mS_j\supset \mE_j$,
\begin{equation}
\begin{aligned}
&\text{GIC}^-(\mS_j)-\text{GIC}^-(\mE_j)=&{\kappa_j}\left\{\ell(\hat{\bbeta}_{\bar{\mS}_j}^0)-\ell(\hat{\bbeta}_{\bar{\mE}_j}^0)\right\}+(|\mS_j|-|\mE_j|)\frac{\alpha_T}{T}.
\end{aligned}
\end{equation}
Same as in \eqref{likedis}, we have
\begin{equation*}
\begin{aligned}
&\ell(\hat{\bbeta}_{\bar{\mS}_j}^0)-\ell(\hat{\bbeta}_{\bar{\mE}_j}^0)
=\underbrace{\frac{1}{T}\int_0^T\left\{\hat{\lambda}_{\mS_j}^0(t)-\hat{\lambda}_{\mE_j}^0(t) \right\}^2\dd t}_{B_1}+\underbrace{\frac{2}{T}\int_0^T\left\{\hat{\lambda}_{\mE_j}^0(t)-\hat{\lambda}_{\mS_j}^0(t)\right\}\left\{\dd N_j(t)-\hat{\lambda}_{\mE_j}^0(t)\right\}}_{B_2}.
\end{aligned}
\end{equation*}
By the definition of $\hat{\bbeta}_{\bar{\mS}_j}^0$ and $\bar{\mS}_j\supset \bar{\mE}_j$, we have $\ell(\hat{\bbeta}_{\bar{\mS}_j}^0)\leq\ell(\hat{\bbeta}_{\bar{\mE}_j}^0)$. Adding $-\frac{2}{T}\int_0^T\lambda_j(t)\dd N_j(t)+\frac{1}{T}\int_0^T\lambda_j^2(t)\dd t$ to both sides of $\ell(\hat{\bbeta}_{\bar{\mS}_j}^0)\leq\ell(\hat{\bbeta}_{\bar{\mE}_j}^0)$, it arrived at that
\begin{equation*}
\frac{1}{T}\int_0^T\{\hat\lambda_{\mS_j}^0(t)-\lambda_{j}(t)\}^2\dd t
\leq \frac{1}{T}\int_0^T\{\hat{\lambda}_{\mE_j}^0(t)-\lambda_j(t)\}^2\dd t+\frac{2}{T}\int_0^T\{\hat{\lambda}_{\mS_j}^0(t)-\hat\lambda_{\mE_j}^0(t)\}\{\lambda_j(t)\dd t-\dd N_j(t)\}.
\end{equation*}
By Lemma~\ref{martingale} with $\epsilon=\frac{2\log p}{T}$, it arrives that
\begin{equation*}
\frac{1}{T}\int_0^T\{\hat\lambda_{\mS_j}^0(t)-\lambda_{j}(t)\}^2\dd t
\leq \frac{1}{T}\int_0^T\{\hat{\lambda}_{\mE_j}^0(t)-\lambda_j(t)\}^2\dd t+8\sqrt{\frac{2\log p}{T}}\left[\frac{\lambda_{\max}}{T}\int_0^T\left\{\hat{\lambda}_{\mS_j}^0(t)-\hat\lambda_{\mE_j}^0(t)\right\}^2\dd t\right]^{1/2},
\end{equation*}
{with probability at least $1-C_3p^{-2}$.}
For the term $B_1$, it holds that
\begin{equation*}
\begin{aligned}
B_1&\leq \frac{1}{T}\int_0^T\left\{\hat\lambda_{\mS_j}^0(t)-\lambda_j(t)\right\}^2\dd t+\frac{1}{T}\int_0^T\left\{\hat\lambda_{\mE_j}^0(t)-\lambda_j(t)\right\}^2\dd t\\
&\leq \frac{2}{T}\int_0^T\left\{\hat\lambda_{\mE_j}^0(t)-\lambda_j(t)\right\}^2\dd t+8\sqrt{\frac{2\log p}{T}}\left[\frac{\lambda_{\max}}{T}\int_0^T\left\{\hat{\lambda}_{\mS_j}^0(t)-\hat\lambda_{\mE_j}^0(t)\right\}^2\dd t\right]^{1/2},
\end{aligned}
\end{equation*}
{with probability at least $1-C_3p^{-2}$.}
If {$B_1\leq \frac{1}{T}\int_0^T\left\{\hat{\lambda}_{\mS_j}^0(t)-\hat{\lambda}_{\mE_j}^0(t) \right\}^2\dd t$}, by \eqref{error1}, it holds with probability at least {$1-2C_3p^{-2}-C_4p^2T\exp(-C_5T^{1/5})$},
\begin{equation*}
B_1\leq {64}\left\{C_1(s+1)^2m_1^{-2d}+\frac{9\lambda_{\max}\log p}{T}\right\}.
\end{equation*}
Otherwise, if {$B_1\geq \frac{1}{T}\int_0^T\left\{\hat{\lambda}_{\mS_j}^0(t)-\hat{\lambda}_{\mE_j}^0(t) \right\}^2\dd t$} and by noting $B_1=\frac{1}{T}\int_0^T\left\{\hat{\lambda}_{\mS_j}^0(t)-\hat{\lambda}_{\mE_j}^0(t) \right\}^2\dd t$, we can get
\begin{equation*}
B_1 \leq \left(2\left[\frac{1}{T}\int_0^T\left\{\hat\lambda_{\mE_j}^0(t)-\lambda_j(t)\right\}^2\dd t\right]^{1/2}+8\sqrt{\frac{2\lambda_{\max}\log p}{T}}\right)\left[\frac{1}{T}\int_0^T\left\{\hat{\lambda}_{\mS_j}^0(t)-\hat\lambda_{\mE_j}^0(t)\right\}^2\dd t\right]^{1/2},
\end{equation*}
which together with \eqref{error1} imply that
\begin{equation*}
\begin{aligned}
B_1 &\leq \left(2\left[\frac{1}{T}\int_0^T\left\{\hat\lambda_{\mE_j}^0(t)-\lambda_j(t)\right\}^2\dd t\right]^{1/2}+8\sqrt{\frac{2\lambda_{\max}\log p}{T}}\right)^2\\
&\leq 256\left\{C_1(s+1)^2m_1^{-2d}+\frac{10\lambda_{\max}\log p}{T}\right\},
\end{aligned}
\end{equation*}
with probability at least {$1-2C_3p^{-2}-C_4p^2T\exp(-C_5T^{1/5})$}. Therefore, we arrive at the result that $B_1\leq 256\left\{C_1(s+1)^2m_1^{-2d}+\frac{10\lambda_{\max}\log p}{T}\right\}$ with probability at least {$1-2C_3p^{-2}-C_4p^2T\exp(-C_5T^{1/5})$}.

For the term $B_2$, it holds with with probability at least {$1-3C_3p^{-2}-2C_4p^2T\exp(-C_5T^{1/5})$},
\begin{equation*}
\begin{aligned}
&B_2=\frac{2}{T}\int_0^T\left\{\hat{\lambda}_{\mE_j}^0(t)-\hat{\lambda}_{\mS_j}^0(t)\right\}\left\{\dd N_j(t)-\lambda_j(t)+\lambda_j(t)-\hat{\lambda}_{\mE_j}^0(t)\right\}\\ 
&\leq 8\left[ \frac{1}{T}\int_0^T\left\{\hat{\lambda}_{\mE_j}^0(t)-\hat{\lambda}_{\mS_j}^0(t)\right\}^2\dd t\right]^{1/2}\left[\sqrt{\frac{2\lambda_{\max}\log p}{T}}+\sqrt{2C_1(s+1)^2m_1^{-2d}+18\lambda_{\max}\frac{\log p}{T}}\right]\\
&\leq 128\sqrt{C_1(s+1)^2m_1^{-2d}+10\lambda_{\max}\frac{\log p}{T}}\left[\sqrt{\frac{\lambda_{\max}\log p}{T}}+\sqrt{C_1(s+1)^2m_1^{-2d}+9\lambda_{\max}\frac{\log p}{T}}\right],
\end{aligned}
\end{equation*}
where the first inequality is due to H\"{o}lder's inequality and Lemma~\ref{martingale} by setting $\epsilon=\frac{2\log p}{T}$ and the second inequality holds due to the previously derived upper bound on $B_1$.
Combining $B_1$ and $B_2$ and noting that $C_1(s+1)^2m_1^{-2d}+\frac{9\lambda_{\max}\log p}{T}=\mathcal{O}(s\log p/T)$, we have $\kappa_j\left\{\ell(\hat{\bbeta}_{\bar{\mS}_j}^0)-\ell(\hat{\bbeta}_{\bar{\mE}_j}^0)\right\}$ is $O_p\left(\frac{s\log p}{T}\right)$. Since $s\log p/\alpha_T=o(1)$, we have $\text{GIC}^-(\mS_j)-\text{GIC}^-(\mE_j)>\alpha_T/(2T)$ with probability at least {$1-5C_3p^{-2}-4C_4p^2T\exp(-C_5T^{1/5})$}.
\eop
}

\subsection{Proof of Theorem~\ref{thm5}}
Under the conditions of Theorem \ref{thm4}, $\hat\bbeta_j$ is selection consistent under $\text{H}_0$, i.e., $\text{supp}_1(\hat\bbeta_j)=\text{supp}_1(\hat\bbeta^1_j)=\mE_j$ with high probability. By the definition of $\hat\bbeta^{H_0}_j$, it also holds that $\text{supp}_1(\hat\bbeta^{H_0}_j)=\mE_j$ with high probability. The proof is divided into three steps. In the first step, we derive the asymptotic expansion of $S_j$ and in the second step, we show the asymptotic distribution of an intermediate term using martingale central limit theorem. In the last step, we use results from step 2 to derive the distribution of $S_j$.

We define some notations. 
Let $\B= 
\begin{bmatrix}
\u & \0_{m_1p}\\
\0_{m_0\times m_1p} & \I_{m_1p\times m_1p} 
\end{bmatrix}\in\mathbb{R}^{(1+m_1p)\times (m_0+m_1p)}$, and $\u=(1,0,\ldots,0)\in\mathbb{R}^{m_0}$. 
Define $\tilde\bgamma_{j}=([\bbeta_{j,0}]_1,(\bbeta_{j,k})_{k\in[p]})\in\mathbb{R}^{1+m_1p}$, where $[\bbeta_{j,0}]_1$ denotes the first element of $\bbeta_{j,0}$.
Let $\G_{\bar\mE_j}\in\mathbb{R}^{(m_0+m_1s)\times (m_0+m_1s)}$ denote the submatrix of $\G$ with rows/columns in $\bar\mE_j$, $\B_{\bar\mE_j}\in\mathbb{R}^{(1+m_1s)\times (m_0+m_1s)}$ denote the submatrix of $\B$ with rows in $\{1\}\bigcup\mE_j$ and columns in $\bar\mE_j$, $\bgamma_{\bar\mE_j}=([\bbeta_{j,0}]_1,(\bbeta_{j,k})_{k\in\mE_j})$ and $\bbeta_{\bar\mE_j}=(\bbeta_{j,0},(\bbeta_{j,k})_{k\in\mE_j})$.
Under the null, we have that $\B^\top\tilde{\bgamma}_{j}=\tilde\bbeta_j$.
We write $\hat\bbeta^{H_0}_j=\B^\top\hat\bgamma_{j}$, where $\hat\bgamma_{j}$ is defined as
\begin{equation}\label{unbiasedest1}
\hat\bgamma_{j}=\arg\min_{\substack{\b_j\in\mathbb{R}^{1+m_1p}\\ \text{supp}_1(\b_j)=\text{supp}_1(\hat\bbeta_j)}}\ell(\B^\top\b_j)\end{equation}

\noindent
\textbf{\bf Step 1:}
Apply Taylor's expansion for $\ell_j(\hat\bbeta^{H_0}_j)$ at the point $\hat{\bbeta}^1_j$ and by noting $\nabla  \ell_j(\hat{\bbeta}^1_j)=\0$, we have
\begin{eqnarray}\label{orgtest1}
S_j=T\left\{\ell_j(\hat\bbeta^{H_0}_j)- \ell_j(\hat{\bbeta}^1_j)\right\}&=&T(\B^\top\hat{\bgamma}_j-\hat{\bbeta}^1_j)^\top\G(\B^\top\hat{\bgamma}_j-\hat{\bbeta}^1_j)\\\nonumber
&=&T(\B_{\bar\mE_j}^\top\hat{\bgamma}_{\bar\mE_j}-\hat{\bbeta}^1_{\bar\mE_j})^\top\G_{\bar\mE_j}(\B_{\bar\mE_j}^\top\hat{\bgamma}_{\bar\mE_j}-\hat{\bbeta}^1_{\bar\mE_j}).
\end{eqnarray}
Define $\ell_j(\bbeta_{\bar\mE_j})=-2\balpha_{\bar\mE_j}\bbeta_{\bar\mE_j}+\bbeta_{\bar\mE_j}^\top\G_{\bar\mE_j}\bbeta_{\bar\mE_j}$, and the following holds
$$
\nabla\ell_j(\hat\bbeta^1_{\bar\mE_j})-\nabla\ell_j(\tilde{\bbeta}_{\bar\mE_j})=2\G_{\bar\mE_j}(\hat\bbeta^1_{\bar\mE_j}-\tilde{\bbeta}_{\bar\mE_j})
$$ 
$$
\nabla\ell_j(\B_{\bar\mE_j}^\top\hat{\bgamma}_{\bar\mE_j})-\nabla\ell_j(\B_{\bar\mE_j}^\top\tilde{\bgamma}_{\bar\mE_j})=2\B_{\bar\mE_j}^\top\G_{\bar\mE_j}\B_{\bar\mE_j}(\hat\bgamma_{\bar\mE_j}-\tilde{\bgamma}_{\bar\mE_j}).
$$ 
Therefore, by noting $\B_{\bar\mE_j}\B_{\bar\mE_j}^\top=\I$, we have 
$$
2(\hat\bbeta^1_{\bar\mE_j}-\tilde{\bbeta}_{\bar\mE_j})=\G_{\bar\mE_j}^{-1}\left\{\nabla\ell_j(\hat\bbeta^1_{\bar\mE_j})-\nabla\ell_j(\tilde{\bbeta}_{\bar\mE_j})\right\}
$$
$$
2\B_{\bar\mE_j}(\hat\bgamma_{\bar\mE_j}-\tilde{\bgamma}_{\bar\mE_j})=\B_{\bar\mE_j}^\top(\B_{\bar\mE_j}^\top\G_{\bar\mE_j}\B_{\bar\mE_j})^{-1}\B_{\bar\mE_j}\B_{\bar\mE_j}^\top\left\{\nabla\ell_j(\B_{\bar\mE_j}^\top\hat{\bgamma}_{\bar\mE_j})-\nabla\ell_j(\B_{\bar\mE_j}^\top\tilde{\bgamma}_{\bar\mE_j})\right\}.
$$
By noting $\B_{\bar\mE_j}^\top\tilde{\bgamma}_{j}=\tilde\bbeta_{\bar\mE_j}$ and $\B_{\bar\mE_j}\B_{\bar\mE_j}^\top=\I$, it is true that $\B_{\bar\mE_j}^\top\nabla\ell_j(\B_{\bar\mE_j}^\top\tilde{\bgamma}_j)=-2\B_{\bar\mE_j}^\top\B_{\bar\mE_j}\balpha_{\bar\mE_j}+2\B_{\bar\mE_j}^\top\B_{\bar\mE_j}\G_{\bar\mE_j}(\B_{\bar\mE_j}^\top\tilde{\bgamma}_{\bar\mE_j})=-2\balpha_{\bar\mE_j}+2\G_{\bar\mE_j}\tilde{\bbeta}_{\bar\mE_j}=\nabla  \ell_j(\tilde{\bbeta}_{\bar\mE_j})$.
Combining the above results, and again by noting $\B_{\bar\mE_j}^\top\tilde{\bgamma}_{\bar\mE_j}=\tilde\bbeta_{\bar\mE_j}$, we get that
\begin{equation}\label{disnull}
\begin{aligned}
2\left\{\hat{\bbeta}^1_{\bar\mE_j}-\B_{\bar\mE_j}^\top\hat{\bgamma}_{\bar\mE_j}\right\}=&-\left(\G_{\bar\mE_j}^{-1}-\B_{\bar\mE_j}^\top(\B_{\bar\mE_j} \G_{\bar\mE_j}\B_{\bar\mE_j}^\top)^{-1}\B_{\bar\mE_j} \right)\nabla \ell(\tilde{\bbeta}_{\bar\mE_j}).
\end{aligned}
\end{equation}

\medskip
\noindent
\textbf{\bf Step 2:}
Next, we derive the distribution of $\widetilde{M}_{T}=\frac{1}{\sqrt{T}}\int_0^T\bPsi_{\bar{\mE}_j}(t)\left\{\lambda_j(t)\dd t-\dd N_j(t)\right\}$.
Denote
$$
U_{T}(t)=\frac{\bPsi_{\bar{\mE}_j} (t)}{\sqrt{T}}\quad\text{and}\quad M_{T}(t)=N_j(t)-\int_{0}^{t}\lambda_j(s)\dd s.
$$
Note that, since $M_{T}(t)$ is a square-integrable martingale and $U_{T}(t)$ is locally bounded and predictable, $\widetilde{M}_{T}=\int_{0}^T U_{T}(t)\dd M_{T}(t)$ is a locally square-integrable martingale. Before we apply the central limit theorem, we check the large jump of $\widetilde{M}_{T}$.
{
Define 
$$
\widetilde{M}_{k,l,\epsilon}(t)=\int_{0}^{t}\left\{\bPsi_{k,l}(s)/\sqrt{T}\right\}1_{\left\{|\bPsi_{k,l}(s)/\sqrt{T}|>\epsilon\right\}}\dd M_{T}(s).
$$
This accumulates all the jumps in $\widetilde{M}_{T}$ before time $t$ that exceeds $\epsilon$. 
The predictable variation process of $\widetilde{M}_{k,l,\epsilon}(t)$ (\cite{andersen2012statistical}, page 78; see also \cite{chen2013inference}) is
\begin{equation}\label{eqn:var1}
\langle\widetilde{M}_{k,l,\epsilon}\rangle(t)=\int_{0}^{t}\left\{\bPsi_{k,l}(s)/\sqrt{T}\right\}^21_{\left\{|\bPsi_{k,l}(s)/\sqrt{T}|>\epsilon\right\}}\lambda_j(s)\dd s.
\end{equation}
By Lenglart's inequality (\cite{andersen2012statistical}, page 86; see also \cite{Lenglart1977li}), we have that 
\begin{equation}\label{eqn:var2}
\mathbb{P}\left(\sup_{t\in (0,T]}|\widetilde{M}_{k,l,\epsilon}(t)|>\eta_0\right)\leq \frac{\delta}{\eta_0^2}+\mathbb{P}\left(\langle\widetilde{M}_{k,l,\epsilon}\rangle(T)>\delta\right).
\end{equation}
By the definition of $\bPsi_{k,l}(t)$ (i.e., $\bPsi_{0,l}(t)=\phi_{0,l}(t)$ is a bounded function and $\bPsi_{k,l}(t)=\int_0^t\phi_{1,l}(t-s)\dd N_k(s)$ where $\phi_{1,l}(t)$ is a bounded function with bounded support) and Theorem \ref{thm2}, it holds that $\bPsi_{k,l}(t)/\sqrt{T}=\mathcal{O}_p(T^{-1/2})$ for any $t$. Plugging this in \eqref{eqn:var1} and we get $\mathbb{P}(\langle\widetilde{M}_{k,l,\epsilon}\rangle(T)>\delta)\rightarrow^P 0$ for all $\delta>0$. 
Taking $\eta_0=\delta^{1/4}$ in \eqref{eqn:var2}, we deduce that $\sup_{t\in (0,T]}|\widetilde{M}_{j,T,\epsilon}|\stackrel{p}{\rightarrow}0$, as $\delta$ can be arbitrarily small.}

Applying the martingale central limit theorem, we can conclude that
\begin{equation}\label{normal}
\V^{-1/2}\widetilde{M}_{T}\stackrel{\mathcal{D}}{\rightarrow}N(\0,\I),
\end{equation}
where
\begin{equation*}
\V=\frac{1}{T}\int_{0}^{T}\mathbb{E}\left(\bPsi_{\bar{\mE}_j}(t)\bPsi_{\bar{\mE}_j}^\top(t)\right)\bar{\lambda}_j(t)\dd t=\bar\lambda_j\mathbb{E}(\G_{\bar{\mE}_j}).
\end{equation*}
where $\bar\lambda_j=\bar{\lambda}_j(t)$ under the null hypothesis.

\medskip
\noindent
\textbf{\bf Step 3:}
By definition of $\ell_j(\tilde\bbeta_{\bar\mE_j})$, we may write $\nabla \ell_j(\tilde{\bbeta}_{\bar\mE_j})=\frac{2}{T}\int_0^T\bPsi_{\bar{\mE}_j}(t)\left\{\tilde\lambda_j(t)\dd t-\dd N_j(t)\right\}$, where we used the fact that $\bPsi_{\bar\mE_j}(t)\tilde\bbeta_{\bar\mE_j}=\tilde\lambda_j(t)$. Therefore, we have
$$
\nabla \ell(\tilde{\bbeta}_{\bar\mE_j})=\frac{2}{\sqrt{T}}\widetilde{M}_{T}+\frac{2}{T}\int_0^T\bPsi_{\bar{\mE}_j}(t)\left\{\tilde\lambda_j(t)-{\lambda}_j(t)\right\}\dd t.
$$ 
Using a similar argument as in \eqref{error0}, we have that 
\begin{equation*}
\left\|\frac{1}{T}\int_{0}^T\bPsi_{0}(t)\left\{\tilde{\lambda}_j(t)-{\lambda}_j(t)\right\}\dd t\right\|_2^2\leq\frac{32{c_4'c_5'}}{m_0}C_1(s+1)^2m_1^{-2d},
\end{equation*}
and using a similar argument as in \eqref{errork1}, we have that
\begin{equation*}
\left|\frac{1}{T}\int_{0}^T\bPsi_{k,l}(t)\left\{\tilde{\lambda}_j(t)-\lambda_j(t)\right\}\dd t\right|^2\leq \frac{2{c_6'}^2C_1\lambda_{\max}^2(s+1)^2}{m_1^{2d+2}}
\end{equation*}
with probability at least {$1-C_3p^{-2}-C_4p^2T\exp(-C_5T^{1/5})$}. 
Thus, using a similar argument as in \eqref{I1bound} and given $s^3T=o(m_1^{2d+1})$, we have that
\begin{equation*}
\left\|\frac{2}{T}\int_{0}^T\bPsi_{\bar{\mE}_j}(t)\left\{\tilde{\lambda}_j(t)-\lambda_j(t)\right\}\dd t\right\|_2=\mathcal{O}_p\left\{(s^3/m_1^{2d+1})^{1/2}\right\}=o_p(T^{-1/2}).
\end{equation*} 
Since $\V^{-1/2}\widetilde{M}_{T}\stackrel{\mathcal{D}}{\rightarrow}N(\0,\I)$, it holds that 
$\Vert\widetilde{M}_{T}\Vert_2=\mathcal{O}_p(1)$, where we used the fact that $\sigma_{\max}(\V^{-1})=\mathcal{O}_p(sm_1)$ by Lemma \ref{lemma:re} and that $\V^{-1/2}\widetilde{M}_{T}$ is $N(\0_{m_0+m_1s},\I_{m_0+m_1s})$.
It is seen that $\nabla \ell(\tilde{\bbeta}_{\bar\mE_j})$ is dominated by the term $\frac{2}{\sqrt{T}}\widetilde{M}_{T}$ and the term $\frac{2}{T}\int_0^T\bPsi_{\bar{\mE}_j}(t)\left\{\tilde\lambda_j(t)-{\lambda}_j(t)\right\}\dd t$ is negligible.
Combined with \eqref{disnull} and \eqref{orgtest1}, we get that, with probability tending to 1,
\begin{equation*}\label{dist0}
\begin{aligned}
S_j&=\widetilde{M}_{T}^\top\bSigma_{T} \widetilde{M}_{T}+o(1).
\end{aligned}
\end{equation*} 
By Corollary~\ref{cor1} and the fact that $\I-\mathbb{E}(\G_{\bar\mE_j})^{1/2}\B_{\bar\mE_j}^\top\left(\B_{\bar\mE_j} \mathbb{E}(\G_{\bar\mE_j})\B_{\bar\mE_j}^\top\right)^{-1}\B_{\bar\mE_j} \mathbb{E}(\G_{\bar\mE_j})^{1/2}$ is an idempotent matrix, we have
\begin{equation*}
\begin{aligned}
\bSigma_T&=\mathbb{E}(\G_{\mathcal{E}_j})^{-1/2}\left\{\I-\mathbb{E}(\G_{\mathcal{E}_j})^{1/2}\B_{\bar\mE_j}^\top\left(\B_{\bar\mE_j} \mathbb{E}(\G_{\bar\mE_j})\B_{\bar\mE_j}^\top\right)^{-1}\B_{\bar\mE_j} \mathbb{E}(\G_{\bar\mE_j})^{1/2}\right\}^2\mathbb{E}(\G_{\mathcal{E}_j})^{-1/2}\\
&=\mathbb{E}(\G_{\mathcal{E}_j})^{-1/2}\left\{\I-\mathbb{E}(\G_{\mathcal{E}_j})^{1/2}\B_{\bar\mE_j}^\top\left(\B_{\bar\mE_j} \mathbb{E}(\G_{\bar\mE_j})\B_{\bar\mE_j}^\top\right)^{-1}\B_{\bar\mE_j} \mathbb{E}(\G_{\bar\mE_j})^{1/2}\right\}\mathbb{E}(\G_{\mathcal{E}_j})^{-1/2}.
\end{aligned}
\end{equation*} 
Correspondingly, it holds that
\begin{equation*}
\widetilde{M}_{T}^\top\bSigma_T\widetilde{M}_{T}=\left(\V^{-1/2}\widetilde{M}_{T}\right)^\top\V^{1/2}\bSigma_T\V^{1/2} \left(\V^{-1/2}\widetilde{M}_{T}\right)=\bar\lambda_j\chi_{m_0-1}^2,
\end{equation*}
where the degree freedom $m_0-1$ is due to that $\mathbb{E}(\G_{\bar\mE_j})^{1/2}\B_{\bar\mE_j}^\top\left(\B_{\bar\mE_j} \mathbb{E}(\G_{\bar\mE_j})\B_{\bar\mE_j}^\top\right)^{-1}\B_{\bar\mE_j} \mathbb{E}(\G_{\bar\mE_j})^{1/2}$ is an idempotent matrix with rank $1+m_1s$ (and $1+m_1s$ nonzero eigenvalues of 1) and therefore $\I-\mathbb{E}(\G_{\bar\mE_j})^{1/2}\B_{\bar\mE_j}^\top\left(\B_{\bar\mE_j} \mathbb{E}(\G_{\bar\mE_j})\B_{\bar\mE_j}^\top\right)^{-1}\B_{\bar\mE_j} \mathbb{E}(\G_{\bar\mE_j})^{1/2}$ is an idempotent matrix with rank $m_0-1$.

\eop

\subsection{Proof of Proposition \ref{pop1}}
{
First, under the conditions of Theorem \ref{thm4}, $\hat\bbeta_j$ is selection consistent under $\text{H}_1$, i.e., $\text{supp}_1(\hat\bbeta_j)=\mE_j$ with high probability. Hence, by the definition of $\bgamma_j$, it also holds that $\text{supp}_1(\B^\top\hat\bgamma_j)=\mE_j$ with high probability. Moreover, under $\text{H}_1$, $\B^\top\hat\bgamma_j$ is an under-fitted model as the background intensity is restricted to a constant. 

Define $L_j(\bbeta)=T\times \ell_j(\bbeta)$. Similar to the proof of Theorem \ref{thm5}, we first apply Taylor's expansion for $L_j(\B^\top\hat{\bgamma}_j)$ at the point $\hat{\bbeta}_j$. By noting $\nabla  L_j(\hat{\bbeta}_j)=\0$, we have
\begin{equation}\label{orgtest}
S_j=L_j(\B^\top\hat{\bgamma}_j)-  L_j(\hat{\bbeta}_j)=T(\B^\top\hat{\bgamma}_j-\hat{\bbeta}_j)^\top\G(\B^\top\hat{\bgamma}_j-\hat{\bbeta}_j).
\end{equation}
Since $\text{supp}_1(\B^\top\hat{\bgamma}_j-\hat{\bbeta}_j)=\mE_j$, by \eqref{eq:bound081} in the proof of Lemma~\ref{lemma:re2}, it holds for some constant $C_7>0$ with probability at least {$1-C_4p^2T\exp(-C_5T^{1/5})$}, 
\begin{equation}\label{eqn:power}
\begin{aligned}
&(\B^\top\hat{\bgamma}_j-\hat{\bbeta}_j)^\top\G(\B^\top\hat{\bgamma}_j-\hat{\bbeta}_j)\geq \frac{C_7}{sm_1}\Vert\A\hat{\bbeta}_{j,0}\Vert_2^2,
\end{aligned}
\end{equation}
where we used the fact that $\A{\bbeta}_{j,0}$ is ${\bbeta}_{j,0}$ without the first element and the first $m_0$ elements of $\B^\top\hat{\bgamma}_j$ is $(\hat\bgamma_{j,1},0,\ldots,0)$ and $\hat\bgamma_{j,1}$ denotes the first element of $\hat\bgamma_j$. 
Next, by Theorem \ref{thm3} and Assumption~\ref{ass6}, we have
\begin{equation*}
\begin{aligned}
\frac{1}{T}\int_0^T\left\{\hat\lambda_{j}(t)-\tilde{\lambda}_j(t)\right\}^2\dd t
&\leq \frac{2}{T}\int_0^T\left\{\hat\lambda_{j}(t)-\lambda_j(t)\right\}^2\dd t+ \frac{2}{T}\int_0^T\left\{\lambda_j(t)-\tilde{\lambda}_j(t)\right\}^2\dd t\\
&\leq 66C_1(s+1)^2m_1^{-2d}+512\lambda_{\max}\frac{s\log p}{T},
\end{aligned}
\end{equation*}
with probability at least {$1-C_3p^{-2}-C_4p^2T\exp(-C_5T^{1/5})$}. 
By \eqref{eqn:b0}, it follows that
\begin{eqnarray*}
\frac{1}{T}\int_0^T\left\{\hat\lambda_{j}(t)-\tilde{\lambda}_j(t)\right\}^2\dd t&\geq&\frac{C_7}{sm_1}\sum_{k\in\mE_j}\left\Vert\hat{\bbeta}_{j,0}-\tilde{\bbeta}_{j,0}\right\Vert_2^2\\
&\ge&\frac{C_7}{sm_1}\sum_{k\in\mE_j}\left\Vert\A\hat{\bbeta}_{j,0}-\A\tilde{\bbeta}_{j,0}\right\Vert_2^2,
\end{eqnarray*}
with probability at least {$1-C_3p^{-2}-C_4p^2T\exp(-C_5T^{1/5})$}, where we used the fact that $\A{\bbeta}_{j,0}$ is ${\bbeta}_{j,0}$ without the first element.
Correspondingly, we have 
\begin{equation}\label{eq:bound081}
\left\Vert\A\hat{\bbeta}_{j,0}-\A\tilde{\bbeta}_{j,0}\right\Vert_2^2=\mathcal{O}(s^2m_1\log p/T),
\end{equation}
where we used the fact that $s^2T/\log p=\mathcal{O}(m_1^{2d})$ as assumed in Theorem \ref{thm4}.

From $\|\A\tilde\bbeta_{j,0}\|_2/(s^2m_1\log p/T)^{1/2}\rightarrow\infty$ and \eqref{eq:bound081}, we have when $T$ is sufficiently large
\begin{equation*}
\begin{aligned}
\left\Vert\A\hat{\bbeta}_{j,0}\right\Vert_2^2&\geq \left\Vert\A\tilde{\bbeta}_{j,0}\right\Vert_2^2-\left\Vert\A\hat{\bbeta}_{j,0}-\A\tilde{\bbeta}_{j,0}\right\Vert_2^2\\
&\geq M'_1\frac{s^2m_1\log p}{T}
\end{aligned}
\end{equation*}
for some constant $M'_1>0$ with probability at least $1-2C_3p^{-2}-3C_4p^2T\exp(-C_5T^{1/5})$. 
Plugging this into \eqref{eqn:power}, we have that $S_j\ge C_7M'_1s\log p/2$ with probability at least {$1-2C_3p^{-2}-3C_4p^2T\exp(-C_5T^{1/5})$}. Combined with the fact that $\bar\lambda_j(t)$ is bounded, we arrive at the desired result. 
}
\eop

\subsection{Plausibility of Assumption \ref{ass6}}
\label{proof:ass6}
{
Let $g_{j,0}(t)=\widetilde{\nu}_j(t)-\nu_j(t)$ and $g_{j,k}(t)=\widetilde{\omega}_{j,k}(t)-\omega_{j,k}(t)$. When $h(x)=x$, we have 
\begin{equation}\label{egn:expand}
\begin{aligned}
&\frac{1}{T}\int_{0}^T \left\{\bPsi^\top(t)\widetilde\bbeta_{j}-\lambda_{j}(t)\right\}^{2}\dd t
=\frac{1}{T}\int_{0}^T \left\{g_{j,0}(t)+\sum_{k=1}^p\int_0^tg_{j,k}(t-s)\dd N_k(s)\right\}^{2}\dd t\\
\leq &(s+1)\left[\frac{1}{T}\int_{0}^T g_{j,0}^2(t)\dd t+\sum_{k=1}^p\frac{1}{T}\int_{0}^T \left\{\int_0^tg_{j,k}(t-s)\dd N_k(s)\right\}^{2}\dd t\right],
\end{aligned}
\end{equation}
where the last inequality holds due to the Cauchy-Schwarz inequality. 
Consider the case where the approximation errors satisfy
\begin{equation*}
\frac{1}{T}\int_{0}^T g_{j,0}^2(t)\dd t=\mathcal{O}(m_0^{-2d})\quad\text{and}\quad \int_0^bg_{j,k}^2(t-s)\dd s=\mathcal{O}(m_1^{-2d}).
\end{equation*}
Next, it holds that
\begin{equation*}
\begin{aligned}
&\frac{1}{T}\int_{0}^T \left\{\int_0^tg_{j,k}(t-s)\dd N_k(s)\right\}^{2}\dd t\\
=&\frac{1}{T}\int_0^T\int_0^t\int_{u_1\neq u_2}g_{j,k}(t-u_1)g_{j,k}(t-u_2)\dd N_k(u_1)\dd N_k(u_2)\dd t+\frac{1}{T}\int_0^T\int_0^tg_{j,k}^2(t-u)\dd N_k(u)\dd u\dd t\\
=&\frac{1}{T}\int_0^T\dd N_k(s_1)\int_0^T\dd N_k(s_2)\int_{\max\{0,s_2-s_1\}}^{b+\min\{0,s_2-s_1\}}g_{j,k}(t)g_{j,k}(t+s_1-s_2)\dd t+\frac{1}{T}\int_0^T\int_u^Tg_{j,k}^2(t-u)\dd t \dd N_k(u).
\end{aligned}
\end{equation*}
Since $g_{j,k}(t)$'s are bounded functions with bounded support (upper bounded by $b$), Theorem~\ref{thm2} is applicable and we have 
\begin{equation*}
\begin{aligned}
&\left|\frac{1}{T}\int_{0}^T \left\{\int_0^tg_{j,k}(t-s)\dd N_k(s)\right\}^{2}\dd t-\mathbb{E}\left[\frac{1}{T}\int_{0}^T \left\{\int_0^tg_{j,k}(t-s)\dd N_k(s)\right\}^{2}\dd t\right]\right|\\
\leq& c_1'T^{-2/5}\sup_{s}\left|\int_{\max\{0,s\}}^{b+\min\{0,s\}}g_{j,k}(t)g_{j,k}(t-s)\dd t\right|+c_1T^{-3/5}\sup_u\left|\int_u^Tg_{j,k}^2(t-u)\dd t\right|\\
\leq& c_1'T^{-2/5}\sup_s\sqrt{\int_0^bg_{j,k}^2(t)\dd t\int_{s}^{s+b} g_{j,k}^2(t-s)\dd t}+c_1T^{-3/5}\sup_u\left|\int_u^{u+b}g_{j,k}^2(t-u)\dd t\right|\\
=&\mathcal{O}(m_1^{-2d}T^{-2/5}),
\end{aligned}
\end{equation*}
with probability at least $1-c_2pT\exp(c_3T^{1/5})-c_2'pT\exp(c_3'T^{1/5})$. Next, we have that 
\begin{equation*}
\begin{aligned}
&\mathbb{E}\left[\frac{1}{T}\int_{0}^T \left\{\int_0^tg_{j,k}(t-s)\dd N_k(s)\right\}^{2}\dd t\right]\\
=&\frac{1}{T}\int_0^T\int_0^t\int_0^tg_{j,k}(t-u_1)g_{j,k}(t-u_2)\mathbb{E}\left\{\dd N_k(u_1)\dd N_k(u_2)\right\}\dd t\\
=&\frac{1}{T}\int_0^T\int_0^t\int_{0}^tg_{j,k}(t-u_1)g_{j,k}(t-u_2)\bar\lambda_{k,k}^{(2)}(u_1,u_2)\dd u_1\dd u_2\dd t+\frac{1}{T}\int_0^T\int_0^tg_{j,k}(t-u)g_{j,k}(t-u)\bar{\lambda}_k(u)\dd u\dd t\\
=&\mathcal{O}(m_1^{-2d}),
\end{aligned}
\end{equation*}
where we used Assumption \ref{ass4} and the fact that $\bar\lambda_k(u)$ is upper bounded.
Putting the above results together, we can conclude that 
\begin{equation*}
\begin{aligned}
&\frac{1}{T}\int_{0}^T \left\{\int_0^tg_{j,k}(t-s)\dd N_k(s)\right\}^{2}\dd t=\mathcal{O}(m_1^{-2d}), 
\end{aligned}
\end{equation*}
with probability at least $1-c_2pT\exp(c_3T^{1/5})-c_2'pT\exp(c_3'T^{1/5})$. Plugging this into \eqref{egn:expand}, we can get that 
$$
\frac{1}{T}\int_{0}^T \left\{\bPsi^\top(t)\widetilde\bbeta_{j}-\lambda_{j}(t)\right\}^{2}\dd t= O\left((s+1)^{2}m_1^{-2d}\right).
$$
with probability at least $1-c_2pT\exp(c_3T^{1/5})-c_2'pT\exp(c_3'T^{1/5})$.
\eop
}

{
\subsection{Proof of Lemma~\ref{lemma:dom}}
\label{sec::dom}
Recall the iterative thinning representation of $\blambda^{(n)}(t)=(\lambda^{(n)}_1(t),\ldots,\lambda^{(n)}_p(t))^\top$ and $\N^{(n)}=(N^{(n)}_{j})_{j\in[p]}$ in 
\eqref{thin}. 
We construct $\blambda^{*(n)}(t)=(\lambda^{*(n)}_1(t),\ldots,\lambda^{*(n)}_p(t))^\top$ and $\N^{*(n)}=(N^{*(n)}_{j})_{j\in[p]}$ in the same manner. 
Let $\lambda_{j}^{*(0)}(t)=0$, $j\in[p]$, and $N_{j}^{*(0)}=\varnothing$. 
For $n\ge1$, construct recursively $\blambda^{*(n)}(t)=(\lambda^{*(n)}_1(t),\ldots,\lambda^{*(n)}_p(t))^\top$ and $\N^{*(n)}=(N^{*(n)}_{j})_{j\in[p]}$ as follows:
\begin{equation}
\label{thin2}
\begin{split}
&\lambda_{j}^{*(n+1)}(t)=\nu^*+\sum_{k=1}^{p}\int_{0}^{t}|\omega_{j,k}(t-u)|\dd N_{k}^{*(n)}(u),\\
&\dd N_{j}^{*(n+1)}(t)=\overline{N}_{j}\left(\left[0,\lambda_{j}^{*(n+1)}(t)\right]\times\dd t\right),\quad j\in[p],
\end{split}
\end{equation}
where $\overline{N}_{j}$ is as defined in \eqref{thin}. 
From the Lipschitz condition of $h(\cdot)$ and the iterative construction in \eqref{thin} and \eqref{thin2}, it is seen that $\lambda_{j}^{*(n)}(t)\ge\lambda_{j}^{(n)}(t)$ for all $n\ge 1$, where $\lambda_{j}^{(n)}(t)$ is as defined in \eqref{thin}. 
See Proposition 2.1 in \citet{costa2018renewal} for a similar result.

It follows from Lemma \ref{thinexp} that $\lambda_{j}^{*(n)}(t)$ is the intensity function of the point process $N_j^{*(n)}(t)$.
Under Assumption 1, the result in \citet{bremaud1996stability} ensures that the sequence $\{\N^{*(n)}\}_{n=1}^{\infty}$ in \eqref{thin2} converges in distribution to the Hawkes process $\N^*$ with intensity function $\blambda^{*}(t)=(\lambda^{*}_1(t),\ldots,\lambda^{*}_p(t))^\top$. 
Combining this with the result in Theorem \ref{thm1}, we arrive at the desired conclusion.
\eop
}

\subsection{Proof of Lemma~\ref{couple}}
\label{sec::couple}
Let $\check{\N}$ be a $p$-variate homogeneous Poisson process with each component process defined on $\mathbb{R}^2$ with intensity $1$, and independent of $\overline{\N}$ defined in \eqref{thin}. For $n\geq 1$ and $j\in[p]$, we construct $\widetilde{\N}^{(n)}$ as 
\begin{eqnarray}\label{tildeN}
&&\widetilde{\lambda}_{j}^{(n+1)}(t)=h\left\{\nu_{j}(t)+\int_{0}^{t}\sum_{k=1}^{p}\omega_{j,k}(t-s)\dd\widetilde{N}_{k}^{(n)}(s)\right\}\\
&&\dd\widetilde{N}_{j}^{(n+1)}(t)=
\begin{cases}
 \check{N}_{j}\left([0,\widetilde{\lambda}_j^{(n+1)}(t)]\times \dd t\right)\quad t\leq z\\\nonumber
\overline{N}_{j}\left([0,\widetilde{\lambda}_j^{(n+1)}(t)]\times \dd t\right)\quad t> z
\end{cases}
\end{eqnarray}
Write $\widetilde\N'(t)= \1_{[t\leq z]}\dd\check{\N}+\1_{[t> z]}\dd\overline{\N}$. 
Note that $\widetilde\N'(t)$ is still a $p$-variate homogeneous Poisson process with each component process defined on $\mathbb{R}^2$ with intensity $1$,
then \eqref{tildeN} can be written as 
\[
\widetilde{\lambda}_{j}^{(n+1)}(t)=h\left\{\nu_{j}(t)+\int_{0}^{t}\sum_{k=1}^{p}\omega_{j,k}(t-s)\dd\widetilde{N}_{k}^{(n)}(s)\right\}
\]
\begin{equation*}
\dd\widetilde{N}_{j}^{(n)}(t)={\widetilde N'_{j}\left([0,\widetilde{\lambda}_j^{(n)}(t)]\times \dd t\right)} \quad j\in[p].
\end{equation*}
From the construction, we can see $\widetilde{\N}$ has the same distribution as $\N$.
To show that $\widetilde{\N}$ is the coupling process of $\N$, we need to show that $\widetilde{\N}$ is independent of $ \mathcal{H}_{z}$ (i.e., $\N$ before time $z$).

We show this through induction. Since $\widetilde\N'$ is independent of $\overline{\N}$ before time $z$, $\widetilde{\N}^{(1)}$ is independent of $\mathcal{H}_{z}^{(1)}$ by construction. Next, we assume that the statement holds true for $n$, that is, $\widetilde{\N}^{(n)}$ is independent of $\mathcal{H}_{z}^{(n)}$. Since $\widetilde{\blambda}^{(n+1)}$ is a function of $\widetilde{\N}^{(n)}$, as defined in \eqref{tildeN}, it is independent of $\mathcal{H}_{z}^{(n)}$. {$\widetilde{\N}^{(n+1)}$ is determined only by $\widetilde{\blambda}^{(n+1)}$ and $\widetilde\N'$, thus it is independent of $\mathcal{H}_{z}^{(n)}$. Also, $\mathcal{H}_{z}^{(n+1)}$ is determined by $\overline{\N}$ and $\mathcal{H}_{z}^{(n)}$. As $\widetilde{\N}^{(n+1)}$ is independent of $\overline{\N}$ before time $z$, it is thus independent of $\mathcal{H}_{z}^{(n+1)}$. Hence, the statement also holds for $n+1$. By induction, we have that $\widetilde{\N}$ is independent of $ \mathcal{H}_{z}$.}

Next, we move to bound the first and second order deviations between $\widetilde{\N}$ and $\N$. We again use the induction method.
First, we show that for any $u>(m-1)b$, $m=1,2,3,\ldots$,
\begin{equation}\label{first}
\mathbb{E}|\dd\widetilde{\N}(z+u)-\dd\N(z+u)|/\dd u\preceq 2\lambda_{\max}\bOmega^{m}\mathbf{1}_{p}.
\end{equation} 
For $m=1$ (i.e., $u>(1-1)b=0$), we have
\begin{align*}
\begin{split}
&\quad \mathbb{E}\left|\dd N_{j}(u+z)-\dd\widetilde{N}_{j}(u+z)\right|/\dd u\\
&={\mathbb{E}\left|\overline{N}_{j}\left([0,\lambda_j(u+z)]\times\dd u\right)-\tilde N'_{j}\left([0,\widetilde{\lambda}_j(u+z)]\times\dd u\right)\right|/\dd u}\\
&=\mathbb{E}|\lambda_{j}(u+z)-\widetilde{\lambda}_{j}(u+z)|\\
&\leq \sum_{k=1}^{p}\int_{0}^{b} |\omega_{j,k}(t)|\times\mathbb{E}\left|\dd N_{k}(u+z-t)-\dd\widetilde{N}_{k}(u+z-t)\right|\\
&\leq 2\lambda_{\max}\sum_{k=1}^{p}\int_{0}^{b}|\omega_{j,k}(t)|\dd t \leq 2 \lambda_{\max}\bOmega_{j,.}^\top \mathbf{1}_{p},
\end{split}
\end{align*}
where the second equality follows from $\dd\widetilde\N'(t)=\dd\overline{\N}(t)$ when $t\geq z$, {the first inequality uses the fact that $h$ is $\theta$-Lipschitz function with $\theta\leq 1$} and the second inequality follows from $
\mathbb{E}|\dd\N_{j}(z+u)-\dd\widetilde{\N}_{j}(z+u)|/\dd u{\leq 2\mathbb{E}|\dd\N_{j}(z+u)|/\dd u}\leq 2\lambda_{\max}$, $j\in[p]$.
Jointly for all $j$ and for $u>0$, we have that 
\[
\mathbb{E}|\dd\widetilde{\N}(z+u)-\dd\N(z+u)|/du\preceq 2\lambda_{\max}\bOmega \mathbf{1}_{p}.
\]
Thus, \eqref{first} holds for $m=1$. 
Next, we assume it holds for $m=n-1$ (i.e., $u> (n-2)b$).
Then, for $m=n$ (i.e., $u> (n-1)b)$, we have that
\begin{align*}
\begin{split}
&\quad \mathbb{E}|\dd N_{j}(u+z)-\dd\widetilde{N}_{j}(u+z)|/\dd u\\
&\leq \sum_{k=1}^{p}\int_{0}^{b} |\omega_{j,k}(t)|\times\mathbb{E}\left|\dd N_{k}(u+z-t)-\dd\widetilde{N}_{k}(u+z-t)\right|\\
&\leq 2\lambda_{\max}\sum_{k=1}^{p}\int_{0}^{b}|\omega_{j,k}(t)|\bOmega^{n-1}_{k,.}\mathbf{1}_{p}\dd t\leq 2\lambda_{\max}\bOmega_{j,.}^\top \bOmega^{n-1}\mathbf{1}_{p},
\end{split}
\end{align*}
where the second inequality is a direct result of \eqref{first} under $m=n-1$.
Hence, jointly for $j\in[p]$, we have that $\mathbb{E}|\dd\widetilde{\N}(z+u)-\dd\N(z+u)|/\dd u\preceq 2\lambda_{\max}\bOmega^{n}\mathbf{1}_{p}$.
By induction, \eqref{first} holds for any $m$ and $u>(m-1)b$. In other words, \eqref{first} holds for any $u>0$ when $m=\lfloor u/b+1\rfloor$.

Next, we show that for any $u>(m-1)b$, $m=1,2,3,\ldots,$
\begin{equation}\label{second}
\mathbb{E}\left|\dd\N(t')\dd\N^\top (u+z)-\dd\widetilde{\N}(t')d\widetilde{\N}^\top (u+z)\right|/(\dd u\dd t') \preceq 2c_{0}^{2}\sum_{i=1}^{m+1}\bOmega^{i-1}\mathbf{1}_{p\times p}(\bOmega^\top )^{m-i+1},
\end{equation}
{where $c_{0}=\max\{\lambda_{\max},\nu+\|\omega\|_{\infty} \}$ with $\nu=\max_j\|\nu_j\|_{\infty}$ and $\|\omega\|_{\infty}=\max_{j,k}\|\omega_{j,k}\|_{\infty}$.}

For $m=1$ (i.e., $u>(1-1)b=0$), we have
\begin{align*}
\begin{split}
&\quad\,\, \mathbb{E}\left|\dd N_{j}(t')\dd N_{k}(u+z)-\dd\widetilde{N}_{j}(t')\dd\widetilde{N}_{k}(u+z)\right|/(\dd u\dd t')\\
&={\mathbb{E}\left| \bar N_{j}\left([0,\lambda_{j}(t')]\times\dd t'\right)\dd N_k(u+z) -\widetilde{N}_{j}\left([0,\lambda_{j}(t')]\times\dd t'\right)\dd \widetilde{N}_{k}(u+z)\right|/(\dd u\dd t')}\\
&= \mathbb{E}|\dd N_{k}(u+z)\lambda_{j}(t')-\dd\widetilde{N}_{k}(u+z){\lambda}_{j}(t')|/\dd u\\
&\leq \nu\mathbb{E}\left|\dd N_{k}(u+z)-\dd\widetilde{N}_{k}(u+z)\right|/\dd u\\
&\quad+\mathbb{E}\left|\sum_{l=1}^{p}\int_{0}^{{t'}}{\omega_{j,l}(t'-t)[\dd N_{l}(t)\dd N_{k}(u+z)-\dd\widetilde{N}_{l}(t)\dd\widetilde{N}_{k}(u+z)]}\right|/\dd u,
\end{split}
\end{align*}
where the second equality uses the fact that $\overline{N}_{j}(t)=\tilde N'_{j}(t)$ when $t\geq z$, and the last inequality uses the triangle inequality and the Lipchitz condition on $h(\cdot)$. The first term in the upper bound above can be bounded using \eqref{first}. For the second term, we have that, for any $u$, 
\begin{align*}
\begin{split}
&\quad \mathbb{E}\left|\dd N_{j}(t)\dd N_{k}(u+z)-\dd\widetilde{N}_{j}(t)\dd\widetilde{N}_{k}(u+z)\right|/(\dd u\dd t)\\
&\leq 2\mathbb{E}\left|\dd N_{j}(t)\dd N_{k}(u+z)\right|/(\dd u\dd t)\\
&=2\mathbb{E}\left|\overline{N}_{j}\left([0,\lambda_j(t)]\times\dd t\right)\overline{N}_{k}\left([0,\lambda_{k}(u+z)]\times\dd u\right)\right|/(\dd u\dd t)\leq 2\lambda_{\max}^{2},
\end{split}
\end{align*}
where the first inequality uses the fact that $\N$ and $\widetilde{\N}$ are independent and have the same distribution.

Next, we have
\[
\begin{split}
&\quad \mathbb{E}\left|\dd N_{j}(t')\dd N_{k}(u+z)-\dd \widetilde{N}_{j}(t')\dd\widetilde{N}_{k}(u+z)\right|/(\dd u\dd t')\\
&\leq\nu\mathbb{E}\left|\dd N_{k}(u+z)-\dd\widetilde{N}_{k}(u+z)\right|/\dd u+|\omega_{j,k}(t'-u-z)|\times\mathbb{E}\left|\dd N_{k}(u+z)-\dd\widetilde{N}_{k}(u+z)\right|/\dd u\\
&+\sum_{l=1}^{p}\int_{0}^{{t'}}|{\omega_{j,l}(t'-t)|\left(1-1_{[l=k,t=u+z]}\right)\mathbb{E}\left|\dd N_{l}(t)\dd N_{k}(u+z)-\dd\widetilde{N}_{l}(t)\dd\widetilde{N}_{k}(u+z)\right|}/\dd u\\
&\leq \left(\nu+\|\omega_{j,k}\|_{\infty}\right)\mathbb{E}\left|\dd N_{k}(u+z)-\dd\widetilde{N}_{k}(u+z)\right|/\dd u\\
&+\sum_{l=1}^{p}\int_{0}^{{t'}}{|\omega_{j,l}(t'-t)|\left(1-1_{[l=k,t=u+z]}\right)\mathbb{E}\left|\dd N_{l}(t)\dd N_{k}(u+z)-\dd\widetilde{N}_{l}(t)\dd\widetilde{N}_{k}(u+z)\right|}/\dd u\\
&\leq 2\lambda_{\max}(\nu+\|\omega_{j,k}\|_{\infty})\bOmega_{k,.}^\top \mathbf{1}_{p} +2\lambda_{\max}^{2}\bOmega_{j,.}^\top \mathbf{1}_{p}\leq 2c_{0}^{2}\bOmega_{k,.}^\top \mathbf{1}_{p}+2c_{0}^{2}\bOmega_{j,.}^\top \mathbf{1}_{p},
\end{split}
\]
where $c_{0}=\max\{\lambda_{\max},\nu+\|\omega\|_{\infty} \}$.
Thus, jointly for $j,k$ and for $u>0$, we have
\begin{align*}
\begin{split}
&\quad \mathbb{E}\left|\dd\N(t')\dd\N^\top (u+z)-\dd\widetilde{\N}(t')\dd\widetilde{\N}^\top (u+z)\right|/(\dd u\dd t')\\
&\leq 2c_{0}^{2}\mathbf{1}_{p}\mathbf{1}_{p}^\top \bOmega^\top +2c_{0}^{2}\bOmega \mathbf{1}_{p}\mathbf{1}_{p}^\top=2c_{0}^{2}\sum_{i=1}^{2}\bOmega^{i-1}\mathbf{1}_{p}\mathbf{1}_{p}^\top (\bOmega^\top )^{2-i},
\end{split}
\end{align*}
and therefore, \eqref{second} holds for $m=1$.

Now we assume it also holds for $m=n-1$ (i.e., $u>(n-2)b$). Then, for $m=n$, we have 
\begin{align*}
\begin{split}
&\quad \mathbb{E}\left|\dd N_{j}(t')\dd N_{k}(u+z)-\dd\widetilde{N}_{j}(t')\dd\widetilde{N}_{k}(u+z)\right|/(\dd u\dd t')\\
&\leq (\nu+\|\omega\|_{\infty})\mathbb{E}\left|\dd N_{k}(u+z)-\dd\widetilde{N}_{k}(u+z)\right|/\dd u\\
&+\sum_{l=1}^{p}\int_{0}^{{t'}}{|\omega_{j,l}(t'-t)|\left(1-1_{[l=k,t=u+z]}\right)\mathbb{E}\left|\dd N_{l}(t)\dd N_{k}(u+z)-\dd\widetilde{N}_{l}(t)\dd\widetilde{N}_{k}(u+z)\right|}/\dd u\\
&\leq 2c_{0}(\nu+\|\omega\|_{\infty})\bOmega_{k,.}^\top \bOmega^{n-1}\mathbf{1}_{p}+2\bOmega_{j,.}^\top \left\{c_{0}^2\sum_{i=1}^{n}\bOmega^{i-1}(\mathbf{1}_{p}\mathbf{1}_{p}^\top)(\bOmega^\top)^{n-i-1}\bOmega_{k,.}\right\}\\
&\leq 2c_{0}^{2}\mathbf{1}_{p}^\top(\bOmega^\top)^{n-1}\bOmega_{k,.}+2c_{0}^{2}\sum_{i=1}^{n}\bOmega_{j,.}^\top \bOmega^{i-1}(\mathbf{1}_{p}\mathbf{1}_{p}^\top)(\bOmega^\top)^{n-i-1}\bOmega_{k,.}.
\end{split}
\end{align*}
Thus jointly for $j,k$ and for $u>(n-1)b$,
\begin{equation*}
\mathbb{E}\left|\dd\N(t')\dd\N^\top (u+z)-\dd\widetilde{\N}(t')\dd\widetilde{\N}^\top (u+z)\right|/(\dd u\dd t')\leq 2c_{0}^{2}\sum_{i=1}^{n+1}\bOmega^{i-1}(\mathbf{1}_{p}\mathbf{1}_{p}^\top)(\bOmega^\top)^{n-i+1}.
\end{equation*}
Or equivalently, for any $u$, letting $m=\lfloor u/b+1\rfloor$ gives
\begin{equation*}
\mathbb{E}\left|\dd\N(t')\dd\N^\top (u+z)-\dd\widetilde{\N}(t')\dd\widetilde{\N}^\top (u+z)\right|/(\dd u\dd t')\leq 2c_{0}^{2}\sum_{i=1}^{m+1}\bOmega^{i-1}\mathbf{1}_{p\times p}(\bOmega^\top)^{m-i+1}.
\end{equation*}
\eop

\subsection{Proof of Lemma~\ref{exptail}}
\label{sec::exptail}
From \eqref{first}, we have that
\begin{equation*}
\mathbb{E}\left|\dd\widetilde{N}_{j}(z+u)-\dd N_{j}(z+u)\right|/\dd u\leq 2\lambda_{\max}\rho_{\bOmega}^{n},
\end{equation*}
where $n=\lfloor u/b+1\rfloor$.
Let $a_{1}=2\lambda_{\max}$ and $a_{2}=-\log(\rho_{\bOmega})/b$, and we have
\begin{equation}\label{firsttail1}
\mathbb{E}\left|\dd\widetilde{N}_{j}(z+u)-\dd N_{j}(z+u)\right|/\dd u\leq a_{1}\exp(-a_{2}u).
\end{equation}
Here we use the fact that $n=\lfloor u/b+1\rfloor\geq u/b$ and $\rho_{\bOmega}^{u/b}\geq \rho_{\bOmega}^{\lfloor u/b+1\rfloor}$.
Similarly, we can get that, for any $u>0$,
\begin{equation}\label{secondtail1}
\begin{split}
&\quad \mathbb{E}\left|\dd\widetilde{N}_{j}(t')\dd\widetilde{N}_{k}(z+u)-\dd N_{j}(t')\dd N_{k}(z+u)\right|/(\dd t'\dd u)\\
&\leq 2c_{0}^{2}\left\{\sum_{i=1}^{n+1}\bOmega^{i-1}(\mathbf{1}_{p}\mathbf{1}_{p}^\top)(\bOmega^\top)^{n-i+1}\right\}_{j,k}\\
&\leq 2\sum_{i=1}^{n+1}\left\|c_{0}\bOmega^{i-1}\mathbf{1}_{p}\right\|_{\infty}\left\|c_{0}\bOmega^{n-i+1}\mathbf{1}_{p}\right\|_{\infty}\\
&\leq 2\sum_{i=1}^{n+1} c^2_{0}\exp\left\{-a_{2}(i-1)b\right\}\exp\left\{-a_{2}(n-i+1)b\right\}\\
&=2c_{0}^{2}(n+1)\exp(-a_{2}nb),
\end{split}
\end{equation}
where $n=\lfloor u/b+1\rfloor$.
Since $\log(n+1)/n=o(1)$, there exists a constant $n_{0}$ such that $\log(n+1)/n\leq a_{2}b/2$ for any $n\geq n_{0}$. We have 
\begin{equation*}
 \mathbb{E}\left|\dd\widetilde{N}_{j}(t')\dd\widetilde{N}_{k}(z+u)-\dd N_{j}(t')\dd N_{k}(z+u)\right|/(\dd t'\dd u)\leq 2c_{0}^{2}\exp(-a_{2}u/2),
\end{equation*}
for $u$ such that $\lfloor u/b+1\rfloor \geq n_{0}$.
Additionally, for $u\leq (n_{0}-1)b$, we have
\begin{align*}
\begin{split}
&\quad\mathbb{E}\left|\dd\widetilde{N}_{j}(t')\dd\widetilde{N}_{k}(z+u)-\dd N_{j}(t')\dd N_{k}(z+u)\right|/(\dd t'\dd u)\\
&\leq 2(n_{0}+1)c_{0}^{2}\exp(-a_{2}u/2)
\end{split}
\end{align*}
Considering the above results together, for any $u\geq 0$, we have
\begin{equation*}
\mathbb{E}\left|\dd\widetilde{N}_{j}(t')\dd\widetilde{N}_{k}(z+u)-\dd N_{j}(t')\dd N_{k}(z+u)\right|/(\dd t'\dd u)\leq a_{3}\exp(-a_{4}u)
\end{equation*}
where $a_{3}=2(n_{0}+1)c_{0}^{2}$ and $a_{4}=a_{2}/2$.
\eop

\subsection{Proof of Lemma~\ref{poisson2}}
\label{exp::poisson2}
Let $M=(nK_{2}/K_{1})^{1/2}$, we have that
\begin{align*}
\begin{split}
& \quad \mathbb{P}(|Z_{1}Z_{2}|>n)\\
&=\mathbb{P}(\{|Z_{2}|\geq M, |Z_{1}Z_{2}|>n \}\cup\{|Z_{2}|< M, |Z_{1}Z_{2}|>n \})\\
&=\mathbb{P}(\{|Z_{2}|\geq M, |Z_{1}Z_{2}|>n \})+\mathbb{P}(\{|Z_{2}|< M, |Z_{1}Z_{2}|>n \})\\
&\leq \mathbb{P}(|Z_{2}|\geq M)+\mathbb{P}(\{|Z_{2}|< M, M|Z_{1}|>n \})\\
&\leq \mathbb{P}(|Z_{2}|\geq M)+\mathbb{P}( M|Z_{1}|>n )\\
&\leq 2\exp(1-(n/(K_{2}K_{1}))^{1/2})\\
&\leq \exp(1-(n/K')^{1/2}),
\end{split}
\end{align*}
where the first inequality is due to $|a+b|\leq|a|+|b|$, the second inequality is due to $\{M|Z_{1}|>n\}\supset\{|Z_{2}|<M, M|Z_{1}|>n\}$, the third inequality is a direct result of Lemma~\ref{poisson}, and the last inequality follows from Lemma A.2 in \cite{fan2014nonparametric}.
\eop

\subsection{Proof of Lemma~\ref{lemma:block}}
\label{sec:block}
{
We first state some useful properties of B-spline basis that are useful in our proof. 
The first property concerns eigenvalues of the matrix formed by normalized B-spline basis.
By Lemmas 6.1-6.2 in \cite{zhou1998local}, there exist positive constants $z_1$, $z_2$ such that 
\begin{equation}\label{eqn:eigen1}
\frac{z_1}{m_1}\leq\sigma_{\min}\left(\int_0^b\bphi_1^\top(s)\bphi_1(s)\dd s\right)\leq\sigma_{\max}\left(\int_0^b\bphi_1^\top(s)\bphi_1(s)\dd s\right)\leq \frac{z_2}{m_1},
\end{equation}
where $b$ is as defined in Assumption \ref{ass2}. 
Let $\text{supp}(\cdot)$ be the support $\{t\in\mX: f(t)\neq 0\}$ for a continuous function $f$ defined on $\mX\in\mathbb{R}$.
By the local property \citep{de1972calculating} of normalized B-spline basis (e.g., the measure of $\text{supp}(\phi_{1,l})$ is $\mathcal{O}(1/m_1)$), it holds for some positive constant $z'_2$ that 
\begin{equation}\label{eqn:integral}
\int_0^b|\phi_{1,l}(s)|\dd s\leq\frac{z'_2}{m_1},\quad \int_0^b\phi_{1,l}^2(t)\dd t\leq\frac{z'_2}{m_1},\quad l\in[m_1].
\end{equation}
The proof is divided into two parts. In the first part, we show $\sigma_{\max}(\G^{(k,k)})\leq \frac{3\gamma_{\max}}{2m_1}$ by first showing $\sigma_{\max}\{\mathbb{E}(\G^{(k,k)})\}\leq \frac{\gamma_{\max}}{m_1}$ and then bounding the difference between $\sigma_{\max}(\G^{(k,k)})$ and $\sigma_{\max}\{\mathbb{E}(\G^{(k,k)})\}$. In the second part, we show $\sigma_{\min}(\G^{(k,k)})\geq \frac{\gamma_{\min}}{2m_1}$ by first showing $\sigma_{\min}\{\mathbb{E}(\G^{(k,k)})\}\geq \frac{\gamma_{\min}}{m_1}$ and then bounding the difference between $\sigma_{\min}(\G^{(k,k)})$ and $\sigma_{\min}\{\mathbb{E}(\G^{(k,k)})\}$.

\noindent
\textbf{Part I.} 
By definition, the matrix $\mathbb{E}(\G^{(k,k)})$ can be expanded as
\begin{equation}\label{decomGk}
\begin{aligned}
&\frac{1}{T}\int_0^T\int_0^t\int_0^t\bphi_1^\top(t-u_1)\bphi_1(t-u_2)\mathbb{E}\left\{\dd N_k(u_1)\dd N_k(u_2)\right\}\dd t\\
=&\underbrace{\small{\frac{1}{T}\int_0^T\int_0^t\int_{0}^t\bphi_1^\top(t-u_1)\bphi_1(t-u_2)\bar{\lambda}_{k,k}^{(2)}(u_1,u_2)\dd u_1\dd u_2\dd t}}_{\mathbb{E}(\G_1^{(k,k)})}+\underbrace{\small{\frac{1}{T}\int_0^T\int_0^t\bphi_1^\top(t-u)\bphi_1(t-u)\bar{\lambda}_k(u)\dd u\dd t}}_{\mathbb{E}(\G_2^{(k,k)})}\\
\end{aligned}
\end{equation}
where $\bar{\lambda}_{k,k}^{(2)}(u_1,u_2)$ is as defined in \eqref{eqn:2nd}. 
Given that $|\bar{\lambda}^{(2)}_{k_1,k_2}(u_1,u_2)|\le\Lambda_{\max}$ in Assumption \ref{ass4} and $\int_0^t|\phi_{1,l}(u)|\dd u\leq\frac{z'_2}{m_1}$ in \eqref{eqn:integral}, we have for $l_1,l_2\in[m_1]$
\begin{equation*}
\begin{aligned}
&\left|\frac{ 1}{T}\int_0^T\int_0^t\int_0^t\phi_{1,l_1}(t-u_1)\phi_{1,l_2}(t-u_2)\bar{\lambda}_{k,k}^{(2)}(u_1,u_2)\dd u_1\dd u_2 \dd t\right|\\
\leq&\frac{ 1}{T}\int_0^T\int_0^t\int_0^t|\phi_{1,l_1}(t-u_1)\phi_{1,l_2}(t-u_2)||\bar{\lambda}_{k,k}^{(2)}(u_1,u_2)|\dd u_1\dd u_2 \dd t\\
\leq& \frac{\Lambda_{\max}}{T}\int_0^T\int_0^t|\phi_{1,l_1}(t-u_1)|\dd u_1\int_0^t|\phi_{1,l_2}(t-u_2)|\dd u_2 \dd t\leq \frac{{z'_2}^2\Lambda_{\max}}{m_1^2}.
\end{aligned}
\end{equation*}
Following the fact that $\|\A\|_2\leq n\|\A\|_{\max}$ for a symmetric matrix $\A\in\mathbb{R}^{n\times n}$, we have
\begin{equation*}
\sigma_{\max}\{\mathbb{E}(\G_1^{(k,k)})\}\leq \frac{{z'_2}^2\Lambda_{\max}}{m_1}.
\end{equation*}
Next, by $0<\bar\lambda_k(u_1)\leq \lambda_{\max}$ in Assumptions \ref{ass2} and \eqref{eqn:eigen1}, we have that
\begin{equation}\label{eqn:maxeg}
\max_{\|\x\|_2=1}\x^\top\mathbb{E}(\G_2^{(k,k)})\x\leq \lambda_{\max}\max_{\|\x\|_2=1}\left(\frac{1}{T}\int_0^T\int_0^t\x^\top\bphi_1(t-u)\bphi_1^\top(t-u)\x\dd u \dd t\right)\leq \frac{z_2\lambda_{\max}}{m_1},
\end{equation}
which gives that $\sigma_{\max}\{\mathbb{E}(\G_2^{(k,k)})\}\leq \frac{{z_2}\lambda_{\max}}{m_1}$.
Putting the above results together, we have
\begin{equation}\label{large}
\begin{aligned}
\sigma_{\max}\left\{\mathbb{E}(\G^{(k,k)})\right\}\leq\sigma_{\max}\left\{\mathbb{E}(\G_1^{(k,k)})\right\}+\sigma_{\max}\left\{\mathbb{E}(\G_2^{(k,k)})\right\}\leq \frac{{z'_2}^2\Lambda_{\max}+z_2\lambda_{\max}}{m_1},
\end{aligned}
\end{equation}
where we used the fact that $\sigma_{\max}\left(\bm{A}+\bm{B}\right)\leq \sigma_{\max}\left(\bm{A}\right)+\sigma_{\max}\left(\bm{B}\right)$. 
Let $\gamma_{\max}={z'_2}^2\Lambda_{\max}+z_2\lambda_{\max}$, we have that 
$\sigma_{\max}\left\{\mathbb{E}(\G^{(k,k)})\right\}\le \frac{\gamma_{\max}}{m_1}$. 
Next, we move to find an upper bound of $\sigma_{\max}\left(\G^{(k,k)}-\mathbb{E}(\G^{(k,k)})\right)$. 
Define $f_1(s)=\int_s^T\Delta_k^\top\bphi_1(t-s)\bphi_1^\top(t-s)\Delta_k\dd t$ and $f_2(s_1,s_2)=\int_{\max\{0,s_2-s_1\}}^{b+\min\{0,s_2-s_1\}}\Delta_k^\top\bphi_1(s)\bphi_1^\top(s+s_1-s_2)\Delta_k\dd s$.
By Theorem~\ref{thm2} and Corollary~\ref{cor1}, for any $\Delta_k\in\mathbb{R}^{m_1}$, with probability at least $1-c_{2}T\exp(-c_{3}T^{1/5})-c_{2}'T\exp(-c_{3}'T^{1/5})$,
\begin{equation*}
\begin{aligned}
\Delta_k^\top\left\{\G^{(k,k)}-\mathbb{E}(\G^{(k,k)})\right\}\Delta_k=&\frac{1}{T}\int_0^Tf_1(s)\left\{\dd N_{k}(s)-\mathbb{E}\dd N_{k}(s)\right\}\\
+&\frac{1}{T}\int_{0}^T\int^T_{s_2\neq s_1}f_2(s_1,s_2)\left\{\dd N_{k}(s_1)\dd N_k(s_2)-\mathbb{E}(\dd N_{k}(s_1)\dd N_k(s_2))\right\}\\
\leq&\|f_{1}\|_{\infty}T^{-3/5}+\|f_{2}\|_{\infty}T^{-2/5}.
\end{aligned}
\end{equation*}
By \eqref{eqn:eigen1} and that the support of $\bphi_1$ is upper bounded by $b$, we have $|f_1(s)|\leq |\int_s^{s+b}\Delta_k^\top\bphi_1(t-s)\bphi_1^\top(t-s)\Delta_k\dd t|\leq \frac{z_2}{m_1}\|\Delta_k\|_2^2$ for any $s>0$. Hence, $\|f_{1}\|_{\infty}\leq \frac{z_2}{m_1}\|\Delta_k\|_2^2$.
Without loss of generality, assume that $s_2<s_1$ and it holds that
\begin{equation*}
\begin{aligned}
&f_2(s_1,s_2)=\int_{0}^{b+s_2-s_1}\Delta_k^\top\bphi_1(s)\bphi_1^\top(s+s_1-s_2)\Delta_k\dd s\\
\leq &\sqrt{\Delta_k^{\top}\int_{0}^{b}\bphi_1(s)\bphi_1^\top(s)\dd s\Delta_k}\sqrt{\Delta_k^{\top}\int_{0}^{b+s_2-s_1}\bphi_1(s+s_1-s_2)\bphi_1^\top(s+s_1-s_2)\dd s\Delta_k}\\
\leq & \Delta_k^{\top}\int_{0}^{b}\bphi_1(s)\bphi_1^\top(s)\dd s\Delta_k\leq\frac{z_2}{m_1}\|\Delta_k\|_2^2,
\end{aligned}
\end{equation*}
where the first inequality holds by the Cauchy-Schwarz inequality, the second inequality holds as {$\Delta_k^{\top}\int_{0}^{b_0}\bphi_1(s)\bphi_1^\top(s)\dd s\Delta_k\leq \Delta_k^{\top}\int_{0}^{b}\bphi_1(s)\bphi_1^\top(s)\dd s\Delta_k$}, for $b_0<b$, and the last inequality holds by $\eqref{eqn:eigen1}$.
Hence, $\|f_{2}\|_{\infty}\leq \frac{z_2}{m_1}\|\Delta_k\|_2^2$.
Consequently, it holds that
\begin{equation}\label{dissample}
\Delta_k^\top\left\{\G^{(k,k)}-\mathbb{E}(\G^{(k,k)})\right\}\Delta_k\leq \frac{2z_2}{m_1}T^{-2/5}\|\Delta_k\|_2^2,
\end{equation}
with probability at least $1-c_{2}T\exp(-c_{3}T^{1/5})-c_{2}'T\exp(-c_{3}'T^{1/5})$. 
For any $k\in[p]$, we have
\begin{equation*}
\sigma_{\max}\left(\G^{(k,k)}\right)\leq \sigma_{\max}\left\{\mathbb{E}(\G^{(k,k)})\right\}+\sigma_{\max}\left\{\G^{(k,k)}-\mathbb{E}(\G^{(k,k)})\right\}\leq \frac{\gamma_{\max}}{m_1}+\frac{2z_2}{m_1}T^{-2/5},
\end{equation*}
with probability at least $1-c_{2}pT\exp(-c_{3}T^{1/5})-c_{2}'pT\exp(-c_{3}'T^{1/5})$. 
When $T$ is sufficiently large, it holds that $\sigma_{\max}\left(\G^{(k,k)}\right)\leq \frac{3\gamma_{\max}}{2m_1}$ with probability at least $1-c_{2}T\exp(-c_{3}T^{1/5})-c_{2}'T\exp(-c_{3}'T^{1/5})$.

\medskip
\noindent
\textbf{Part II.}
By definition, the matrix $\mathbb{E}(\G^{(k,k)})$ can be further expanded as
\begin{equation}\label{decomGk1}
\begin{aligned}
&\frac{1}{T}\int_0^T\int_0^t\int_0^t\bphi_1^\top(t-u_1)\bphi_1(t-u_2)\mathbb{E}\left\{\dd N_k(u_1)\dd N_k(u_2)\right\}\dd t\\
=&\underbrace{\small{\frac{1}{T}\int_0^T\int_0^t\int_{0}^t\bphi_1^\top(t-u_1)\bphi_1(t-u_2)C_{k,k}^0(u_1,u_2)\dd u_1\dd u_2\dd t}}_{\mathbb{E}(\G_1^{(k,k)})}+\underbrace{\small{\frac{1}{T}\int_0^T\int_0^t\bphi_1^\top(t-u)\bphi_1(t-u)\bar{\lambda}_k(u)\dd u\dd t}}_{\mathbb{E}(\G_2^{(k,k)})}\\
&+\underbrace{\frac{1}{T}\int_0^T\left\{\int_0^t\bphi_1^\top(t-u_1)\bar{\lambda}_k(u_1)\dd u_1\right\}\left\{\int_0^t\bphi_1(t-u_2)\bar{\lambda}_k(u_2)\dd u_2\right\}\dd t}_{\mathbb{E}(\G_3^{(k,k)})}.\\
\end{aligned}
\end{equation}
where $C_{k,k}^0(u_1,u_2)$ is as defined in \eqref{eqn:c}. 
Since $\C^0(u_1,u_2)$ is valid as assumed in Assumption \ref{ass4}, it holds that $\sigma_{\min}\left(\mathbb{E}(\G_1^{(k,k)})\right)\geq 0$. It is straightforward to see that $\sigma_{\min}\left(\mathbb{E}(\G_3^{(k,k)})\right)\geq 0$.
For $\mathbb{E}(\G_2^{(k,k)})$, we have
\begin{equation}\label{mineigenk}
\begin{aligned}
\sigma_{\min}\left(\mathbb{E}(\G_2^{(k,k)})\right)
\geq&\sigma_{\min}\left(\frac{1}{T}\int_0^b\int_0^t\bphi_1^\top(t-u)\bphi_1(t-u)\bar{\lambda}_k(u)\dd u\dd t\right)\\
&+\sigma_{\min}\left(\frac{1}{T}\int_b^T\int_0^t\bphi_1^\top(t-u)\bphi_1(t-u)\bar{\lambda}_k(u)\dd u\dd t\right),
\end{aligned}
\end{equation}
which holds due to $\sigma_{\min}\left(\bm{A}+\bm{B}\right)\geq \sigma_{\min}\left(\bm{A}\right)+\sigma_{\min}\left(\bm{B}\right)$. By $\bar{\lambda}_k(u)\geq\Lambda_{\min}>0$ as assumed in Assumption \ref{ass4} and \eqref{eqn:eigen1}, a similar argument as in \eqref{eqn:maxeg} leads to
$$
\sigma_{\min}\left(\frac{1}{T}\int_0^b\int_0^t\bphi_1^\top(t-u)\bphi_1(t-u)\bar{\lambda}_k(u)\dd u\dd t\right)\geq 0
$$ 
and 
$$
\sigma_{\min}\left(\frac{1}{T}\int_b^T\int_0^t\bphi_1^\top(t-u)\bphi_1(t-u)\bar{\lambda}_k(u)\dd u\dd t\right)\geq\frac{z_1\Lambda_{\min}}{2m_1}
$$
when $T$ is sufficiently large. Combining $\mathbb{E}(\G_1^{(k,k)})$, $\mathbb{E}(\G_2^{(k,k)})$ and $\mathbb{E}(\G_3^{(k,k)})$, we arrive at the desired conclusion that $\sigma_{\min}\left(\mathbb{E}(\G^{(k,k)}) \right)\geq \frac{\gamma_{\min}}{m_1}$ with $\gamma_{\min}=\frac{z_1\Lambda_{\min}}{2}$.
Lastly, by \eqref{dissample}, we have, for $k\in[p]$, 
\begin{equation*}
\sigma_{\min}\left(\G^{(k,k)}\right)\geq \sigma_{\min}\left\{\mathbb{E}(\G^{(k,k)})\right\}-\sigma_{\max}\left\{\G^{(k,k)}-\mathbb{E}(\G^{(k,k)})\right\}\geq \frac{\gamma_{\min}}{m_1}-\frac{2z_2}{m_1}T^{-2/5},
\end{equation*}
with probability at least {$1-c_{2}T\exp(-c_{3}T^{1/5})-c_{2}'T\exp(-c_{3}'T^{1/5})$}. When $T$ is sufficiently large, it holds that $\sigma_{\min}\left(\G^{(k,k)}\right)\geq\frac{\gamma_{\min}}{2m_1}$ with probability at least {$1-c_{2}T\exp(-c_{3}T^{1/5})-c_{2}'T\exp(-c_{3}'T^{1/5})$}.	
\eop
}

\subsection{Proof of Lemma~\ref{lemma:re}}
\label{sec:re}
{
The proof is divided into two steps. In the first step, we show the restricted eigenvalue condition for the population matrix $\mathbb{E}(\G)$ and in the second step, we bound the difference between the sample matrix $\G$ and the population matrix $\mathbb{E}(\G)$.

Similar as \eqref{decomGk1}, we expand $\mathbb{E}(\G^{(k_1,k_2)})$ for $k_1\neq k_2\in [p]$ as
\begin{equation}\label{decomGkk}
\begin{aligned}
\mathbb{E}(\G^{(k_1,k_2)})=&\frac{1}{T}\int_0^T\int_0^t\int_{0}^t\bphi_1^\top(t-u_1)\bphi_1(t-u_2)C_{k_1,k_2}^0(u_1,u_2)\dd u_1\dd u_2\dd t\\
&+\frac{1}{T}\int_0^T\left\{\int_0^t\bphi_1^\top(t-u_1)\bar{\lambda}_{k_1}(u_1)\dd u_1\right\}\left\{\int_0^t\bphi_1(t-u_2)\bar{\lambda}_{k_2}(u_2)\dd u_2\right\}\dd t.
\end{aligned}
\end{equation}	
Moreover, we have $\mathbb{E}(\G^{(0,0)})=\frac{1}{T}\int_0^T\bphi_0(t)\bphi_0^\top(t)\dd t$ and $\mathbb{E}(\G^{(0,k)})=\frac{1}{T}\int_0^T\bphi_0(t)\int_0^t\bphi_1^\top(t-u)\bar{\lambda}_{k}(u)\dd u\dd t$ for $k\in[p]$. 
For any $\Delta=(\Delta_0,\Delta_1,\ldots,\Delta_p)\in\mathbb{R}^{m_0+pm_1}$, we can write
\begin{equation*}
\begin{aligned}
&\Delta^\top\mathbb{E}(\G)\Delta=\Delta_0^\top\mathbb{E}(\G^{(0,0)})\Delta_0+2\sum_{k=1}^p\Delta_0^\top\mathbb{E}(\G^{0,k})\Delta_k+\sum_{k_1=1}^p\sum_{k_2=1}^p\Delta_{k_1}^\top\mathbb{E}(\G^{(k_1,k_2)})\Delta_{k_2}\\
=&\underbrace{\Delta_0^\top\frac{1}{T}\int_0^T\bphi_0(t)\bphi_0^\top(t)\dd t\Delta_0}_{B_1}+\underbrace{2\sum_{k=1}^p\Delta_0^\top\frac{1}{T}\int_0^T\bphi_0(t)\int_0^t\bphi_1^\top(t-u)\bar{\lambda}_k(u)\dd u\dd t\Delta_k}_{B_2}\\
&+\underbrace{\sum_{k_1=1}^p\sum_{k_2=1}^p \Delta_{k_1}^\top\frac{1}{T}\int_0^T\left\{\int_0^t\bphi_1(t-u_1)\bar{\lambda}_{k_1}(u_1)\dd u_1\right\}\left\{\int_0^t\bphi_1^\top(t-u_2)\bar{\lambda}_{k_2}(u_2)\dd u_2\right\}\dd t\Delta_{k_2}}_{B_3}\\
&+\underbrace{\sum_{k_1=1}^p\sum_{k_2=1}^p \Delta_{k_1}^\top\frac{1}{T}\int_0^T\int_0^t\int_{0}^t\bphi_1(t-u_1)\bphi_1^\top(t-u_2)C_{k_1,k_2}^0(u_1,u_2)\dd u_1\dd u_2\dd t\Delta_{k_2}}_{B_4}\\
&+\underbrace{\sum_{k=1}^p\Delta_k^\top\frac{1}{T}\int_0^T\int_0^t\bphi_1(t-u)\bphi_1^\top(t-u)\bar{\lambda}_k(u)\dd u\dd t\Delta_k}_{B_5}.
\end{aligned}
\end{equation*}
Similar as in \eqref{eqn:eigen1}, it holds that $B_1\geq \frac{z_3}{m_1}\|\Delta_0\|_2^2$ for some $z_3>0$. 
Define $C_0=\frac{4z_2^2(1+c)^2\lambda_{\max}^2}{z_1\Lambda_{\min}}$. For $B_2$, it holds that
\begin{equation*}
\begin{aligned}
B_2&\leq \frac{C_0s}{C_0s+1}\cdot\Delta_0^\top\frac{1}{T}\int_0^T\bphi_0(t)\bphi_0^\top(t)\dd t\Delta_0+\frac{C_0s+1}{C_0s}\cdot\frac{1}{T}\int_0^T\left\{\sum_{k=1}^p\Delta_k^\top\int_0^t\bphi_1(t-u)\bar{\lambda}_k(u)\dd u\right\}^2\dd t\\
&=\frac{C_0s}{C_0s+1}B_1+\frac{C_0s+1}{C_0s}B_3,
\end{aligned}
\end{equation*}
where we used the fact that $2xy\leq ax^2+a^{-1}y^2$ for $a>0$. By $\bar{\lambda}_k(u)\leq \lambda_{\max}$ as assumed in Assumption \ref{ass2}, \eqref{eqn:integral} and Cauchy-Schwarz inequality, it holds that
\begin{equation*}
\begin{aligned}
&\Delta_{k_1}^\top\frac{1}{T}\int_0^T\left\{\int_0^t\bphi_1^\top(t-u_1)\bar{\lambda}_{k_1}(u_1)\dd u_1\right\}\left\{\int_0^t\bphi_1(t-u_2)\bar{\lambda}_{k_2}(u_2)\dd u_2\right\}\dd t\Delta_{k_2}\\
\leq& \lambda_{\max}^2\sqrt{\frac{1}{T}\int_0^T\left\{\Delta_{k_1}^\top\int_0^t\bphi_1^\top(t-u_1)\dd u_1\right\}^2\dd t}\sqrt{\frac{1}{T}\int_0^T\left\{\Delta_{k_2}^\top\int_0^t\bphi_1^\top(t-u_2)\dd u_2\right\}^2\dd t}\\
\leq &\frac{{z'_2}^2\lambda_{\max}^2}{m_1}\|\Delta_{k_1}\|_2\|\Delta_{k_2}\|_2,
\end{aligned}
\end{equation*}
where we used the fact that $\left\{\Delta_{k_1}^\top\int_0^t\bphi_1^\top(t-u_1)\dd u_1\right\}^2\leq\|\int_0^t\bphi_1(t-u_1)\dd u_1\|_2^2\|\Delta_{k_1}\|_2^2\leq\frac{{z'_2}^2}{m_1}\|\Delta_{k_1}\|_2^2$. Correspondingly, $B_3$ can be bounded as
\begin{equation}\label{term3bound}
\begin{aligned}
B_3\leq \frac{{z'_2}^2\lambda_{\max}^2}{m_1}\left(\sum_{k=1}^p\|\Delta_{k}\|_2\right)^2\leq \frac{{z'_2}^2\lambda_{\max}^2(1+c)^2}{m_1}\left(\sum_{k\in \mathcal{E}_j}\|\Delta_{k}\|_2\right)^2\leq \frac{{z'_2}^2\lambda_{\max}^2(1+c)^2s}{m_1}\sum_{k\in \mathcal{E}_j}\|\Delta_{k}\|_2^2,
\end{aligned}
\end{equation}
where the second inequality uses the fact that $\sum_{k\in \mathcal{E}_j^c}\|\Delta_{k}\|_2\leq c\sum_{k\in \mathcal{E}_j}\|\Delta_{k}\|_2$ and the last inequality uses the Cauchy-Schwarz inequality. Considering term $B_4$, since $\C^0(u_1,u_2)$ is valid, it holds that
\begin{equation*}
B_4=\frac{1}{T}\int_0^T\int_0^t\int_{0}^t\f^\top(u_1)\C^0(u_1,u_2)\f(u_2)\dd u_1\dd u_2\dd t\geq 0
\end{equation*}
where $\f(u)=(\Delta_{1}^\top\bphi_1(t-u),\ldots,\Delta_{p}^\top\bphi_1(t-u))$. By \eqref{mineigenk}, we have $B_5\geq \frac{z_1\Lambda_{\min}}{2m_1}\sum_{k=1}^p\|\Delta_k\|_2^2$. Plugging the above results on $B_1$ to $B_5$ into $\Delta^\top\mathbb{E}(\G)\Delta$, we have
\begin{equation*}
\begin{aligned}
\Delta^\top\mathbb{E}(\G)\Delta&\geq B_1-\frac{C_0s}{C_0s+1}B_1-\frac{C_0s+1}{C_0s}B_3+B_3+B_4+B_5\\
&\geq \frac{z_3}{m_1(C_0s+1)}\|\Delta_0\|_2^2-\frac{{z'_2}^2\lambda_{\max}^2(1+c)^2s}{C_0sm_1}\sum_{k\in \mathcal{E}_j}\|\Delta_{k}\|_2^2+\frac{z_1\Lambda_{\min}}{2m_1}\sum_{k=1}^p\|\Delta_k\|_2^2.
\end{aligned}
\end{equation*}
From the definition of $C_0$, we have $\frac{z_2^2\lambda_{\max}^2(1+c)^2s}{C_0sm_1}=\frac{z_1\Lambda_{\min}}{4}$. Since it was defined that $\gamma_{\min}=\frac{z_1\Lambda_{\min}}{2}$, we have 
$$
\Delta^\top\mathbb{E}(\G)\Delta\geq  \frac{z_3}{m_1(C_0s+1)}\|\Delta_k\|_0^2+ \frac{\gamma_{\min}}{2m_1}\sum_{k=1}^p\|\Delta_k\|_2^2\geq \frac{\gamma_{\min}}{2m_1}\sum_{k=1}^p\|\Delta_k\|_2^2.
$$

Next, we show that $\Delta^\top\G\Delta\geq \frac{\gamma_{\min}}{4m_1}\sum_{k=1}^p\|\Delta_k\|_2^2$.
To do this, we move to find the upper bound of $\Delta^\top\{\G-\mathbb{E}(\G)\}\Delta$. 
Noting $\mathbb{E}(\G^{(0,0)})=\G^{(0,0)}$ ($\G^{(0,0)}$ does not involve $\N$), it holds that
\begin{equation*}
\begin{aligned}
\Delta^\top\{\G-\mathbb{E}(\G)\}\Delta
=\sum_{k_1=1}^p\sum_{k_2=1}^p\Delta_{k_1}^\top\{\G^{(k_1,k_2)}-\mathbb{E}(\G^{(k_1,k_2)})\}\Delta_{k_2}+2\sum_{k=1}^p\Delta_0^\top\{\G^{(0,k)}-\mathbb{E}(\G^{(0,k)})\}\Delta_k.\\
\end{aligned}
\end{equation*}
Consider the term $\Delta_{k_1}^\top\{\G^{(k_1,k_2)}-\mathbb{E}(\G^{(k_1,k_2)})\}\Delta_{k_2}$. When $k_1= k_2\in[p]$, it can be bounded using \eqref{dissample}. Define $f_2(s_1,s_2)=\int_{\max\{0,s_2-s_1\}}^{b+\min\{0,s_2-s_1\}}\Delta_{k_1}\bphi_1(s)\bphi_1^\top(s+s_1-s_2)\Delta_{k_2}\dd s$. When $k_1\neq k_2\in[p]$, applying the result in Theorem \ref{thm2} gives 
\begin{equation*}
\begin{aligned}
&\Delta_{k_1}^\top\left\{\G^{(k_1,k_2)}-\mathbb{E}(\G^{(k_1,k_2)})\right\}\Delta_{k_2}\\
=&\frac{1}{T}\int_0^T\int_0^Tf_2(s_1,s_2)\left\{\dd N_{k_1}(s_1)\dd N_{k_2}(s_2)-\mathbb{E}(\dd N_{k_1}(s_1)\dd N_{k_2}(s_2))\right\}
\leq \|f_{2}\|_{\infty}T^{-2/5},
\end{aligned}
\end{equation*}
with probability at least {$1-c_{2}'T\exp(-c_{3}'T^{1/5})$}. 
Without loss of generality, assume that $s_2<s_1$ and it holds that
\begin{equation*}
\begin{aligned}
&f_2(s_1,s_2)=\int_{0}^{b+s_2-s_1}\Delta_{k_1}^\top\bphi_1(s)\bphi_1^\top(s+s_1-s_2)\Delta_{k_2}\dd s\\
\leq &\sqrt{\Delta_{k_1}^{\top}\int_{0}^{b}\bphi_1(s)\bphi_1^\top(s)\dd s\Delta_{k_1}}\sqrt{\Delta_{k_2}^{\top}\int_{0}^{b+s_2-s_1}\bphi_1(s+s_1-s_2)\bphi_1^\top(s+s_1-s_2)\dd s\Delta_{k_2}}\\
\leq & \sqrt{\Delta_{k_1}^{\top}\int_{0}^{b}\bphi_1(s)\bphi_1^\top(s)\dd s\Delta_{k_1}}\sqrt{\Delta_{k_2}^{\top}\int_{0}^{b}\bphi_1(s)\bphi_1^\top(s)\dd s\Delta_{k_2}}\leq\frac{z_2}{m_1}\|\Delta_{k_1}\|_2\|\Delta_{k_2}\|_2,
\end{aligned}
\end{equation*}
where the first inequality holds by the Cauchy-Schwarz inequality, the second inequality holds as {$\Delta_k^{\top}\int_{0}^{b_0}\bphi_1(s)\bphi_1^\top(s)\dd s\Delta_k=\Delta_k^{\top}\int_{0}^{b}\bphi_1(s)\bphi_1^\top(s)\dd s\Delta_k$}, for $b_0<b$, and the last inequality holds by $\eqref{eqn:eigen1}$. 
Hence, $\|f_{2}\|_{\infty}\leq \frac{z_2}{m_1}\|\Delta_{k_1}\|_2\|\Delta_{k_2}\|_2$.
It then arrives that
\begin{equation}\label{dissample1}
\Delta_{k_1}^\top\left\{\G^{(k_1,k_2)}-\mathbb{E}(\G^{(k_1,k_2)})\right\}\Delta_{k_2}\leq \frac{z_2}{m_1}T^{-2/5}\|\Delta_{k_1}\|_2\|\Delta_{k_2}\|_2,
\end{equation}
with probability at least {$1-c_{2}'T\exp(-c_{3}'T^{1/5})$}.
By \eqref{dissample} and \eqref{dissample1}, it holds that
\begin{equation*}
\begin{aligned}
&\sum_{k_1=1}^p\sum_{k_2=1}^p\Delta_{k_1}^\top\{\G^{(k_1,k_2)}-\mathbb{E}(\G^{(k_1,k_2)})\}\Delta_{k_2}\leq\frac{2z_2}{m_1}T^{-2/5}\sum_{k_1=1}^p\sum_{k_2=1}^p\|\Delta_{k_1}\|_2\|\Delta_{k_2}\|_2\\
=&\frac{2z_2}{m_1}T^{-2/5}\left\{\sum_{k_1\in\mathcal{E}_j}\sum_{k_2\in\mathcal{E}_j}\|\Delta_{k_1}\|_2\|\Delta_{k_2}\|_2+\sum_{k_1\in\mathcal{E}_j^c}\sum_{k_2\in\mathcal{E}_j^c}\|\Delta_{k_1}\|_2\|\Delta_{k_2}\|_2+2\sum_{k_1\in\mathcal{E}_j}\sum_{k_2\in\mathcal{E}_j^c}\|\Delta_{k_1}\|_2\|\Delta_{k_2}\|_2\right\}\\
\leq & \frac{2z_2}{m_1}T^{-2/5} \left\{\left(\sum_{k\in\mathcal{E}_j}\|\Delta_k\|_2\right)^2+c^2\left(\sum_{k\in\mathcal{E}_j}\|\Delta_k\|_2\right)^2+2c\left(\sum_{k\in\mathcal{E}_j}\|\Delta_{k}\|_2\right)^2\right\}\\
\leq &\frac{2z_2(1+c)^2s}{m_1T^{2/5}}\sum_{k\in\mathcal{E}_j}\|\Delta_k\|_2^2,
\end{aligned}
\end{equation*}
with probability at least {$1-c_2pT\exp(-c_3T^{1/5})-c_{2}'p^{2}T\exp(-c_{3}'T^{1/5})$}, where the second inequality holds from $\sum_{k\in\mathcal{E}_j^c}\|\Delta_k\|_2\leq c\sum_{k\in \mathcal{E}_{j}}\|\Delta_k\|_2$ and the last inequality uses $(\sum_{k\in\mathcal{E}_j}\|\Delta_{k}\|_2)^2\leq s\sum_{k\in\mathcal{E}_j}\|\Delta_k\|_2^2$. 

Next, we use a similar argument to bound the term $\sum_{k=1}^p\Delta_0^\top(\G^{(0,k)}-\mathbb{E}(\G^{(0,k)}))\Delta_k$. By Theorem \ref{thm2}, for any $k\in[p]$, it holds with probability at least {$1-c_{2}T\exp(-c_{3}T^{1/5})$} that
\begin{equation*}
\begin{aligned}
&\Delta_0^\top\left\{\G^{(0,k)}-\mathbb{E}(\G^{(0,k)})\right\}\Delta_k
=&\frac{1}{T}\int_0^Tf_1(s)\left\{\dd N_{k}(s)-\mathbb{E}\dd N_{k}(s)\right\}
\leq \|f_{1}\|_{\infty}T^{-3/5},
\end{aligned}
\end{equation*}
where $f_1(s)=\int_s^T\Delta_0^\top\bphi_0(t)\bphi_1^\top(t-s)\Delta_k\dd t$. 
For any $s>0$, by Cauchy-Schwarz inequality and that the support of $\bphi_1$ is upper bounded by $b$, we have
\begin{equation*}
\begin{aligned}
f_{1}(s)=&\int_{s}^{s+b}\Delta_0^\top\bphi_0(t)\bphi_1^\top(t-s)\Delta_k\dd t\\
\leq &\sqrt{\Delta_0^{\top}\int_{s}^{b+s}\bphi_0(t)\bphi_0^\top(t)\dd t\Delta_0}\sqrt{\Delta_k^{\top}\int_{s}^{b+s}\bphi_1(t-s)\bphi_1^\top(t-s)\dd t\Delta_k}\leq\sqrt{\frac{z_2z_4}{m_1}}\|\Delta_0\|_2\|\Delta_k\|_2,
\end{aligned}
\end{equation*}
where the last inequality uses \eqref{eqn:eigen1} and $\Delta_0^{\top}\int_{s}^{b+s}\bphi_0(t)\bphi_0^\top(t)\dd t\Delta_0\leq z_4\|\Delta_0\|_2^2$ for some positive constant $z_4$. Such a result can be shown following the same argument used in Lemma 6.1 of \cite{zhou1998local}. 
Then we have that 
\begin{equation}\label{dissample2}
\Delta_0^\top\left\{\G^{(0,k)}-\mathbb{E}(\G^{(0,k)})\right\}\Delta_k\leq \sqrt{\frac{z_2z_4}{m_1}}T^{-3/5}\|\Delta_0\|_2\|\Delta_k\|_2,
\end{equation}
with probability at least {$1-c_{2}T\exp(-c_{3}T^{1/5})$}. By $\sum_{k\in\mathcal{E}_j^c}\|\Delta_k\|_2\leq c\sum_{k\in \mathcal{E}_{j}}\|\Delta_k\|_2$, it holds with probability at least $1-c_{2}pT\exp(-c_{3}T^{1/5})$ that
\begin{equation*}
\begin{aligned}
2\sum_{k=1}^p\Delta_0^\top(\G^{(0,k)}-\mathbb{E}(\G^{(0,k)}))\Delta_k&\leq 2\sqrt{\frac{z_2z_4(1+c)^2}{m_1T}}T^{-3/5}\sum_{k\in\mathcal{E}_j}\|\Delta_0\|_2\|\Delta_k\|_2\\
&\leq \frac{z_3}{2m_1(C_0s+1)}\|\Delta_0\|_2^2+\frac{2z_2z_4(1+c)^2(C_0s+1)}{z_3T^{6/5}}(\sum_{k\in\mathcal{E}_j}\|\Delta_k\|_2)^2,
\end{aligned}
\end{equation*}
where we used the fact that $2xy\leq ax^2+a^{-1}y^2$. 
Finally, we have
\begin{equation*}
\begin{aligned}
&\Delta^\top\G\Delta\geq\Delta^\top\mathbb{E}(\G)\Delta-\Delta^\top\{\G-\mathbb{E}(\G)\}\Delta\\
\geq& \frac{z_3}{m_1(C_0s+1)}\|\Delta_0\|_2^2+\frac{\gamma_{\min}}{2m_1}\sum_{k}^p\|\Delta_k\|_2^2-\frac{z_3}{2m_1(C_0s+1)}\|\Delta_0\|_2^2\\
&-\frac{2z_2z_4(1+c)^2(C_0s+1)s}{z_3T^{6/5}}\sum_{k\in\mathcal{E}_j}\|\Delta_k\|_2^2-\frac{2z_2(1+c)^2s}{m_1T^{2/5}}\sum_{k\in\mathcal{E}_j}\|\Delta_k\|_2^2,
\end{aligned}
\end{equation*}
with probability at least {$1-2c_{2}pT\exp(-c_{3}T^{1/5})-c_{2}'p^2T\exp(-c_{3}'T^{1/5})$}. Since $sT^{-2/5}=o(1)$ and $sm_1=\mathcal{O}(T^{4/5})$, the term $\frac{2z_2(1+c)^2s}{m_1T^{2/5}}+\frac{2z_2z_4(1+c)^2(C_0s+1)s}{z_3T^{6/5}}\leq \frac{\gamma_{\min}}{4m_1}$ when $T$ is sufficiently large. As such, it holds that  
\begin{equation}\label{eqn:b0}
\Delta^\top\G\Delta\geq \frac{z_3}{2m_1(C_0s+1)}\|\Delta_0\|_2^2+\frac{\gamma_{\min}}{4m_1}\sum_{k\in\mathcal{E}_j}\|\Delta_k\|_2^2\geq\frac{\gamma_{\min}}{4m_1}\sum_{k\in\mathcal{E}_j}\|\Delta_k\|_2^2,
\end{equation}
with probability at least {$1-2c_{2}pT\exp(-c_{3}T^{1/5})-c_{2}'p^2T\exp(-c_{3}'T^{1/5})$}. 
\eop
}

\subsection{Proof of Lemma~\ref{lemma:re2}}
\label{sec:re2}
{
Similar as in the proof of Lemma~\ref{lemma:re}, we may write
\begin{equation*}
\Delta^\top\mathbb{E}(\G)\Delta=B_1+B_2+B_3+B_4+B_5.
\end{equation*}
For $B_2$, it can be bounded as
\begin{equation*}
\begin{aligned}
B_2&\leq \frac{T^{1/2}-1}{T^{1/2}}\Delta_0^\top\frac{1}{T}\int_0^T\bphi_0(t)\bphi_0^\top(t)\dd t\Delta_0+\frac{T^{1/2}}{T^{1/2}-1}\frac{1}{T}\int_0^T\left\{\sum_{k=1}^p\Delta_k^\top\int_0^t\bphi_1(t-u)\bar{\lambda}_k(u)\dd u\right\}^2\dd t\\
&=\frac{T^{1/2}-1}{T^{1/2}}B_1+\frac{T^{1/2}}{T^{1/2}-1}B_3,
\end{aligned}
\end{equation*}
where we used the fact that $2xy\leq ax^2+a^{-1}y^2$ for any $a>0$. Similar as in \eqref{term3bound}, $B_3$ can be bounded as
\begin{equation*}
\begin{aligned}
B_3\leq \frac{{z'_2}^2\lambda_{\max}^2}{m_1}\left(\sum_{k=1}^p\|\Delta_{k}\|_2\right)^2\leq \frac{{z'_2}^2\lambda_{\max}^2(1+c)^2}{m_1}\left(\sum_{k\in \mS}\|\Delta_{k}\|_2\right)^2\leq \frac{{z_2'}^2\lambda_{\max}^2(1+c)^2|\mS|}{m_1}\sum_{k\in \mS}\|\Delta_{k}\|_2^2.
\end{aligned}
\end{equation*}
Plugging the above terms into $\Delta^\top\mathbb{E}(\G)\Delta$ and using a similar argument as in the proof of Lemma~\ref{lemma:re}, we have
\begin{equation*}
\begin{aligned}
\Delta^\top\mathbb{E}(\G)\Delta&\geq B_1-\frac{T^{1/2}-1}{T^{1/2}}B_1-\frac{T^{1/2}}{T^{1/2}-1}B_3+B_3+B_4+B_5\\
&\geq \frac{z_3}{m_1T^{1/2}}\|\Delta_0\|_2^2-\frac{{z_2'}^2\lambda_{\max}^2(1+c)^2|\mS|}{m_1(T^{1/2}-1)}\sum_{k\in \mathcal{E}_j}\|\Delta_{k}\|_2^2+\frac{z_1\Lambda_{\min}}{2m_1}\sum_{k=1}^p\|\Delta\|_2^2.
\end{aligned}
\end{equation*}
Due to $|\mS|=o(T^{1/2})$, $\frac{z_1\Lambda_{\min}}{2}-\frac{{z'_2}^2\lambda_{\max}^2(1+c)^2|\mS|}{T^{1/2}}\geq\frac{z_1\Lambda_{\min}}{4}\geq \frac{\gamma_{\min}}{2}$ when T is sufficiently large, as it was defined that $\gamma_{\min}=\frac{z_1\Lambda_{\min}}{2}$. This implies that
$$
\Delta^\top\mathbb{E}(\G)\Delta\geq \frac{z_3}{m_1T^{1/2}}\|\Delta_0\|_2^2+\frac{\gamma_{\min}}{2m_1}\sum_{k=1}^p\|\Delta_k\|_2^2.
$$

Next, we expand $\Delta^\top(\G-\mathbb{E}(\G))\Delta$ as
\begin{equation*}
\begin{aligned}
\Delta^\top(\G-\mathbb{E}(\G))\Delta
=\sum_{k_1=1}^p\sum_{k_2=1}^p\Delta_{k_1}^\top\{\G^{(k_1,k_2)}-\mathbb{E}(\G^{(k_1,k_2)})\}\Delta_{k_2}+2\sum_{k=1}^p\Delta_0^\top\{\G^{(0,k)}-\mathbb{E}(\G^{(0,k)})\}\Delta_k.\\
\end{aligned}
\end{equation*}
By \eqref{dissample}, \eqref{dissample1} and \eqref{conineq2} with $\epsilon_1=T^{-1/2}$, we have
\begin{equation*}
\begin{aligned}
&\sum_{k_1=1}^p\sum_{k_2=1}^p\Delta_{k_1}^\top\{\G^{(k_1,k_2)}-\mathbb{E}(\G^{(k_1,k_2)})\}\Delta_{k_2}\leq\frac{2z_2}{m_1}T^{-1/2}\sum_{k_1=1}^p\sum_{k_2=1}^p\|\Delta_{k_1}\|_2\|\Delta_{k_2}\|_2\\
=&\frac{2z_2}{m_1}T^{-1/2}\left\{\sum_{k_1\in\mS}\sum_{k_2\in\mS}\|\Delta_{k_1}\|_2\|\Delta_{k_2}\|_2+\sum_{k_1\in\mS^c}\sum_{k_2\in\mS^c}\|\Delta_{k_1}\|_2\|\Delta_{k_2}\|_2+2\sum_{k_1\in\mS}\sum_{k_2\in\mS^c}\|\Delta_{k_1}\|_2\|\Delta_{k_2}\|_2\right\}\\
\leq & \frac{2z_2}{m_1}T^{-1/2} \left\{(\sum_{k\in\mS}\|\Delta_k\|_2)^2+c^2(\sum_{k\in\mS}\|\Delta_k\|_2)^2+2c(\sum_{k\in\mS}\|\Delta_{k}\|_2)^2\right\}\leq \frac{2z_2(1+c)^2|\mS|}{m_1T^{1/2}}\sum_{k\in\mS}\|\Delta_k\|_2^2,
\end{aligned}
\end{equation*}
with probability at least {$1-c_{2}pT\exp(-c_{3}T^{1/6})-c_{2}'p^2T\exp(-c_{3}'T^{1/6})$}. The second inequality is from $\sum_{k\in\mS^c}\|\Delta_k\|_2\leq c\sum_{k\in \mS}\|\Delta_k\|_2$ and the last inequality uses that $(\sum_{k\in\mS}\|\Delta_{k}\|_2)^2\leq |\mS|\sum_{k\in\mS}\|\Delta_k\|_2^2$. 
By \eqref{conineq11} with $\epsilon_1=T^{-2/3}$, with probability at least {$1-c_{2}T\exp(-c_{3}T^{1/6})$},
\begin{equation*}
\begin{aligned}
&\Delta_0^\top\left\{\G^{(0,k)}-\mathbb{E}(\G^{(0,k)}\right\}\Delta_k
=&\frac{1}{T}\int_0^Tf_1(s)\left\{\dd N_{k}(s)-\mathbb{E}\dd N_{k}(s)\right\}
\leq \|f_{1}\|_{\infty}T^{-2/3},
\end{aligned}
\end{equation*}
where $f_1(s)=\int_s^T\Delta_0^\top\bphi_0(t)\bphi_1^\top(t-s)\Delta_k\dd t$. Similar as \eqref{dissample2}, we have that 
\begin{equation}\label{dissample21}
\Delta_0^\top\left\{\G^{(0,k)}-\mathbb{E}(\G^{(0,k)})\right\}\Delta_k\leq \sqrt{\frac{z_2z_4}{m_1T^{4/3}}}\|\Delta_0\|_2\|\Delta_k\|_2,
\end{equation}
with probability at least {$1-c_{2}T\exp(-c_{3}T^{1/6})$}. Then it arrives at
\begin{equation*}
\begin{aligned}
2\sum_{k=1}^p\Delta_0^\top\left\{\G^{(0,k)}-\mathbb{E}(\G^{(0,k)})\right\}\Delta_k&\leq 2\sqrt{\frac{z_2z_4}{m_1T^{4/3}}}\sum_{k=1}^p\|\Delta_0\|_2\|\Delta_k\|_2\\
&\leq \frac{z_3}{2m_1T^{1/2}}\|\Delta_0\|_2^2+\frac{2z_2z_4(1+c)^2|\mS|}{z_3T^{5/6}}\sum_{k\in\mS}\|\Delta_k\|_2^2.
\end{aligned}
\end{equation*}
with probability at least {$1-c_{2}pT\exp(-c_{3}T^{1/6})$}, where the last inequality follows the fact that $2xy\leq ax^2+a^{-1}y^2$. Together, we have that
\begin{equation*}
\begin{aligned}
&\Delta^\top\G\Delta\geq\Delta^\top\mathbb{E}(\G)\Delta-\Delta^\top(\G-\mathbb{E}(\G))\Delta\\
\geq& \frac{z_3}{m_1T^{1/2}}\|\Delta_0\|_2^2+\frac{\gamma_{\min}}{2m_1}\sum_{k}^p\|\Delta_k\|_2^2-\frac{z_3}{2m_1T^{1/2}}\|\Delta_0\|_2^2-\frac{2z_2z_4(1+c)^2|\mS|}{z_3T^{5/6}}\sum_{k=1}^p\|\Delta_k\|_2^2\\
&-\frac{2z_2(1+c)^2|\mS|}{m_1T^{1/2}}\sum_{k\in\mathcal{E}_j}\|\Delta_k\|_2^2,
\end{aligned}
\end{equation*}
with probability at least {$1-2c_{2}pT\exp(-c_{3}T^{1/6})-c_{2}'p^2T\exp(-c_{3}'T^{1/6})$}. Since $|\mS|m_1T^{-5/6}=o(1)$ and $|\mS|T^{-1/2}=o(1)$, we have 
\begin{equation*}
\Delta^\top\G\Delta\geq \frac{z_3}{2m_1T^{1/2}}\|\Delta_0\|_2^2+\frac{\gamma_{\min}}{4m_1}\sum_{k\in\mS}\|\Delta_k\|_2^2=\frac{\gamma_{\min}}{4m_1}\sum_{k\in\mS}\|\Delta_k\|_2^2,
\end{equation*}
with probability at least {$1-2c_{2}pT\exp(-c_{3}T^{1/6})-c_{2}'p^2T\exp(-c_{3}'T^{1/6})$}.
} 
\eop

\newpage
\renewcommand{\thesection}{B}
\renewcommand{\thesubsection}{B\arabic{subsection}}
\setcounter{section}{0}
\setcounter{subsection}{0}  

\section{Computational Details}
\subsection{Selecting the numbers of B-splines}\label{sec:m0m1}
{The choices of $m_0$ and $m_1$ for constructing B-splines is based on a heuristic BIC type criterion that balances model fitting and model complexity. 
A similar procedure was considered in \citet{kozbur2020inference}.
We consider a BIC type criterion as the procedure first selects an initial set $\hat\mE^{\text{initial}}_j$. Given the initial set $\hat\mE^{\text{initial}}_j$, we no longer need to estimate a high dimensional vector $\bbeta_j$ when selecting $m_0$ and $m_1$.

To select $m_0$ and $m_1$, we consider all combinations among a set of working values for $m_0$ and $m_1$. 
First, fixing $m_0$ and $m_1$ at their respective maximum values, we estimate $\hat\bbeta_j^{\text{initial}}$ with $\eta^{\text{initial}}_j$ selected using the GIC criterion in \eqref{bic}. The initial estimate $\hat\bbeta_j^{\text{initial}}$ correspondingly gives an initial set $\hat\mE^{\text{initial}}_j$. 
Next, for each $(m_0,m_1)$ combination, $\hat\bbeta^{(m_0,m_1)}_{\hat\mE^{\text{initial}}_j}$ is estimated conditioning on the initial set $\hat\mE^{\text{initial}}_j$ and we calculate
\begin{equation}\label{bicm}
\text{BIC}(m_0,m_1)=\ell_j(\hat\bbeta^{(m_0,m_1)}_{\hat\mE^{\text{initial}}_j})\cdot\kappa_j+\hat p_{m_0,m_1}\log(T)/T,
\end{equation}
where $\ell_j(\hat\bbeta^{(m_0,m_1)}_{\hat\mE^{\text{initial}}_j})$ is the loss \eqref{loss} conditioning on $\hat\mE^{\text{initial}}_j$, $\kappa_j$ is as defined in \eqref{bic} and $\hat p_{m_0,m_1}=\Vert\hat\bbeta^{(m_0,m_1)}_{\hat\mE^{\text{initial}}_j}\Vert_0$. 
Finally, we select the $(m_0,m_1)$ combination that minimizes $\text{BIC}(m_0,m_1)$.
In practice, we find such a procedure gives a satisfactory performance, which is demonstrated numerically in the next section. 
}

\subsection{Numerical experiments on $m_0$ and $m_1$}\label{sec:varym}
{
In this section, we evaluate the performance of the procedure proposed in Section \ref{sec:m0m1} and the sensitivity of our estimation output to $m_0$ and $m_1$. We consider Setting 1.2 in Simulation 1 over 100 data replications. To estimate the background intensities and transfer functions, we use quadratic B-splines with equally spaced knots. 
We note that for quadratic B-splines, given the two boundary knots, the minimum number of splines is 3, corresponding to 0 internal knots. 
Table \ref{tab:bic} shows the BIC values calculated from \eqref{bicm} over a range of $m_0$ and $m_1$ values. 
It is seen that $m_0$ is selected to be 4 and $m_1$ is selected to be 3. 
Upon further empirical investigation, such numbers of splines are indeed sufficient to well approximate the background intensities and transfer functions in Setting 1.2.
Next, Table \ref{tab:varym} shows the estimation errors $\text{MSE}(\nu)$, $\text{MSE}(\omega)$ and selection accuracy measured in F$_1$ score over a range of $m_0$ and $m_1$ values. It is seen that $(m_0,m_1)=(4,3)$ indeed gives a good comparative performance, and the estimation and selection accuracies are not overly sensitive the choices of $m_0$ and $m_1$. 
\begin{table}[!t]
	\centering
	\begin{tabular}{c|cccccc} 
		\hline  & $m_1=3$ & $m_1=4$  & $m_1=5$  & $m_1=6$  \\ \hline
		$m_0=3$ & -38.155 & -38.131 & -38.093 & -38.038 \\
		$m_0=4$ & \textbf{-42.538} & -42.522 & -42.496 & -42.463 \\
		$m_0=5$ & -42.425 & -42.409 & -42.385 & -42.352 \\
		$m_0=6$ & -42.382 & -42.367 & -42.343 & -42.309\\	\hline
		\end{tabular}
	\caption{BIC values when $m_0,m_1=3,4,5,6$. }
	\label{tab:bic}
\end{table}

\begin{table}[!t]
\setlength{\tabcolsep}{3pt}
	\centering
	\begin{tabular}{c|cccc|cccc|cccc} \hline
		& \multicolumn{4}{c}{$\text{MSE}(\nu)$} \vline & \multicolumn{4}{c}{$\text{MSE}(\omega)$} \vline & \multicolumn{4}{c}{$\text{F}_1$ score}\\ \cline{2-13}
  & {\footnotesize $m_1=3$} & {\footnotesize $m_1=4$} & {\footnotesize $m_1=5$} & {\footnotesize $m_1=6$} 
  & {\footnotesize $m_1=3$} & {\footnotesize $m_1=4$} & {\footnotesize $m_1=5$}  & {\footnotesize $m_1=6$}
  & {\footnotesize $m_1=3$} & {\footnotesize $m_1=4$} & {\footnotesize $m_1=5$}  & {\footnotesize $m_1=6$}\\ \hline
{\footnotesize $m_0=3$}  &14.918 & 14.940 & 14.982 & 14.984 & 0.320 & 0.322 & 0.323 & 0.326 & 0.727 & 0.720 & 0.725 & 0.720 \\
{\footnotesize $m_0=4$}  & 5.053 & 5.106 & 5.156 &  5.104 & 0.359 & 0.360 & 0.362 & 0.356 & 0.909 & 0.886 & 0.859 & 0.822 \\
{\footnotesize $m_0=5$}  & 6.091 & 6.164 & 6.138 & 6.175 & 0.355 & 0.355 & 0.357 & 0.355 & 0.905 & 0.883 & 0.845 & 0.817 \\
{\footnotesize $m_0=6$}  & 5.614 & 5.786 & 5.750 & 5.704 & 0.358 & 0.360 & 0.361 & 0.357 & 0.905 & 0.879 & 0.862 &0.822 \\ \hline
	\end{tabular}
	\caption{$\text{MSE}(\nu)$, $\text{MSE}(\omega)$ and $\text{F}_1$ scores when $m_0,m_1=3,4,5,6$.}
	\label{tab:varym}
\end{table}}

\subsection{Additional simulation results}\label{sec:simadd}
{In Simulation 2, under the Erdos-Renyi network model with edge probability $p_e=0.025$, we additionally considered $p=100, 200$ and $T=20,40$. 
The background intensities and transfer functions are the same as in Setting 2. 
The B-spline bases used to estimate the background intensities and transfer functions are selected following the same procedure as in Simulation 1.
Table~\ref{tab21} reports the false negative rate (FNR), false positive rate (FPR) and F$_1$ score for \texttt{NStaHawkes}, with standard errors in the parentheses, over 100 data replications. It is seen that \texttt{NStaHawkes} achieves satisfactory performance even when $p=200$. Moreover, the selection accuracy improves with $T$. Specifically, when $T=40$, the FNR is 0 for both $p=100$ and $p=200$ while the FPR remain very close to 0.

\begin{table}[!t]
	\centering
	\begin{tabular}{c|ccc|ccc} \hline\hline
		Erdos-Renyi network & \multicolumn{3}{c}{$T=20$} \vline & \multicolumn{3}{c}{$T=40$}\\ \cline{2-7}
		$p_e=0.025$&FNR &FPR &F$_1$ score  &FNR &FPR &F$_1$ score\\ \hline
		$p=100$   & 0.003 & 0.002 & 0.931   & 0.000 & 0.001 & 0.975\\ 
		& (0.001) & (0.000) & (0.002) & (0.000) & (0.000) & (0.001) \\ \hline 
		$p=200$   & 0.036 & 0.001 & 0.926 & 0.000 & 0.001 & 0.942    \\ 
		 & (0.001) & (0.000)  & (0.001)  & (0.000) & (0.000)  & (0.001) \\ \hline \hline  
	\end{tabular}
	\caption{The false negative rate (FNR), false positive rate (FPR) and F$_1$ score for \texttt{NStaHawkes} with varying $p$ and $T$. The standard errors are shown in the parentheses.}\label{tab21}
\end{table}
}

\end{document}